\documentclass[12pt]{article}
\usepackage{graphicx}
\usepackage{XoohmE}
\usepackage{booktabs}

\def\bo{{\raise.005ex\hbox{\large$\Box$}}}

\def\Tilde#1{\widetilde{#1}}
\def\equals{\!\!\!=\!\!\!}
\def\SO{\mathop{\mathrm{SO}}\nolimits}

\def\4{\oplus}
\def\8{\otimes}

{}{}{}
\font\cfnt=lcircle10 at 9pt
\def\lplus{\mathop{\kern2pt
            \raise1.275ex\hbox to0pt{\cfnt\char"07\hss}\kern-.6pt+}}
\def\YT#1#2{\vcenter{\hbox{\vbox{\baselineskip0pt\parskip=\medskipamount%
             \def\B{$\sqcap$\llap{$\sqcup$}\kern-1.9pt}
              \def\Bd{\hbox{\kern2.4pt\raise.4pt\hbox{$\cdot$}\kern-5.7pt\B\kern0pt}}
              \def\4{\raise.25pt\hbox to0pt{\hss\kern2pt--\hss}}
              \def\Z{\hfil\vskip-5.9pt}\lineskiplimit0pt\lineskip0pt%
               \setbox0=\hbox{#1}\hsize\wd0\parindent=0pt#2}\,}}}
\def\Ft#1{\footnote{#1}}
\newdimen\parshift\parshift=\parindent
\catcode`@=11
 \long\def\@footnotetext#1{\insert\footins{\reset@font\footnotesize\interlinepenalty%
  \interfootnotelinepenalty\splittopskip\footnotesep\splitmaxdepth\dp\strutbox%
   \floatingpenalty\@MM\hsize\columnwidth\addtolength{\hsize}{-2\parindent}
    \@parboxrestore\protected@edef\@currentlabel{\csname p@footnote\endcsname\@thefnmark}
      \color@begingroup
       \@makefntext{\rule\z@\footnotesep\ignorespaces#1\@finalstrut\strutbox}
        \color@endgroup}}
 \long\def\@makefntext#1{\hglue\parshift
                         \vbox{\noindent\hb@xt@0em{\hss\@makefnmark\,}#1}}
\catcode`@=12
 \def\bpl{\Big(}
 \def\bpr{\Big)}
 \def\ba{\left(\begin{array}}
 \def\ea{\end{array}\right)}
 
 \def\der{\partial}
 \def\brr{\begin{eqnarray}}
 \def\err{\end{eqnarray}}
 
 \def\R{\mathbb R}
 \def\S{\mathbb S}

 \newcommand{\fr}[2]{{\textstyle\frac{#1}{#2}}}

 \HarvTitles 
 \seceq

\begin{document}

 \begin{flushright}
 UMDEPP 06-011
 \end{flushright}

 \begin{center}
{\Large\bf
Adinkras and the Dynamics of Superspace Prepotentials
}\\[3mm]
{\bf C.F.\,Doran$^a$, M.G.\,Faux$^b$, S.J.\,Gates, Jr.$^c$,
     T.\,H\"{u}bsch$^d$, K.M.\,Iga$^e$ and G.D.\,Landweber$^f$}\\[1mm]
{\small\it
  $^a$Department of Mathematics,
      University of Washington, Seattle, WA 98105
  {\tt  doran@math.washington.edu}
  \\
  $^b$Department of Physics,
      State University of New York, Oneonta, NY 13825\\
  {\tt  fauxmg@oneonta.edu}
  \\
  $^c$Department of Physics,
      University of Maryland, College Park, MD 20472\\
  {\tt  gatess@wam.umd.edu}
  \\
  $^d$Department of Physics and Astronomy,
      Howard University, Washington, DC 20059\\
  {\tt  thubsch@howard.edu}
  \\
  $^e$Natural Science Division,
      Pepperdine University, Malibu, CA 90263\\
  {\tt  Kevin.Iga@pepperdine.edu}
  \\
 $^f$Mathematics Department,
     University of Oregon, Eugene, OR 97403-1222\\
{\tt  greg@math.uoregon.edu}
 }\\[3mm]
{\bf ABSTRACT}\\[2mm]
\parbox{5.5in}{\parindent=2pc\noindent
We demonstrate a method for describing one-dimensional $N$-extended
supermultiplets and building
supersymmetric actions in terms of unconstrained
prepotential superfields, explicitly working with the Scalar supermultiplet.
The method uses intuitive manipulations of Adinkras and ${\cal GR}({\rm d},N)$
algebras, a variant of Clifford algebras.  In the process we clarify the
relationship between Adinkras, ${\cal GR}({\rm d},N)$ algebras, and superspace.}
\end{center}


 \section{Introduction}
During the second half of the twentieth century, quantum field theory in general
and Yang-Mills theories in particular provided a fertile and important arena for
speculating about fundamental laws of nature.  A fundamental
aspect of these theories is that the constituent fields comprise representations
of various symmetries.
In these theories, most notably the standard model of particle
physics, which has garnered spectacular experimental verification,
the elementary fields describe off-shell representations of internal
symmetries, where the important qualifier {\it off-shell}\/
indicates that the symmetry representation of the fields is
independent of their four-momentum configuration.  The most
important contemporaneous arena for attempting to reconcile particle
physics with gravitation is perturbative string theory, its
developing non-perturbative generalizations, such as {\it M}-theory,
and its effective descriptions in terms of supergravity theories,
all of which involve supersymmetry in one way or another.  However,
there remains a noteworthy fundamental structural distinction
between string-inspired physics and Yang-Mills theories concerning
the way the respective inherent symmetries are represented.
Superstring theories, and their effective descriptions in terms of
ten- or eleven-dimensional supergravity, are formulated {\it
on-shell}\/: the supersymmetry is realized only when the basic
fields satisfy classical equations of motion. This discrepancy
indicates that the current understanding of supersymmetry is as yet
incomplete, and it motivates the investigation of how off-shell
supersymmetry can be realized generally in quantum field theories.

A traditional approach to classifying irreducible supersymmetry
representations relies on the fact that all known off-shell
supermultiplets can be formulated using Salam-Strathdee superfields
subject to differential constraints and gauge transformations.
Distinctions between supermultiplets can be encoded using different
ways to pose such restrictions.  This approach has an appealing
elegance to it. Unfortunately, for cases with more than a few
supersymmetries, the range of possible constraints is large, and no
compelling rhyme nor reason has emerged as a means for organizing
these.  For instance, the importantly influential ${N} = 4$ Super
Yang-Mills theory in four dimensions has never been described
off-shell, and if there exists a constrained ${N} = 4$ superfield
description of this supermultiplet, this has not yet been
discovered. We believe that in order to resolve this dilemma, we
must go beyond ordinary superspace techniques, instead developing
new approaches based on emerging facts about the mathematical
underpinnings of supersymmetry.

In previous papers\cite{DFGHIL01,DFGHIL00,FG1} we described a
re-conceptualization for organizing the mathematics associated with
supersymmetry representation theory which is complementary to, but
logically independent of, the popular Salam-Strathdee superspace
methods.  Our approach provides fresh insight and additional
leverage from which to attack the off-shell problem. One of our
motivating desires is to 
determine an off-shell field theory description of four-dimensional
${N} = 4$ Super Yang-Mills theory and the ten- and
eleven-dimensional supergravity theories. Our investigations are
predicated on two related themes: The first purports that the
mathematical content of supersymmetry in field theories of arbitrary
spacetime dimension is fully encoded in the seemingly restricted
context of one-dimensional field theories, {\it i.e.}, within
supersymmetric quantum mechanics. The second is the observation that
the representations of one-dimensional superalgebras admit a
classification in terms of graph theory, using diagrams called
``Adinkras'' which we have been incrementally developing.

Our two most recent previous papers on this subject\cite{DFGHIL01,DFGHIL00}
have concentrated on formal mathematical
aspects of this approach and were devoted to developing precise
terminology, developing mathematical theorems associated with our
Adinkra diagrams, and describing part of a supermultiplet classification
scheme using the language of Adinkras.  In this
paper we use these techniques to elucidate instead some of the
physics of supersymmetry rather than the mathematics.  In particular we
address the question of how the the special class of irreducible
one-dimensional arbitrary $N$-extended supermultiplets known as
Scalar supermultiplets can be described in terms of unconstrained
superfields\Ft{Theorem 7.6 of Ref.\cite{DFGHIL01} guarantees this, and gives
a general algorithm to this effect; part of our present task then is to
show the concrete details of this construction.},
known as prepotentials, and how these can be used to
build supersymmetric action functionals for these supermultiplets. We
focus on Scalar supermultiplets because these provide the simplest
non-trivial context in which to illustrate our techniques.  Similar
techniques can be brought to bear on a wide class of interesting
supermultiplets; we intend to produce followup papers in the near future
addressing some of these questions.

A familiarity with the basic techniques described in\cite{DFGHIL01,FG1}, which
in turn are predicated on developments appearing in\cite{GR1,GR2,enuf}, is an
absolute prerequisite for following our subsequent discussion.  Central to these are
the relevance of ${\cal GR}({\rm d},N)$ algebras to supersymmetry representations, the
meaning and the significance of Adinkra diagrams, and the basic idea concerning
how automorphisms on the space of supermultiplets may be coded in terms of raising and
lowering operations on Adinkras.

This paper is structured as follows:

In Section~\ref{introadinkra} we describe a special class of Adinkra
diagrams, known as Base Adinkras, which are the graphical
counterparts of Clifford algebra superfields.  We explain how these
diagrams, which are in general reducible, can be used as fundamental
tools for constructing irreducible supermultiplets
via geometric vertex raising operations. In Section~\ref{topads} we
review another special class of Adinkras, known as Top Adinkras,
which are the graphical counterparts of Salam-Strathdee superfields;
these provide the connection between our technology and more
traditional techniques. We explain how Top Adinkras can be obtained
from Base Adinkras via extreme application of vertex raising
operations, and we review a relationship between superspace
differentiation and vertex raising. We explain how Top Adinkras can be
used to organize the construction of superspace operators useful for
projecting onto subspaces corresponding to irreducible
representations; this method supplies a graphical counterpart to the
organization of superspace differential projection operators. In
Section~\ref{Clifford} we review the concept of garden algebras, and
we explain in algebraic terms what is meant by a Clifford algebra
superfield. This section describes algebraically many of the
diagrammatic facts appearing in Section~\ref{introadinkra}.  In
Section~\ref{scammed} we review the definition of Scalar supermultiplets
and we develop the rudiments of an algorithm for discerning a
prepotential description of these, which is implemented in the
balance of the paper.  In Section~\ref{n2n2} we focus on the special
case of $N=2$ Scalar supermultiplets and methodically develop the
corresponding prepotential superfields and a manifestly
supersymmetric action built as a superspace integral involving
these.  This allows for a clean exposition regarding superspace
gauge structures endemic to similar prepotential descriptions in the
context of general $N$-extended supersymmetry. In Section~\ref{prepsca}
we describe the main computational result of the
paper, by generalizing the $N=2$ analysis presented in Section~\ref{n2n2} to the case of general Scalar superfields for any value
of $N$.  An important output of this analysis is that we provide the
first descriptions of 1D, $N$-arbitrary superprojectors which
naturally are associated with Scalar supermultiplets.
 We then briefly summarize our results with concluding remarks.

 \section{Adinkrammatics}
 \label{introadinkra}
The mathematical data of one-dimensional $N$-extended
supermultiplets can often be conveniently described in terms of
bipartite graphs known as Adinkras, introduced in\cite{FG1}. The
vertices of these graphs correspond to the component fields of the
supermultiplet, while the edges encode the supersymmetry
transformations. In addition, each vertex of an Adinkra comes with
an integral height assignment corresponding to twice the engineering
dimension \Ft{A field with engineering dimension $\delta$ has units
of $(\,{\rm mass}\,)^\delta$ in a system where $\hbar=c=1$.} of the
corresponding component field.  A subset of vertices, called sinks,
correspond to local maxima; these connect via edges {\em {only}} to
vertices with lower height.\Ft{In\cite{FG1}, the edges were directed
with arrows indicating the placement of time derivatives in the
supersymmetry transformations. Here, we assume that the fields have
well-defined engineering dimensions, and that the Adinkra vertices
have a height assignment. In this case, all arrows point towards the
vertex of greater height. With this convention, our use of sources
and sinks agrees with the standard usage in graph theory.} Another
subset of vertices, known as sources, correspond to local minima;
these connect via edges {\em {only}} to vertices with greater
height.  In\cite{DFGHIL01} we proved the so-called ``Hanging Gardens
Theorem'', which states that an Adinkra is fully determined by
specifying the underlying graph together with the set of sinks and
the heights of those sinks. An Adinkra can then be envisioned as a
latticework or a macram\'{e}, hanging from its sinks, and we
alternatively refer to the sinks as ``hooks''.

By performing various geometric operations on their Adinkras, we can
transform one supermultiplet into another. In earlier papers we have
described some of these operations as part of an ongoing endeavor to
describe a mathematically rigorous supersymmetry representation
theory.  In a ``vertex raising'' operation\Ft{This operation is
related to the ``automorphic duality'' transformation
of\cite{FG1,enuf}, which we also referred to as a ``vertex raise''
in\cite{DFGHIL01}. The corresponding transformation of
supermultiplets is called ``dressing'' in\cite{KRT}.}, we take a
vertex which is a local minimum or source and increase its height by
two, physically lifting it two levels on the page. In terms of the
supermultiplet, such an operation replaces the corresponding
component field with a new component field of engineering dimension
one greater. In\cite{DFGHIL01} we explained how such vertex raising
operations, when applied to superfields, can be implemented via
superspace derivatives.

A central construct in this system is a particular supermultiplet
known as the ``Clifford algebra superfield'', whose component fields
correspond to a basis of the Clifford algebra $\mathrm{Cl}(N)$. This
supermultiplet, which in general is reducible, has $2^{N-1}$ boson
fields sharing a common engineering dimension and $2^{N-1}$ fermion
fields of common engineering dimension one-half unit greater than
the bosons. All supermultiplets can be obtained from Clifford
algebra superfields via a combination of operations analogous to
vertex raising and quotients or projections derived from the
symmetries of the corresponding Adinkra.

 In the balance of this section we provide a graphical
 description of the Clifford algebra superfield and some
 related and relevant diagrammatic operations.
 It should be understood that there exists an algebraic
 context for these methods, and that a full appreciation requires a
 synergistic understanding of the diagrams and their underlying algebraic
 structure.  The algebraic context for the diagrams are reviewed below in
 Section~\ref{Clifford}.  (Subsection~\ref{ppnodes} in particular provides an especially
 useful context for understanding the ways in which vertices may be
 coalesced.)

 \subsection{The Base Adinkra}
 \label{baabaa}
The Adinkra corresponding to the Clifford algebra superfield is
called the ``Base Adinkra''. For example, the ${N} = 4$ Clifford
algebra superfield includes eight bosons sharing a common
engineering dimension and eight fermions with engineering dimension
one-half unit greater than the bosons. The ${N} = 4$ Base Adinkra
can be drawn as follows:
 \brr
 \includegraphics[width=3in]{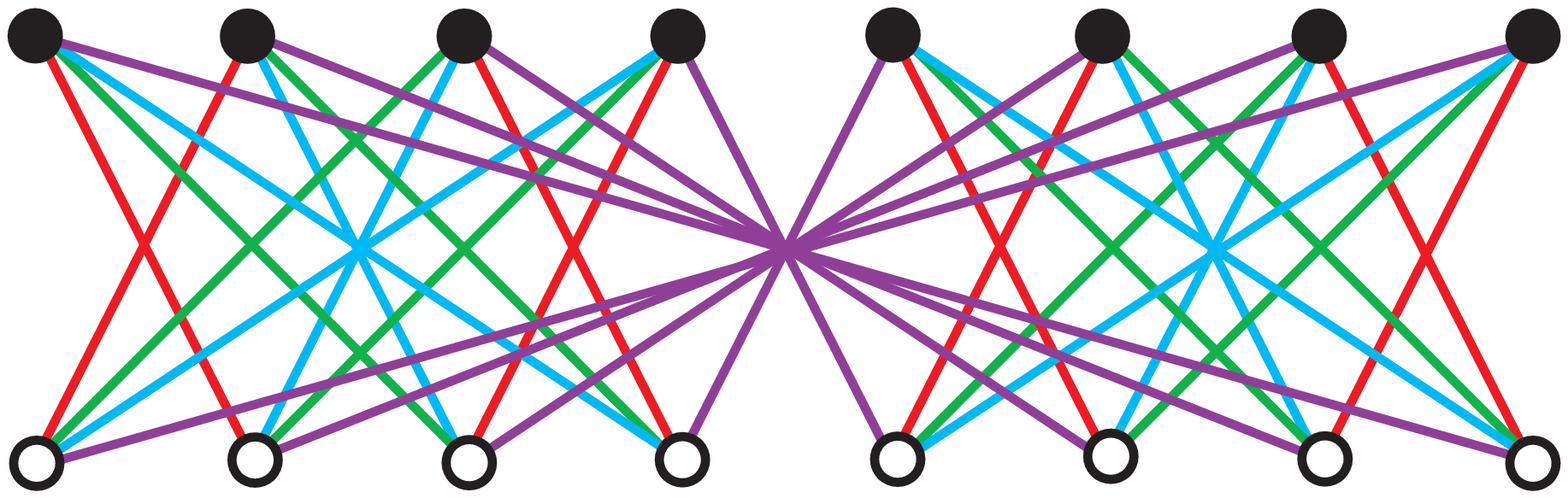} 
 \label{acc4}\err
where each of the four supersymmetries corresponds to a unique edge
color, each bosonic vertex corresponds to a boson field while
each fermionic vertex corresponds to a fermion field. We use
throughout this paper the convention promulgated in\cite{DFGHIL01}
whereby the vertical placement of Adinkra vertices correlates
faithfully with the height assignment. Thus, higher components, having larger
engineering dimension, appear closer to the top of the diagram.  The
existence of the height assignment to each vertex implies that
Adinkras also provide a natural realization of an abelian symmetry,
whose generator (denoted by {\bf d}) is realized on each vertex by
multiplication of the vertex by one-half times the height assignment
of the vertex.  This is the basis of the filtration discussed in\cite{DFGHIL00}.

In~(\ref{acc4}), a left-right symmetry is apparent: the Adinkra
remains unchanged if it is reflected about a vertical axis passing
through its center.  This symmetry of the Adinkra gives rise to a
projection of the $N=4$ Clifford algebra superfield onto an
irreducible submultiplet corresponding to the  $N = 4$ Scalar
Adinkra. This projection is described in detail in Subsection
\ref{irreps} below.

Another way to draw the ${N} = 4$ Base Adinkra is as follows:
 \brr
 \includegraphics[width=3in]{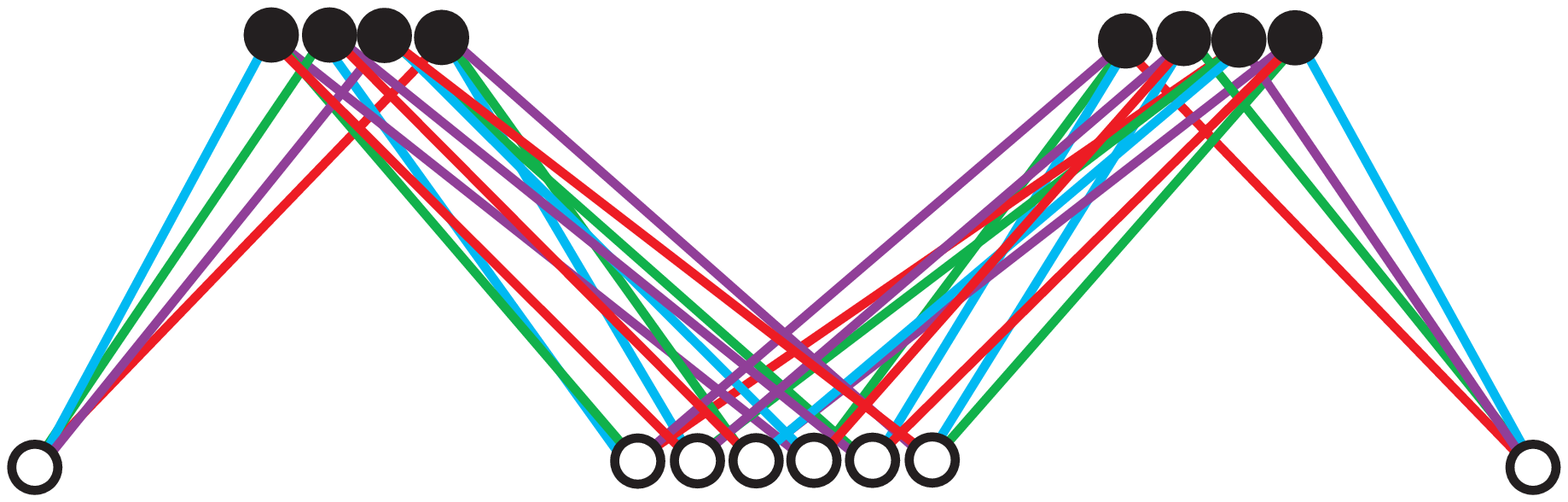}
 \label{fold4}\err
This can be obtained from~(\ref{acc4}) by moving vertices while
maintaining the inter-vertex edge connections.  This second
presentation of the Base Adinkra can be obtained by selecting one
bosonic vertex of the Base Adinkra, call it $\phi$, and putting it
on the left side of the diagram.  We then gather together those
vertices in the Base Adinkra that are one edge away from $\phi$, and
these are placed at the same height as they were in the Base Adinkra
(one level above $\phi$) but slightly to the right of $\phi$. We
then take the vertices that are two edges away from $\phi$, and
these six vertices are placed slightly to the right of that, at the
correct height (the same height as $\phi$), and so on.

The meaning of this arrangement comes about when we consider that
each vertex in the Base Adinkra can be obtained by applying a finite
antisymmetrized sequence of supersymmetry generators $Q_I$ to $\phi$.
Thus, we can label the vertices that are one edge away from $\phi$ using
a single index $I$ ranging from $1$ to $N$.  Likewise, the vertices that
are two edges away from $\phi$ can be labelled with two indices that
are antisymmetrized, and so on.  In general, vertices that are $p$ edges
away from $\phi$ are labelled as antisymmetric $p$-tensors with indices
ranging from $1$ to $N$.

This is not merely suggestive formalism.  Once we fix the field
$\phi$, the supersymmetry generators map $\phi$ to other fields.
The group $\SO(N)$ of $R$-symmetries, which acts naturally on the
$N$ supersymmetry generators, thus also acts on the fields. The
orbits are precisely the clumps of vertices in the above diagram,
and form a representation of $\SO(N)$ that is isomorphic to the
corresponding exterior tensor power of the standard representation
of $\SO(N)$.

We can abbreviate~(\ref{fold4}) as
 \brr
 \includegraphics[width=1.5in]{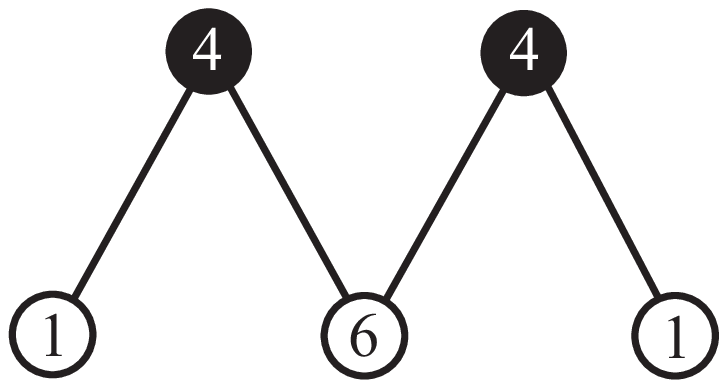}
 \label{C4}\err
Here the vertex multiplicity is indicated by a numeral, and the edges
have been coalesced.  (In general, a black edge encodes the bundled
action of $N$ supersymmetries in a manner which is well defined.)%
\Ft{See~(\ref{generic}) below for an algebraic clarification
of this point.}

More generally, for any $N$, the Base Adinkra
characteristically admits an accordion-like presentation, for example
as shown here:
 \brr
 \includegraphics[width=3.3in]{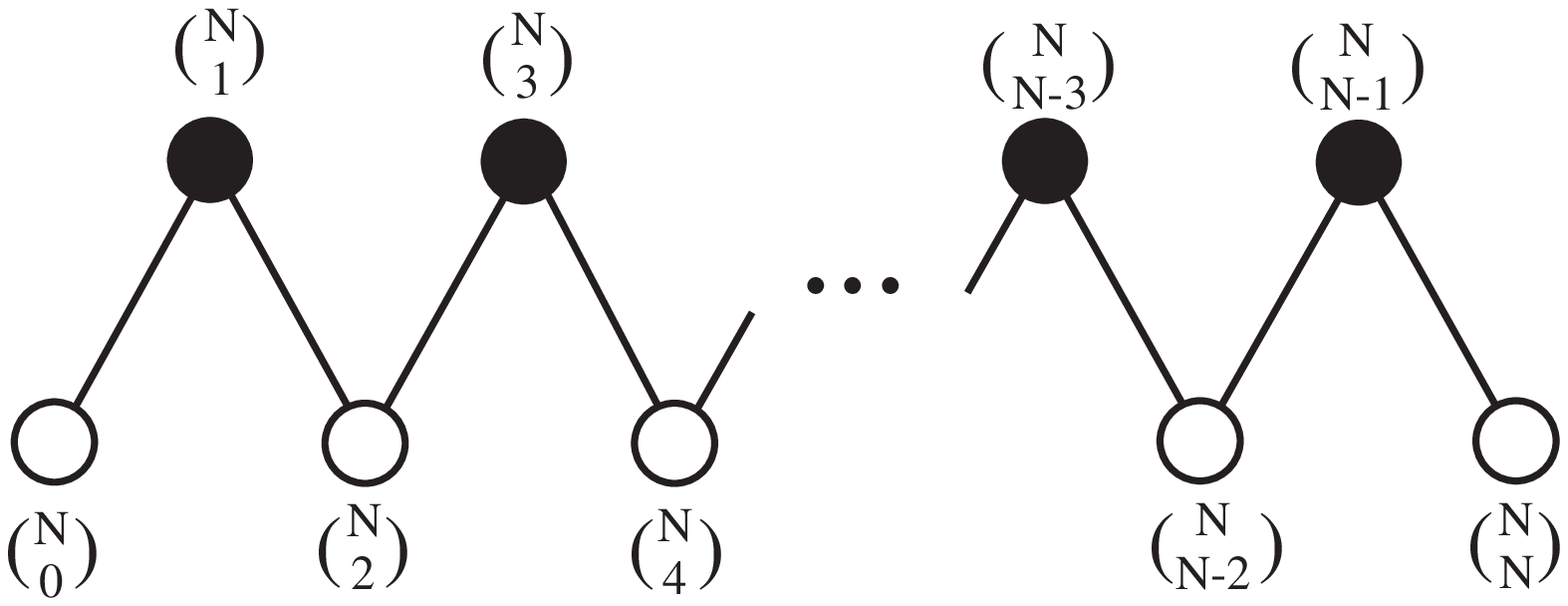}
 \label{BN}\err
where the indicated vertex multiplicities are the binomial
coefficients ${N\choose p}$, where $p=0,1,\dots,N$ sequentially
labels the compound vertices starting from the left.  This reflects
the fact that these vertices coincide with rank $p$ antisymmetric
tensor representations of $\SO(N)$, each of which describes
${N\choose p}$ degrees of freedom.  For the sake of brevity, we
refer to such representations as ``$p$-forms''. Thus, the left-most
bosonic vertex (white circle) is a zero-form, and the other vertices
are collected into a chain, such that the second compound vertex is
a one-form fermion, the third compound vertex is a two-form boson,
and so forth.  The Adinkra shown in~(\ref{BN}) is the general Base
Adinkra for cases in which $N$ is even; in cases where $N$ is odd,
the chain terminates on a fermionic vertex (black circle) rather
than on a bosonic vertex (white circle),
 \brr
 \includegraphics[width=3.3in]{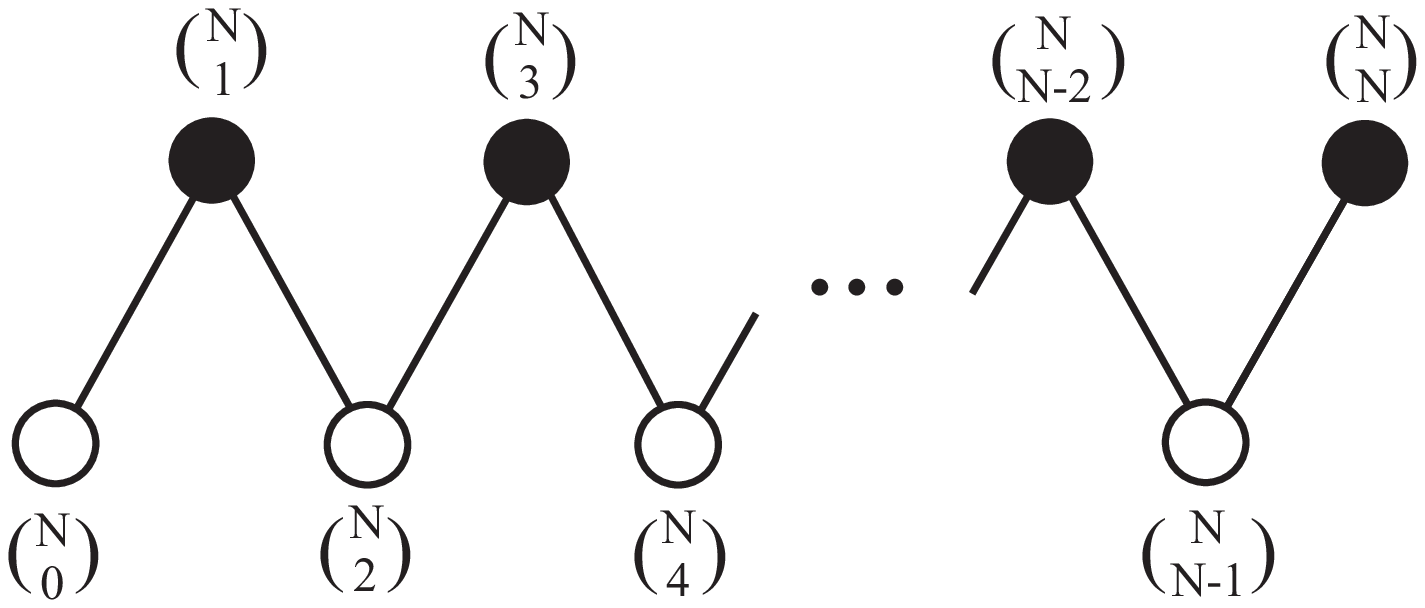} 
 \label{BNodd}\err
 This Adinkra codifies particular transformation rules, which are exhibited
 below in Subsection~\ref{CAS}.

 \subsection{The Dual Base Adinkra}
We have explained how the components of the Base Adinkra can be organized into
$p$-form representations of $\SO(N)$, such that the bosons are even-forms and the fermions are
odd-forms.  Alternatively, the vertices of the Base Adinkra can be organized differently
such that the fermions are even-forms and the bosons are odd-forms.  To see this, we
start with our original rendering of the Base Adinkra, shown in~(\ref{acc4}), but this time
re-organize the same vertices into the following configuration,
 \brr
 \includegraphics[width=3in]{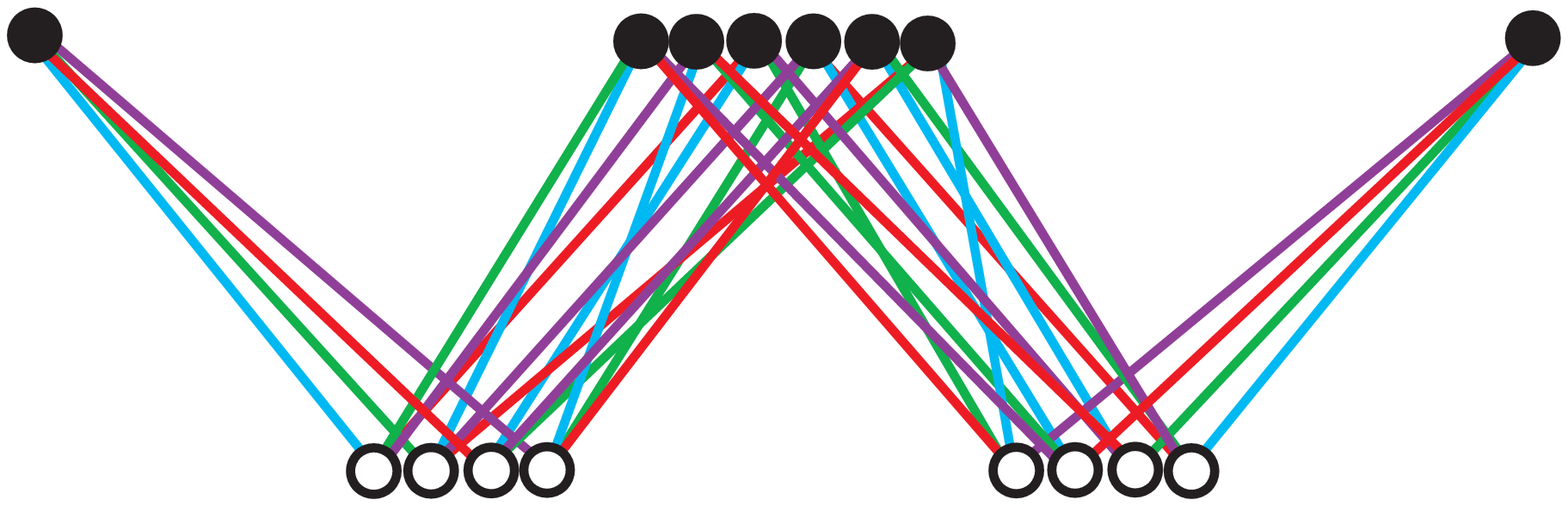} 
 \label{shuffle4}
 \err
This Adinkra is equivalent to~(\ref{acc4}) and also
to~(\ref{fold4}), and these can be transformed into each other
merely by grouping the vertices together in different ways. In
particular, (\ref{shuffle4}) is obtained from (\ref{acc4}) by
starting with a $0$-form fermion, and collecting vertices according
to their distance from this fermion. We can
abbreviate~(\ref{shuffle4}) as
 \brr
 \includegraphics[width=1.5in]{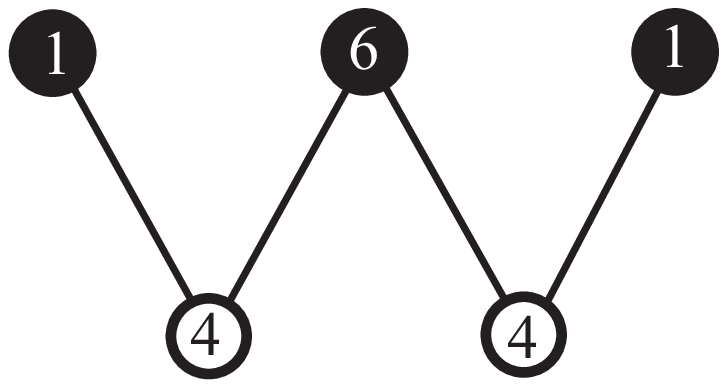} 
 \label{C42}\err
The difference between this Adinkra and~(\ref{C4}) lies in the way the $\SO(N)$ structure
has been imposed on the skeletal Adinkra~(\ref{acc4}).  A similar alternative grouping can
be applied to the Base Adinkra for any $N$, resulting in the following Dual Base Adinkra for even $N$:
 \brr
 \includegraphics[width=3.3in]{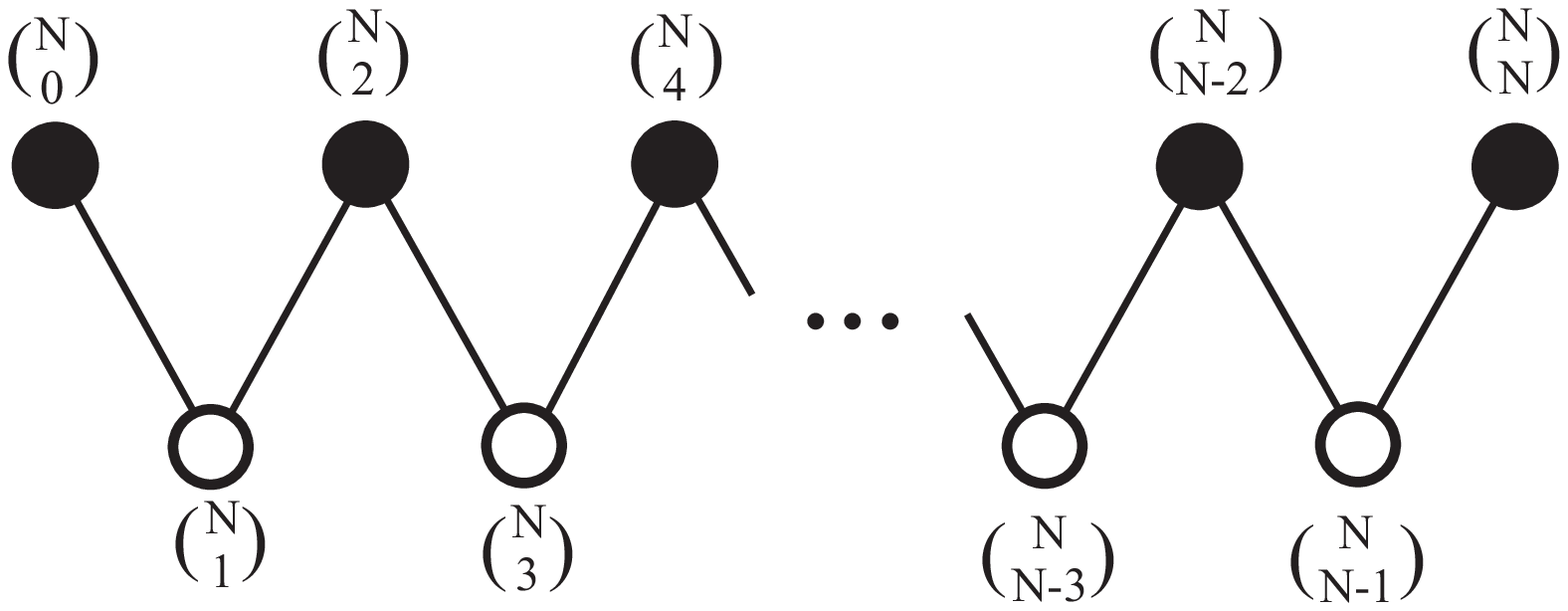} 
 \label{cz2}\err
 The difference between~(\ref{cz2}) and~(\ref{BN}) lies in
 the manner in which the vertices have been grouped into
 representations of $\SO(N)$.  An algebraic description of this
 concept is described in Subsection~\ref{coucou}.

 \subsection{The Conjugate Base Adinkra and its dual}
Every supermultiplet has a counterpart obtained by toggling the statistics of all of its
component fields, replacing each boson with a fermion, and vice-versa; this operation
is known as a Klein flip\Ft{In the mathematical supersymmetry literature\cite{DF,PD}, this is called parity reversal and is denoted $\Pi$.} (see also\cite{GK}).  The
Klein flipped analog of the Clifford algebra superfield is
called the Conjugate Clifford superfield; the corresponding Adinkra, called the Conjugate
Base Adinkra, is (for $N$ even)
 \brr
 \includegraphics[width=3.3in]{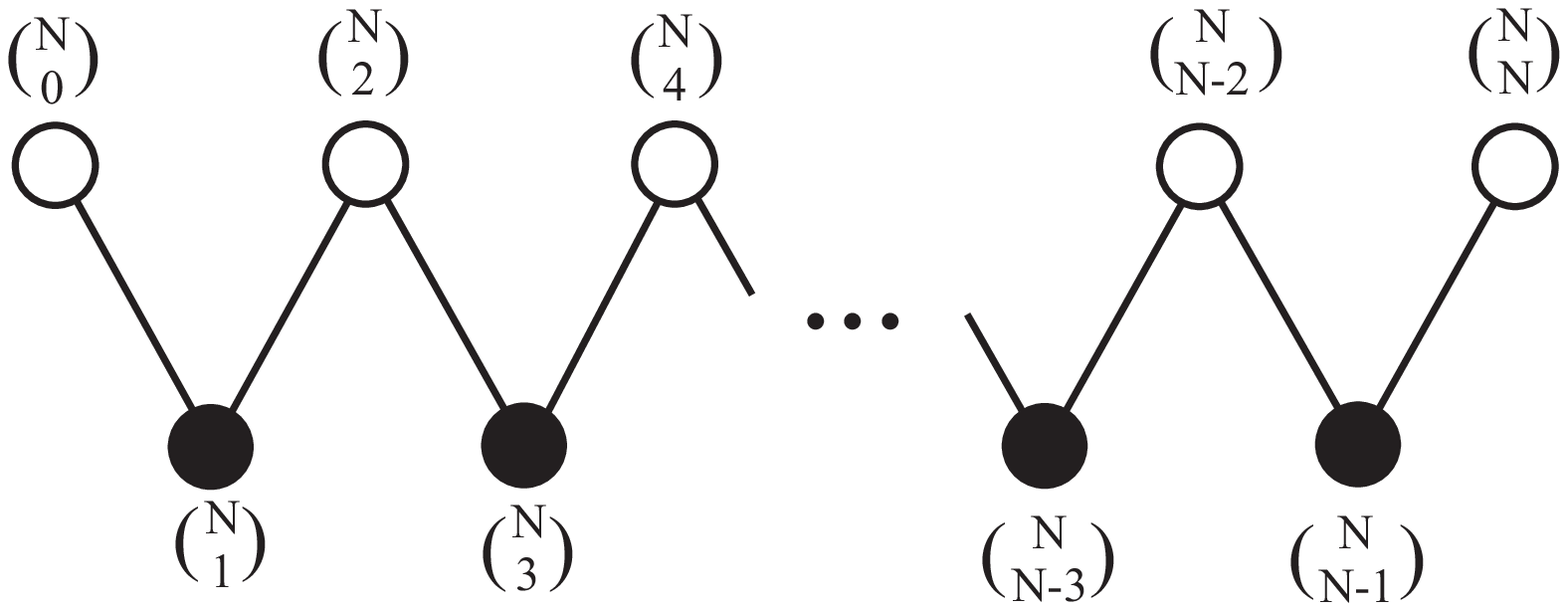} 
 \label{CBN}\err
This is obtained from~(\ref{cz2}) by replacing all boson vertices with fermion vertices and
vice-versa.\Ft{By using~(\ref{BN}) as an intermediary,  we can describe another operative way to connect~(\ref{cz2}) with~(\ref{CBN}). According to
this alternate scheme, we first transform~(\ref{cz2}) into~(\ref{BN}) by merely re-grouping vertices in the manner described above.  Then, via a sequence of
vertex raises, we can map~(\ref{BN}) into~(\ref{CBN}).}

Finally, there is an alternate way to group the vertices of the Conjugate Base Adinkra into
$\SO(N)$ tensors; in a manner similar to that described above, we can re-group the
compound vertices in~(\ref{CBN}) so that the bosons are  odd-forms while the fermions are
even-forms (for even $N$):
 \brr
 \includegraphics[width=3.3in]{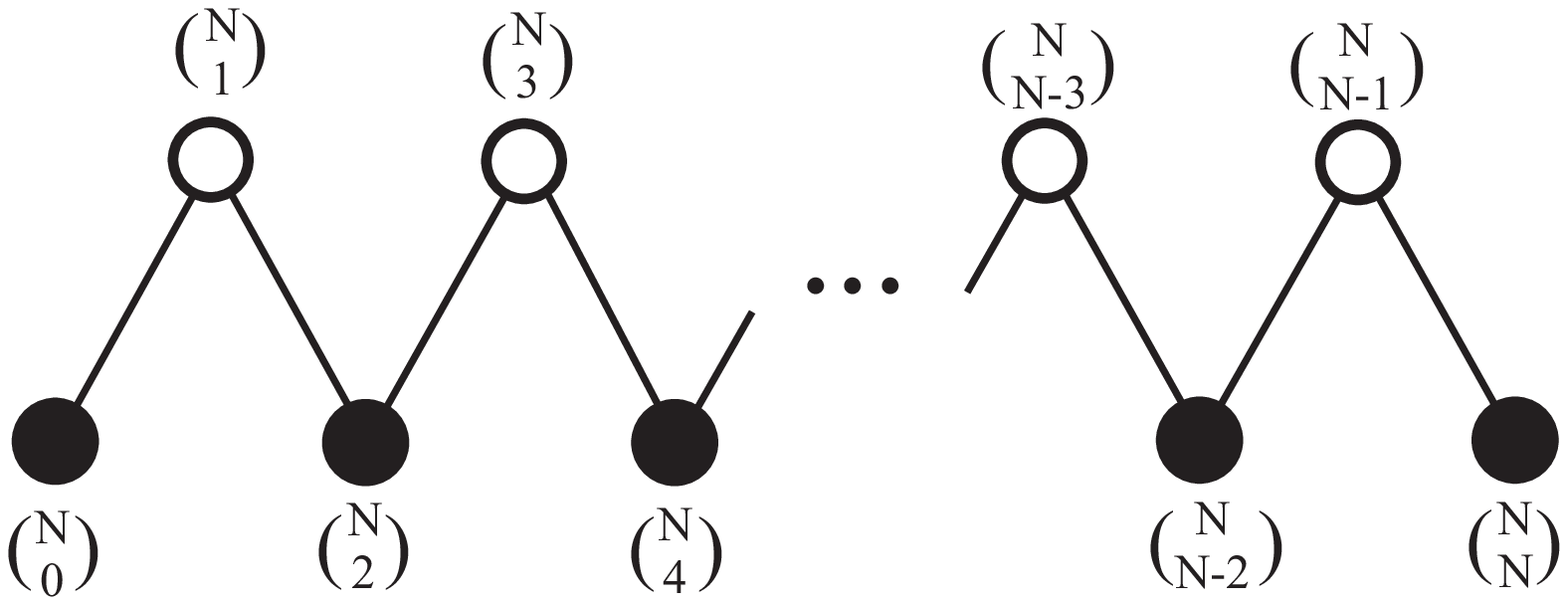} 
 \label{CBN2}\err
 This Adinkra may also be obtained as the Klein flip of~(\ref{BN}).
 \begin{figure}
 \begin{center}
 \includegraphics[width=5in]{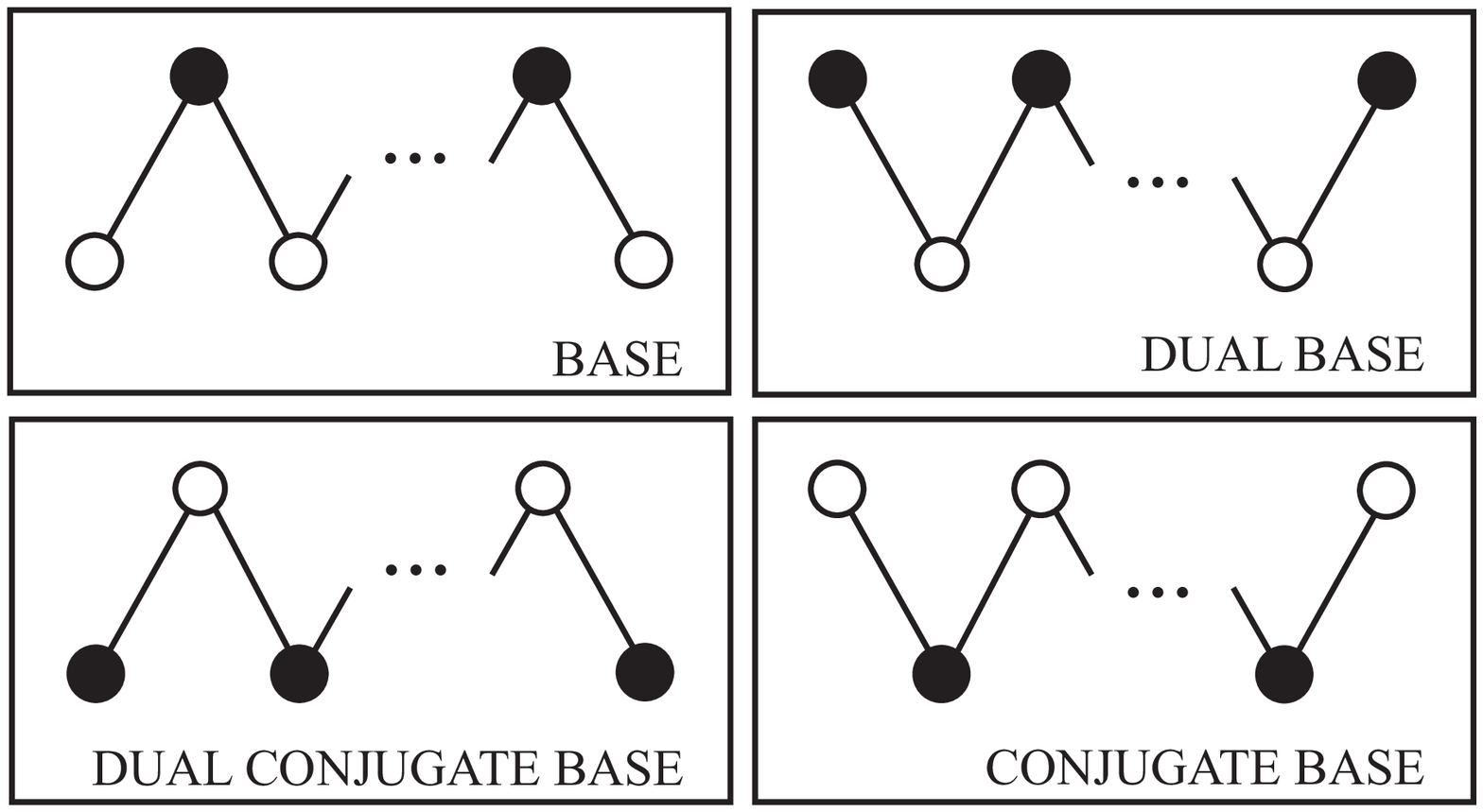}
 \caption{The Base Adinkra describes a zig-zag chain of $\SO(N)$ tensor fields
 starting with a lower component zero-form boson.  The Dual Base, the Conjugate
 Base, and the Dual Conjugate Base Adinkras describe analogous constructions
 distinguished by whether the zero-form is a boson or a fermion or by whether it
 is the bosons or fermions which have lower height. We display here the Adinkras
 for even $N$; the ones for odd $N$ are analogous, starting with the Base
 Adinkra~(\ref{BNodd}).}
 \label{chart}
 \end{center}
 \end{figure}
The Base Adinkra~(\ref{BN}), its dual~(\ref{cz2}), its conjugate~(\ref{CBN}), and the Conjugate
Base Adinkra~(\ref{CBN2}) are depicted together in Figure \ref{chart}.  Each of these
constructions can be used as a launching point for describing more general supermultiplets.

 \subsection{The Scalar Adinkra}
 \label{irreps}
 The Base Adinkra is central concept in the representation of one-dimensional $N$-extended supersymmetry.
 Many of the irreducible representations can be obtained from the
 Base Adinkra by applying various operations.  One such operation
 was mentioned above in Subsection~\ref{baabaa}: by imposing consistent vertex
 identifications we can project a given Adinkra onto
 sub-Adinkras corresponding to smaller representations.  For
 example, we can make pairwise identifications of those vertices in
~(\ref{acc4}) mapped into each other by a left-right folding operation.
 What results is the following Adinkra:
\brr
\includegraphics[width=1.8in]{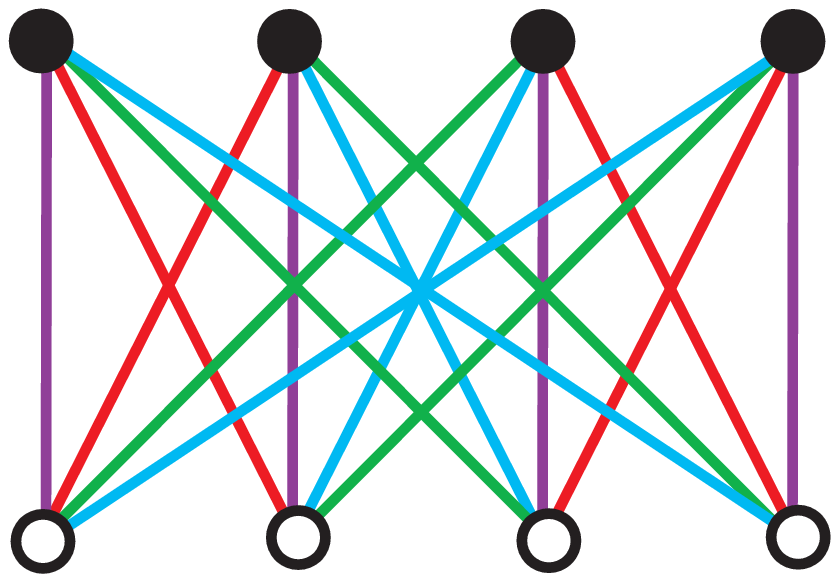} 
\label{ScalarA}\err which corresponds to the irreducible ${N} = 4$
Scalar supermultiplet.  A geometric way to understand this
projection is to identify the underlying graph of the $N=4$ Base
Adinkra with the vertices and edges of a $4$-dimensional hypercube
(or tesseract).  We then take a quotient of this hypercube,
identifying antipodal vertices and edges (see\cite{FG1}). In
general, the relationship between the Clifford algebra superfield
and the Scalar superfield, for each value of $N$, can be understood
in terms of quotients of cubical Adinkras.

  \subsection{Node raising and other supermultiplets}
The generalized Base Adinkras, shown in Figure \ref{chart}, and the
Scalar Adinkra, shown in~(\ref{ScalarA}), share the feature that
their vertices span only two different height assignments, or
equivalently that the corresponding supermultiplets have component
fields of only two engineering dimensions. To construct Adinkras
corresponding to supermultiplets with fields of more than two
engineering dimensions, we can start with these Adinkras and operate
on them by vertex raising operations. To raise a vertex in an
Adinkra, we take a source vertex and increase its height assignment
by two. At the level of supermultiplets, a vertex raising operation
replaces a component field with a new component field given by the
$\tau$ derivative of the original component field.  For example, if
a given vertex corresponds to the field $\phi(\tau)$, then we can
define a new field via $\tilde{\phi}:=\der_\tau\phi$.
 If $\phi$ corresponds to a
 source vertex, then the supersymmetry transformations continue to
 involve only local superspace operators, and we thereby obtain
 a new supermultiplet.
Since the operator $\der_\tau$
carries one unit of engineering dimension, it follows that
$\tilde{\phi}(\tau)$ describes a  higher component.
In\cite{DFGHIL01} we discussed these operations at length,
explaining relationship between vertex raising and superspace
derivation.

For example, suppose we start with the Base Adinkra, as drawn in~(\ref{BN}).  If we
raise the fifth bosonic vertex, counting from the left, what results is the following new
Adinkra:
  \brr
 \includegraphics[width=3.3in]{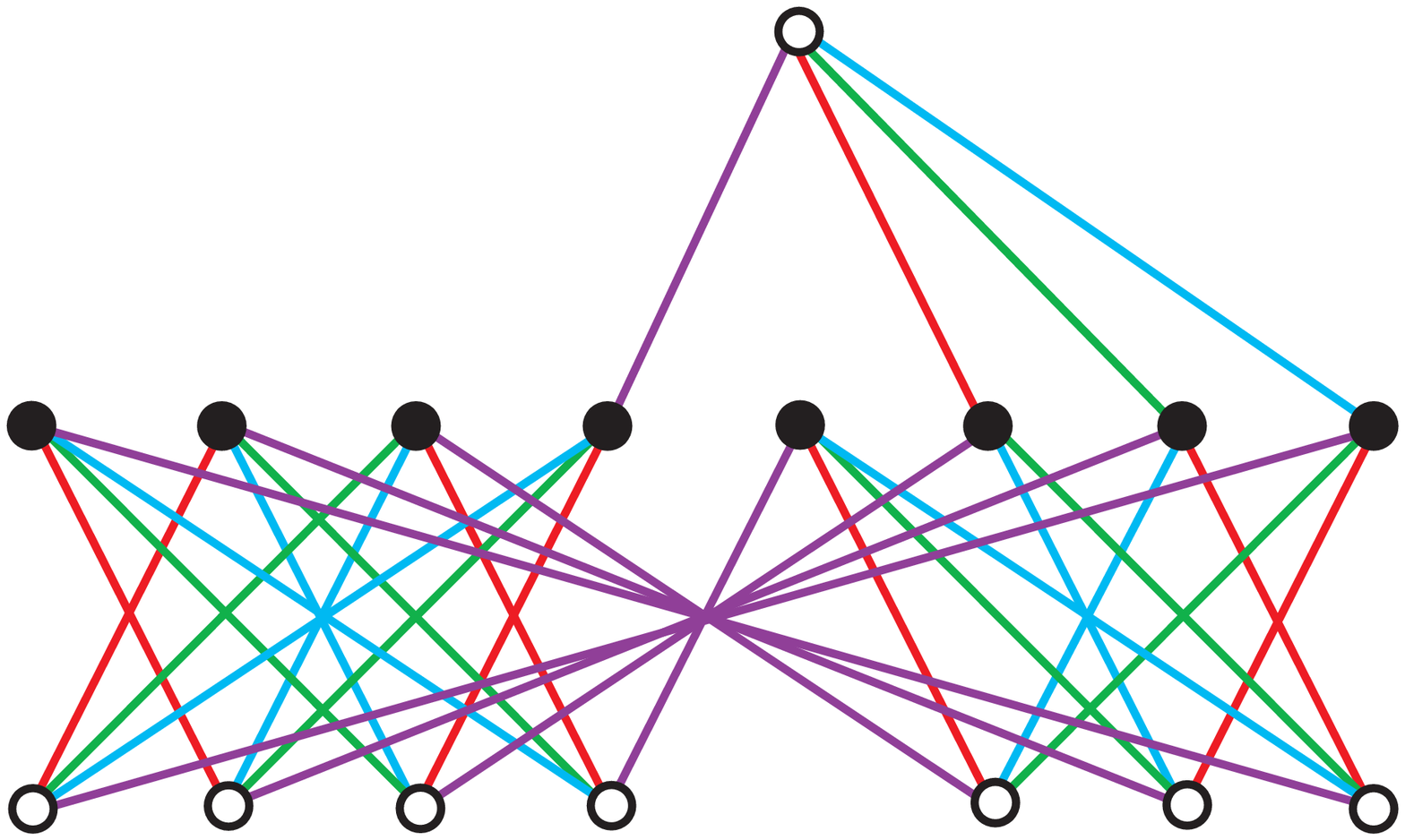} 
 \label{tent}\err
This Adinkra describes a supermultiplet which is distinct from the Clifford algebra
superfield, as evidenced by the fact that three different height assignments
are represented.

If we raise {\em en masse} a collection of vertices which have been coalesced into an
$\SO(N)$ $p$-form, then the resulting Adinkra respects the $\SO(N)$ structure
in the sense that the components of each $\SO(N)$ tensor continue to share
a common engineering dimension.  For instance, if we start with the ${N} = 4$
Base Adinkra as shown in~(\ref{C4}), we can raise the multiplicity-six compound
two-form vertex to obtain
  \brr
 \includegraphics[width=1.5in]{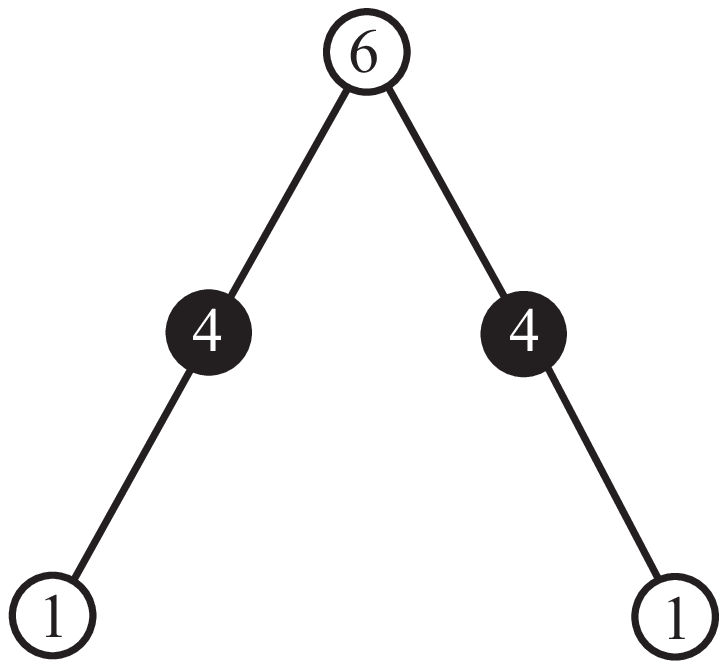} 
 \err
As another possibility, we could start again with the Base Adinkra~(\ref{C4}) and
 raise the singlet four-form vertex, to obtain
 \brr
 \includegraphics[width=1.5in]{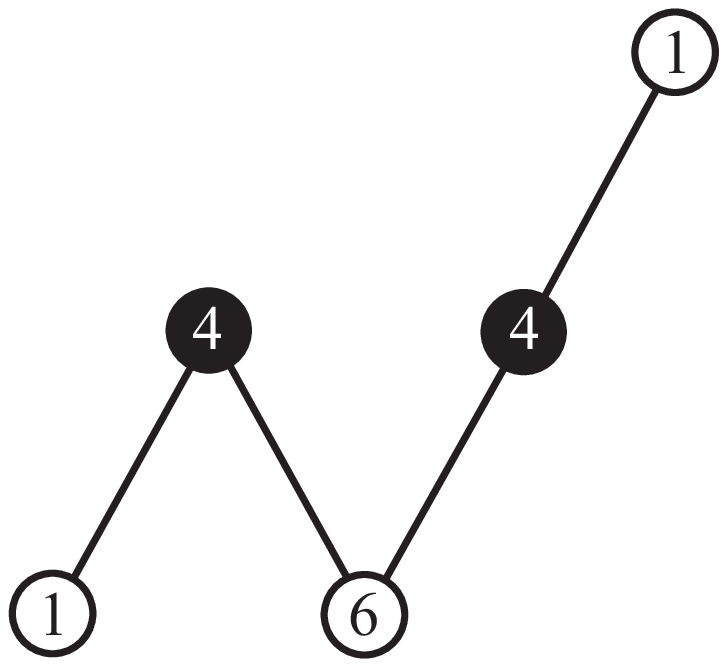} 
 \err
There are many other possibilities.  A subset of the possible
vertex raising operations maintains the height-equivalence of all
components of each $\SO(N)$ tensor, while the complementary set
breaks this height-equivalence feature.  As an example of a raising
operation in the latter class, we could start with~(\ref{C4}), and
then raise the multiplicity-one four-form vertex and also one of the
vertices out of the six in the multiplicity-six compound vertex, as
follows:
 \brr
 \includegraphics[width=1.5in]{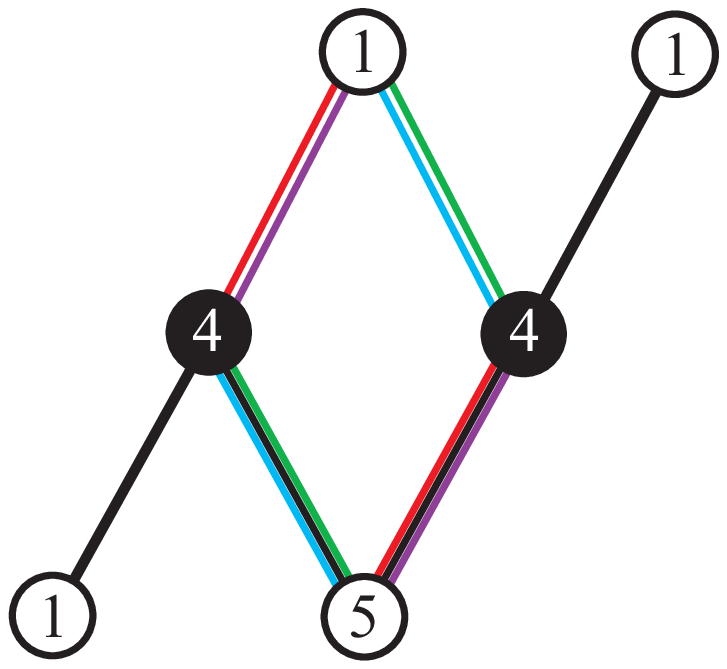} 
 \err
In this Adinkra, the black edges correspond, as above, to a bundling
of all four supersymmetries.  However, the bicolored edges indicate
vertex interrelationships involving only two of the four
supersymmetries.  The edges with combined black and colored edges
describe a bundling of all four supersymmetries, but some of the
implicit connectivity, namely that associated with the bicolored
edges, is missing from this bundling.  Here we  see that the six
components of the bosonic two-form do not share a common height
assignment.  Thus, the two-form does not have a collectively
unambiguous engineering dimension.  In this case we say that the
supermultiplet has a {\it skew} $R$-charge, as opposed to a conventional
$R$-charge.

Recall that the discussion of $p$-forms in Adinkras depends on choosing
a starting vertex to be the $0$-form, and note that the skewness of the
$R$-charge may depend on this choice of $0$-form vertex.  Indeed, the
$R$-symmetries act on the supersymmetry generators, not the supermultiplet, unless
we fix a choice of a vertex.  We will therefore
define a supermultiplet or Adinkra to have a conventional $R$-charge if there exists some
choice of $0$-form vertex so that the vertices for each $p$-form all have the
same engineering dimension, and a supermultiplet is said to be skew otherwise.

In the case of an Adinkra with a conventional
$R$-charge, the generator {\bf d} introduced in Subsection~\ref{baabaa}
commutes with the generator of the $\SO(N)$ $R$-charge.
For any Adinkra with a skew $R$-charge, these generators do not
commute.

Also note that the presence of a skew $R$-charge does not preclude the existence
of an invariant functional, built using the components of such a supermultiplet, which
is both supersymmetric and $\SO(N)$-invariant.  Multiplets having a conventional
$R$-charge form a class which is distinct from those in which the $\SO(N)$-structure is skew, and may prove interesting to model building.  However,
in the balance of this paper we consider only supermultiplets having conventional
$R$-charge.

 \section{Top Adinkras and Salam-Strathdee Superfields}
 \label{topads}
The Base Adinkra and its kin described in the previous section comprise an
extreme class of supermultiplets in the sense that the component fields span a
minimal number of distinct engineering dimensions, namely two.  Another
extreme class of supermultiplets are those described by a connected Adinkra
involving $2^N$ total component fields (vertices) spanning a {\it maximal}
number of distinct height assignments.  This class of supermultiplets can be
obtained from the generalized Base Adinkras by raising vertices until the
chains depicted in Figure \ref{chart} are fully extended rather than maximally
compressed.  For instance, if we start with the Base Adinkra~(\ref{BN}), we
can lift vertices while maintaining $R$-charge until we obtain the following
Adinkra:
 \brr \includegraphics[width=.6in]{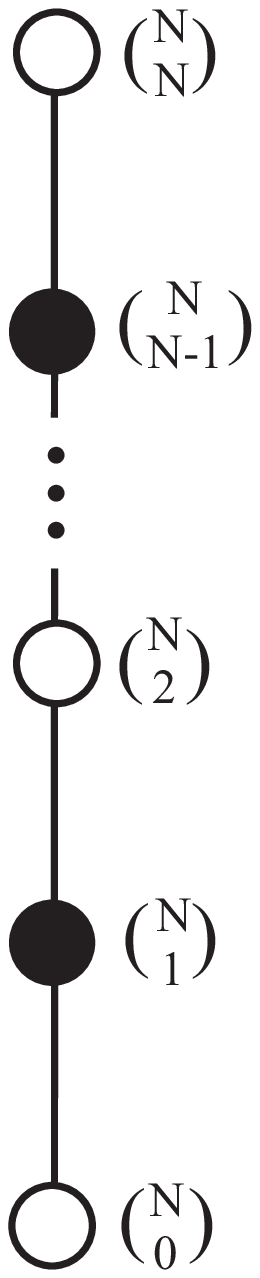} 
 \label{TN}\err
This is the unique fully-extended Adinkra having a zero-form boson as its
lowest component, and is called the Top Adinkra.  This supermultiplet spans $N+1$
different height assignments, and corresponds directly with the scalar
Salam-Strathdee superfield, $\Phi(\tau,\theta^I)$, where $\theta^I$ are the
fermionic superspace coordinates.  The lowest vertex in~(\ref{TN}) corresponds
to the lowest component of the superfield, $\Phi\,|$, {\it i.e.}, that component
which survives projection to the $\theta^I\to 0$ submanifold of the superspace,
sometimes called the {\it body} of the superfield.  The next highest vertices in~(\ref{TN}) correspond to the body of the derivative superfield, $D_I\Phi\,|$ . In
general, the $p$-form vertices in a Top Adinkra are proportional to $D_{[I_1}\cdots
D_{I_p]}\Phi\,|$.\Ft{A superspace derivative is defined as $D_I:=
\der_I+i\,\theta_I\,\der_\tau$, whereby $\{D_I\,,\,D_J\}=2\,i\,\der_\tau$\,.  It follows
that the product of two superspace derivatives can be
decomposed as $D_I\,D_J=D_{[I}\,D_{J]}+i\,\delta_{IJ}\,\der_\tau$.  Because of
this,  a complete set of differential operators on superspace
is generated by the antisymmetric operator products
$D_{[I_1}\cdots D_{I_n]}$, and by the time derivatives $\der_\tau^p$ following
the action of these.}
 The component transformation rules associated with the generators
of supersymmetry transformations on superspace are identical with those codified
by the Adinkra, as spelled out in\cite{DFGHIL01,DFGHIL00,FG1}.

 Starting with the Top Adinkra~(\ref{TN}), one can raise the lowermost vertex
 to obtain the following distinct Adinkra:
 \brr \includegraphics[width=1.4in]{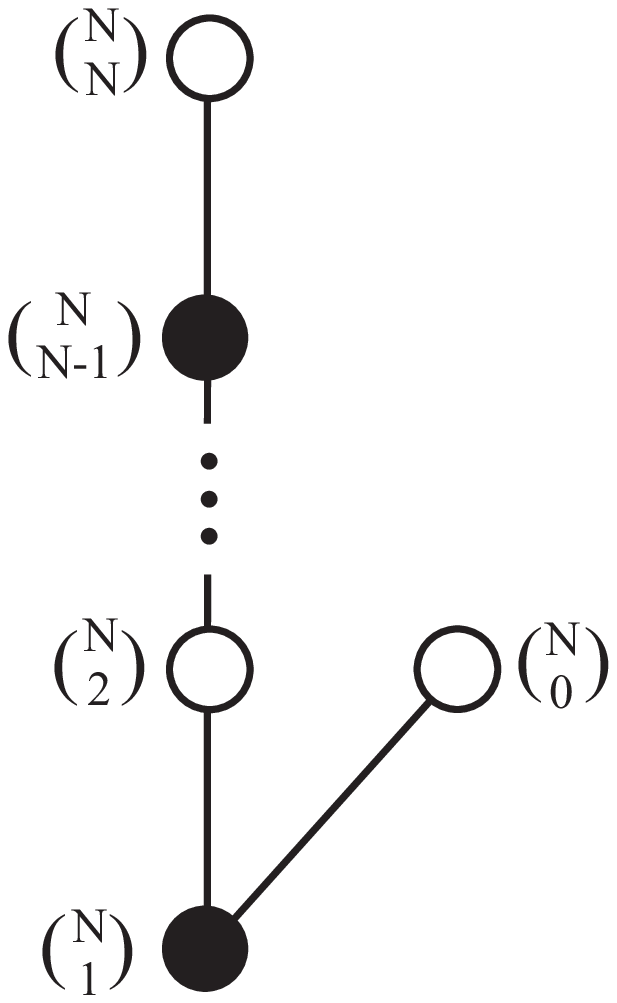} 
 \label{Lift1}\err
This new Adinkra corresponds to the superderivative superfield $D_I\Phi$,
modulo the zero-mode of the singlet field labeled ``${N\choose0}$''.
 In this way, the superspace derivative
operation is mirrored on the Top Adinkra by a vertex raising.  To be more precise,
the singlet vertex ``${N\choose0}$'' in the Adinkra~(\ref{Lift1}) describes the
 $\tau$ derivative of the corresponding singlet field in $\Phi$, its lowest component.
 This begs an interesting and relevant question:
Does there exist a superspace description of the supermultiplet
described by~(\ref{Lift1}) in which the Adinkra vertices correlate
one-to-one, without derivatives, to the components of some
unconstrained superfields?  As it turns out, such a construction
does exist, is related to the result of Theorem 7.6 of
Ref.\cite{DFGHIL01}, and we describe it in detail below.  This
construction requires not one, but two unconstrained superfields,
$\Phi_1$ and $\Phi_2$, called ``prepotentials'' and which are
associated to the two local maxima (sinks) of the
Adinkra~(\ref{Lift1}): those labeled ``${N\choose N}$'' and
``${N\choose0}$''.
 A particular linear combination of superspace derivatives of these contains precisely the supermultiplet described
by~(\ref{Lift1}).  This linear combination comprises a superfield
subject to a constraint. There are twice as many component fields
collectively described by $\Phi_1$ and $\Phi_2$ as there are
described by~(\ref{Lift1}). The excess component fields correspond
to {\it gauge} degrees of freedom and do not appear
in~(\ref{Lift1}).

 Next, starting with~(\ref{Lift1}) we can raise the one-form vertices to obtain:
 \brr \includegraphics[width=1.4in]{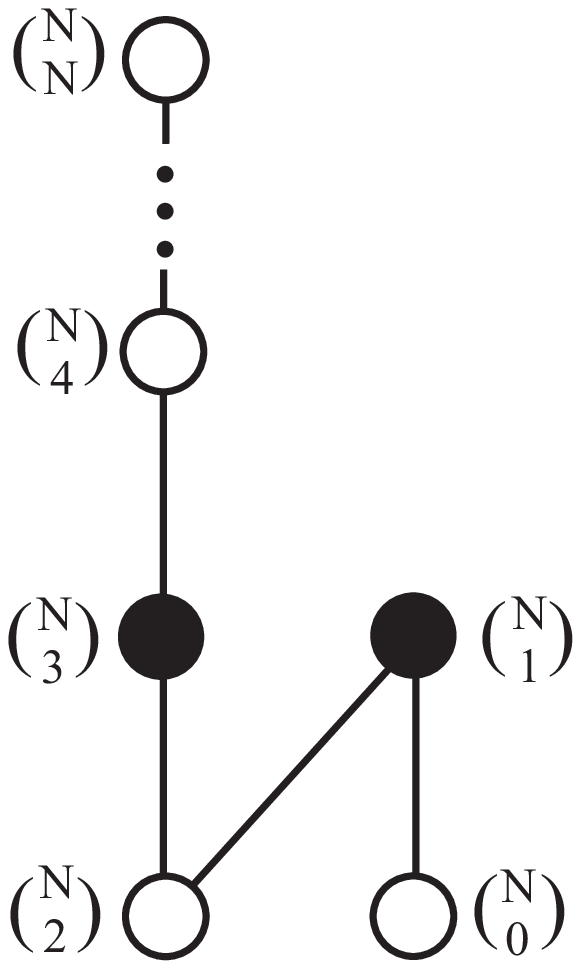} 
 \label{Lift2}\err
This Adinkra does not correspond directly to a single unconstrained
superfield. What we mean by this is that the particular component
transformation rules encoded by~(\ref{Lift2}) do not coincide with
the transformation rules associated with the components of any
particular unconstrained superfield for which the superfield
components and the Adinkra vertices are in one-to-one
correspondence.  This does not mean that the supermultiplet
described by~(\ref{Lift2}) does not have a superspace
interpretation. Indeed, the source-sink reversal of Theorem~7.6 of
Ref.\cite{DFGHIL01} guarantees that a corresponding superfield
exists, and provides a general algorithm for its determination.  In
this case, for example, the supermultiplet in question can be
described by an unconstrained scalar superfield along with a set of
$N$ unconstrained fermionic superfields, respectively corresponding
to the vertex labeled ``${N\choose N}$'' and the multiple vertex
labeled ``${N\choose1}$''. These $N+1$ total unconstrained
prepotential superfields once again correspond to the $N+1$ local
maxima (sinks) in the Adinkra~(\ref{Lift2}), and they involve a
total of $(N+1)\cdot 2^N$ total components, significantly more than
the $2^N$ vertices appearing in~(\ref{Lift2}). The extra degrees of
freedom appearing in the superspace description are associated with
gauge degrees of freedom; there exist linear combinations of the
prepotential superfields and derivatives thereof which involve only
those degrees of freedom corresponding to the Adinkra vertices
in~(\ref{Lift2}). The superfield corresponding to such a linear
combination is subject to constraints. It has been a historically
interesting question to attempt to classify the possible realizable
superfield constraints which give rise to irreducible
supermultiplets. The paradigm we are espousing speaks pointedly to
this endeavor. We expand on these ideas, and include
 relevant algebraic details, later in this paper.

Starting with~(\ref{Lift2}), one can raise the singlet vertex labeled
``${N\choose0}$'' to obtain:
 \brr \includegraphics[width=1.4in]{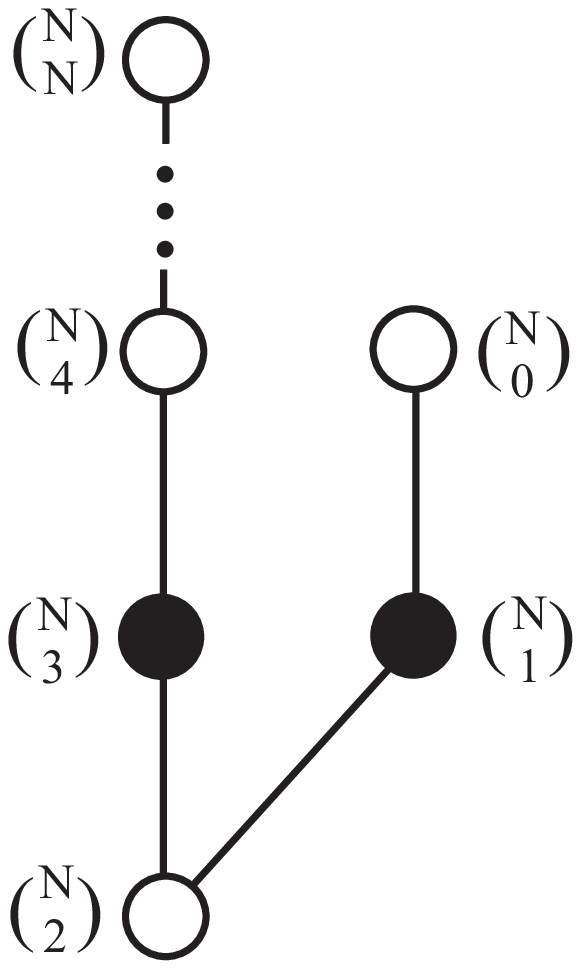} 
 \label{Lift3}\err
This Adinkra corresponds to the superderivative superfield $D_{[I}D_{J]}\Phi$
modulo several modes associated with the lifted vertices.  More precisely, the
one-form vertex in~(\ref{Lift3}) describes the $\tau$ derivative of the corresponding
one-form fermion in $\Phi$ and the singlet vertex in~(\ref{Lift3}) describes
the second derivative $\der_\tau^2$ of the corresponding singlet boson in $\Phi$, its lowest component.

\begin{figure}
 \begin{center}
 \includegraphics[width=5.5in]{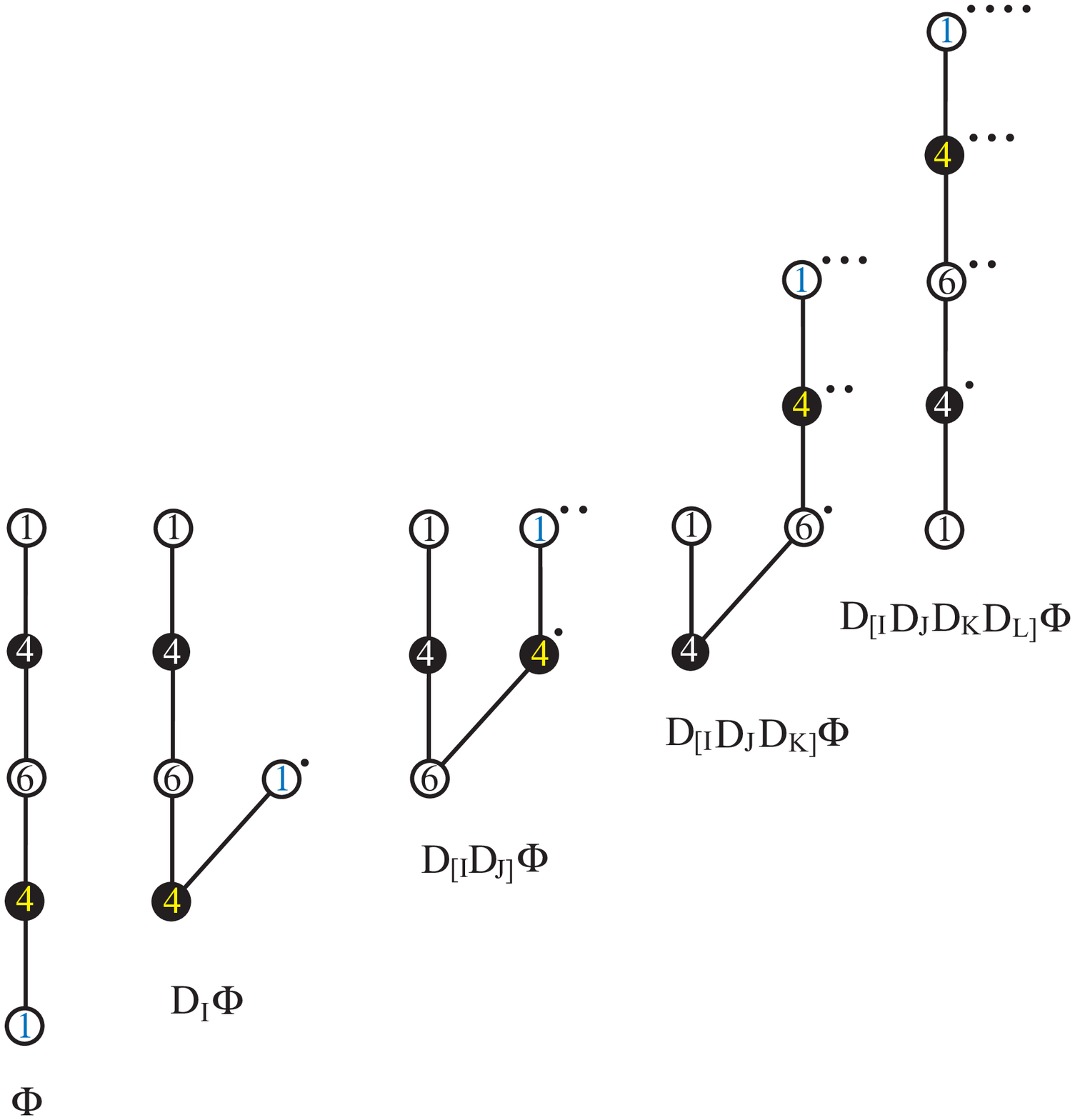}
 \caption{The ${N} = 4$ Top Adinkra, corresponding to the unconstrained superfield
 $\Phi$, and a sequence of related Adinkras obtained as antisymmetric products of
 superspace derivatives acting on $\Phi$. We have drawn distinctions between the
 two multiplicity-four fermion vertices and the two singlet vertices by using different
 coloring on the multiplicity labels appearing on these vertices.}
 \label{DLift}
 \end{center}
 \end{figure}

The Adinkras~(\ref{Lift1}) and~(\ref{Lift3}) share a feature not exhibited
by~(\ref{Lift2}) or by the majority of Adinkras: namely, these Adinkras have exactly
two sinks and exactly one compound source, meaning that the source vertices
combine into a particular $\SO(N)$ $p$-form.  These Adinkras are obtained from the
Top Adinkra by lifting its lowest vertex upward, dragging other vertices behind, as if one were raising a chain.  Adinkras
with this feature correspond to the antisymmetric product of superderivatives acting
on an unconstrained superfield $\Phi$, which itself corresponds to the Top Adinkra.
This concept is illustrated by Figure \ref{DLift} in the particular case of ${N} = 4$
supersymmetry.
The dots which appear on some vertices in Figure \ref{DLift} indicate the
relationship between these vertices and the vertices in the leftmost, Top, Adinkra.  For
example, the topmost vertex in the rightmost Adinkra, the one corresponding to $
D_{[I}D_J D_K D_{L]}\Phi$, has a blue numeral and has four dots.  These dots
indicate that this vertex describes the fourth derivative $\der_\tau^4$ of the field
corresponding to the lowermost vertex in the Top Adinkra. (This is the unique vertex
in the Top Adinkra having a blue numeral.) Notice that the top derivative $
D^N=\fr{1}{4!}\,\ve^{I_1\cdots I_N}\,D_{I_1}\cdots D_{I_N}$ completely swivels the
Top Adinkra about its hook, so that its source becomes a sink, albeit differentiated,
and its sink becomes a source.

The Top Adinkra, which corresponds to an unconstrained superfield $\Phi$, and
its elemental derivatives,
 \brr\Xi_p^q=D_{[I_1}\cdots D_{I_p]}\,\der_\tau^q\,\Phi \,,
 \err
describe building blocks from which more general superfields may be constructed by forming linear
combinations.  The question of which superfield constraints correspond to which
irreducible supermultiplets can be re-phrased as a question of which linear combinations
of the basic building blocks $\Xi_p^q$ correspond to the irreducible supermultiplets.

Note that the basic building blocks can be visualized in terms of sets, such as those
pictured in Figure \ref{DLift} plus versions of such diagrams raised by global
differentiation, by which we mean similar diagrams obtained by adding a common
number $q$ of  $\tau$ derivatives to each vertex.  Figure \ref{DLift} enumerates
the set $\{\,\Xi_p^0\,\}$ in the case ${N} = 4$, for the cases $p=0,1,2,3,4$.  Additional
diagrams $\Xi_p^{q\ne 0}$ are obtained by differentiating all vertices $q$ times.
Each $\tau$ derivative lifts the entire diagram by one engineering dimension, which
corresponds to two height units since the height is twice the engineering
dimension.  For example, the relationship between the diagrams $\Xi_1^0=D_I\,\Phi$
and $\Xi_1^1=D_I\,\dot{\Phi}$ is seen as follows:
 \brr \includegraphics[width=2.5in]{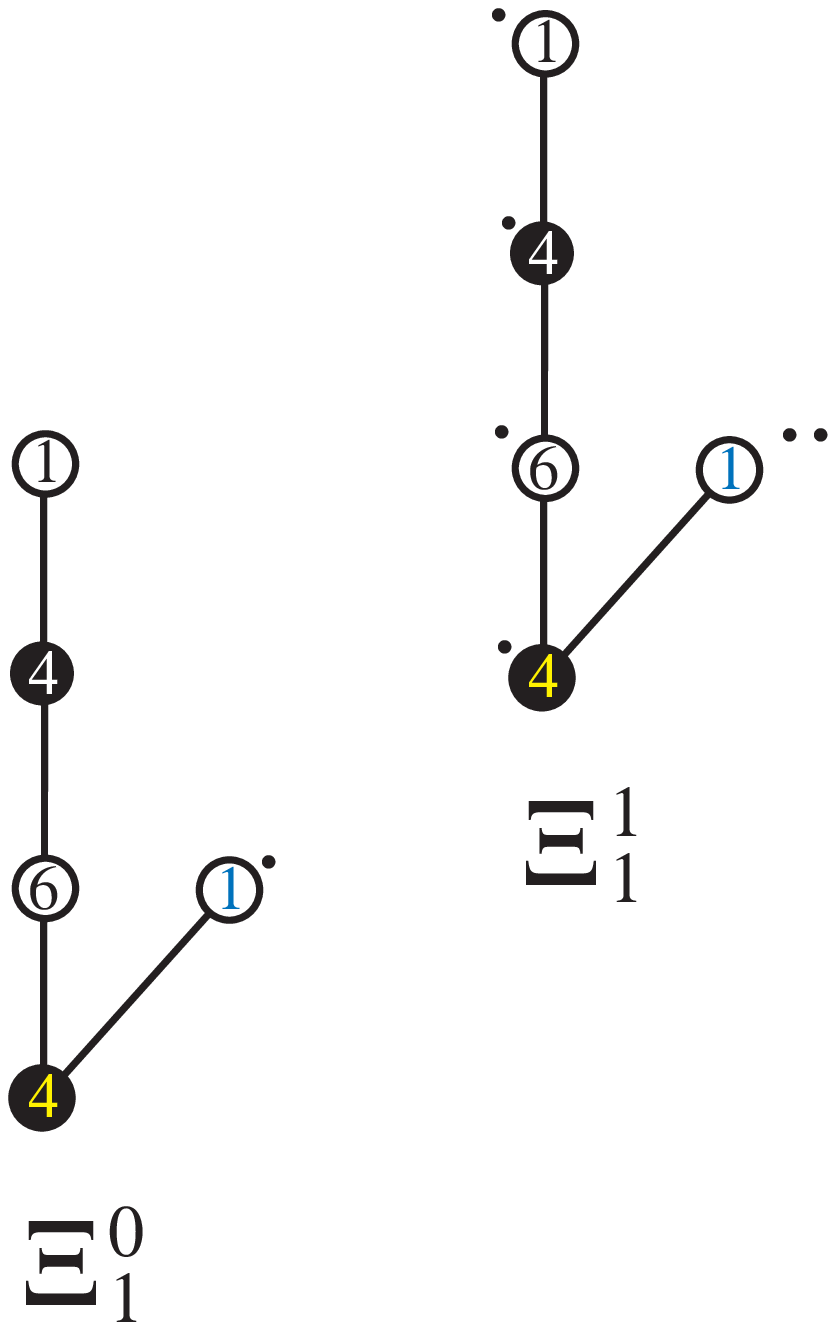} 
 \label{Bip1}\err
The fact that the vertices of $\Xi_1^1$ have two height units greater than their
counterparts in $\Xi_1^0$ is manifested by the raised placement of the second
diagram relative to the first.

 \section{Garden Algebras, and Clifford Algebra Superfields}
 \label{Clifford}
 Adinkra diagrams provide a concise and elegant way to represent supermultiplets.
This is loosely analogous to the way Feynman diagrams represent integrals appearing
in field theory calculations.  But Adinkra diagrams have their own magic; these can be
manipulated in a way which mirrors various algebraic tasks associated with superfields
or associated with component field calculations.  Some of these have been described
above.  The core algebraic underpinning of Adinkras lies, however, in the realm of $
{\cal GR}({\rm d},N)$ algebras, introduced in\cite{GR1} and\cite{GR2}, which have emerged
as vitally important for supersymmetry
representation theory.  In this section we review these algebras and their relevance to
one-dimensional supersymmetry.  We use these to describe the algebraic counterpart
to the pictorial presentation in Section~\ref{introadinkra}.

A supersymmetry transformation $\d_Q(\e)$ is parameterized by $\e^I$, where $I=
1,...,N$ is an $\SO(N)$ vector index. Two supersymmetry transformations commute into
a time translation according to
 \brr [\,\d_Q(\e_1)\,,\,\d_Q(\e_2)\,]=
      -2\,i\,\e^I_1\,\e^I_2\,\der_\tau \,.
 \label{nqq}\err
The representations of~(\ref{nqq}) can be classified using the so-called ``garden
algebra'' ${\cal GR}({\rm d},N)$, generated by two sets  $(\,L_I\,)_i\,^{\hat{\jmath}}$ and $(\,R_I\,)_{\hat{\imath}}\,^j$ of $N$
$d\times d$ matrices, known as ``garden matrices'', subject to the relations
\brr (\,L_I\,R_J+L_J\,R_I\,)_i\,^j &\equals&
      -2\,\delta_{IJ}\,\delta_i\,^j
      \nonumber\\[.1in]
       (\,R_I\,L_J+R_J\,L_I\,)_{\hat{\imath}}\,^{\hat{\jmath}} &\equals&
      -2\,\delta_{IJ}\,\delta_{\hat{\imath}}\,^{\hat{\jmath}}
      \nonumber\\[.1in]
      L_I &\equals& -R_I^T \,.
 \label{garden}\err
We note that garden algebras are closely related to Clifford algebras.
Indeed, a choice of garden matrices generating ${\cal GR}({\rm d},N)$ contains the same mathematical information as a representation of the
Clifford algebra $\mathrm{Cl}(N)$ on a $d+d$-dimensional super vector space,
with the Clifford generators acting by odd, skew-adjoint operators.

\begin{figure}
\begin{center}
\includegraphics[width=4.0in]{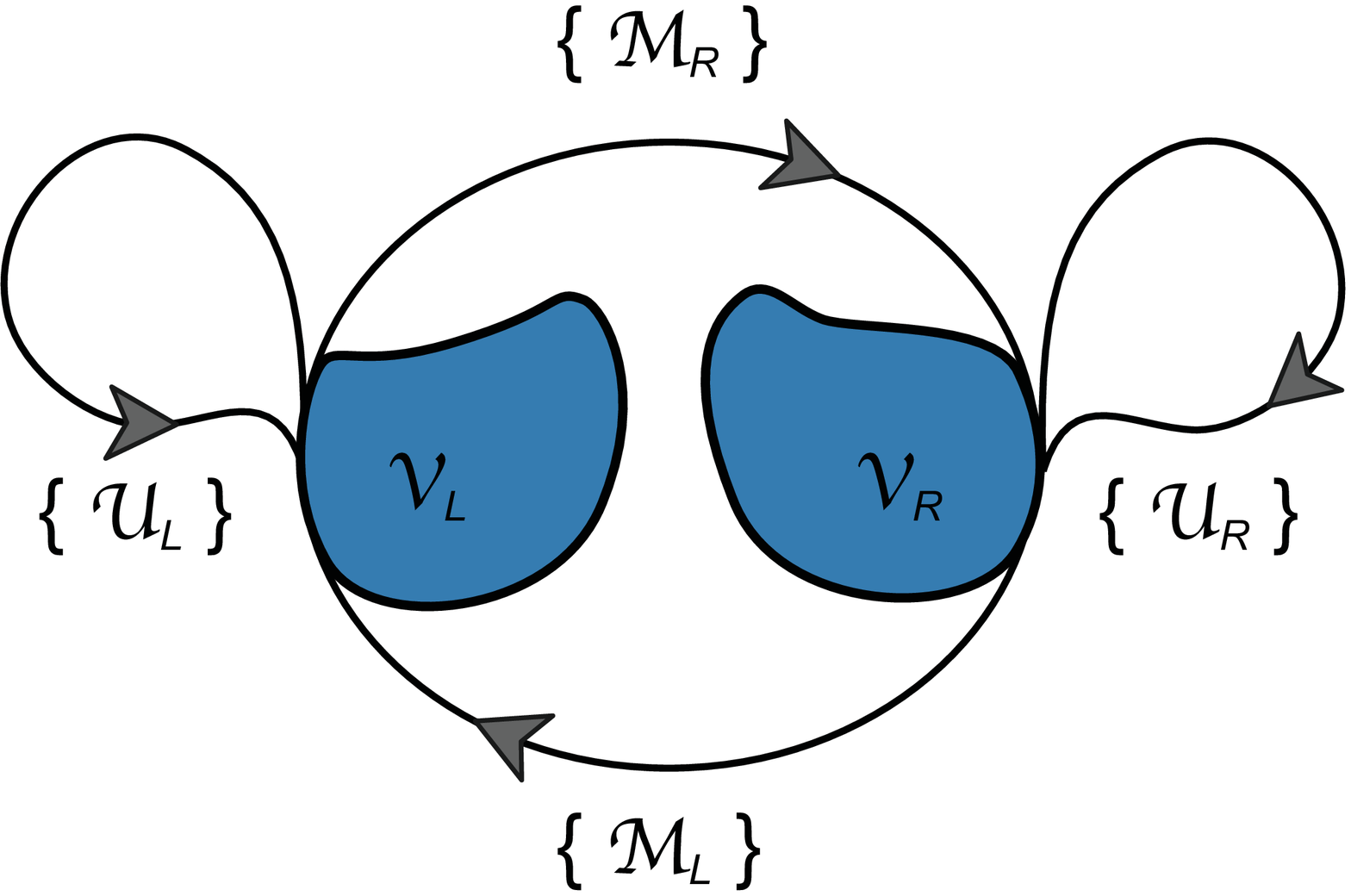}
\caption{The PpG diagram.}
 \label{ppgfig}
 \end{center}
 \end{figure}

For fixed values of $N$, there are
multiple values of $d$ for which these matrices exist.  But there is one $d_N$ that is the
least integer for which $d\times d$ garden matrices exist.  The value of this integer for
every $N$ is tabulated in\cite{FG1}.  There are different sorts of indices adorning
$(\,L_I\,)_i\,^{\hat{\jmath}}$ and $(\,R_I\,)_{\hat{\imath}}\,^j$.  The non-hatted indices
$i,j,...$ span a vector space ${\cal V}_L\cong \R^d$ while the hatted indices
$\hat{\imath}, \hat{\jmath},...$ span another vector space ${\cal V}_R\cong \R^d$.  These indices
adorn the matrices $L_I$ and $R_I$, the first index labels the row (thus, the range) and the second index
labels the column (thus, the domain).
Thus, the $L_I$ describe linear maps from ${\cal V}_R$ to ${\cal V}_L$, and the $R_I$ describe
linear maps from ${\cal V}_L$ to ${\cal V}_R$.  The compositions of these will then be
maps on ${\cal V}_L
\oplus{\cal V}_R$ that can be furthermore classified according to their domain and range
as follows:
 \brr  &&\{\,{\cal M}_L\,\}\,:\,
      {\cal V}_R\to{\cal V}_L
      \hspace{.3in}
      \{\,{\cal U}_L\,\}\,:\,
      {\cal V}_L\to{\cal V}_L
      \nonumber\\[.1in]
      &&\{\,{\cal M}_R\,\}\,:\,
      {\cal V}_L\to{\cal V}_R
      \hspace{.3in}
       \{\,{\cal U}_R\,\}\,:\,
      {\cal V}_R\to{\cal V}_R \,.
 \label{spaces}\err
This formalism produces a visualization  of these concepts in a
coordinate-independent manner. We use Venn diagrams to represent the
sets ${\cal V}_L$ and ${\cal V}_R$.  The set of linear operators
that act between and on these sets may be represented by a set of
directed arrows as shown in the Placement-putting Graph \cite{enuf},
shown in Figure \ref{ppgfig}

In this paper the action of a matrix is defined in terms of left multiplication.  Thus,
$(\,L_I\,)_i\,^{\hat{\jmath}}\in {\cal M}_L$ and $(\,R_I\,)_{\hat{\imath}}\,^j\in{\cal M}_R$.
The ``normal part of the enveloping algebra'', denoted $\wedge {\cal EGR}({\rm d},N)$,
is generated by the wedge products involving $L_I$ and $R_I$,
 \brr (\,f_I\,)_i\,^{\hat{\jmath}} &\equals& (\,L_I\,)_i\,^{\hat{\jmath}}
      \hspace{1.1in}
      (\,\tilde{f}_{I}\,)_{\hat{\imath}}\,^j=
      (\,R_I\,)_{\hat{\imath}}\,^j
      \nonumber\\[.1in]
      (\,f_{IJ}\,)_i\,^j &\equals& (\,L_{[I}\,R_{J]}\,)_i\,^j
      \hspace{.7in}
      (\,\tilde{f}_{IJ}\,)_{\hat{\imath}}\,^{\hat{\jmath}}=
      (\,R_{[I}\,L_{J]}\,)_{\hat{\imath}}\,^{\hat{\jmath}}
      \nonumber\\[.1in]
      (\,f_{IJK}\,)_i\,^{\hat{\jmath}} &\equals&
      (\,L_{[I}\,R_{J}\,L_{K]}\,)_i\,^{\hat{\jmath}}
      \hspace{.3in}
      (\,\tilde{f}_{IJK}\,)_{\hat{\imath}}\,^j=
      (\,R_{[I}\,L_J\,R_{K]}\,)_{\hat{\imath}}\,^j \,,
 \label{fdef}\err
and so forth.  The subsets of the vector spaces ${\cal U}_{L,R}$ and
${\cal M}_{L,R}$, defined in~(\ref{spaces}), which are also in
$\wedge {\cal EGR}({\rm d},N)$, are called, respectively, ${\cal
U}_{L,R}^{(n)}$ and ${\cal M}_{L,R}^{(n)}$, where the superscript
$(n)$ indicates the ``normal part''.

 \subsection{The Placement-Putting of Adinkra Nodes}
 \label{ppnodes}
 The component fields of one-dimensional supermultiplets
 naturally admit a structural organization associated with the vector spaces
 described above.
 For example, the components of a Clifford algebra superfield, or
 equivalently the vertices of a Base Adinkra,
 are valued in subsets of $\wedge {\cal EGR}({\rm d},N)$, such as
 ${\cal U}_L^{(n)}$ and ${\cal M}_R^{(n)}$, whereas the components
 of Scalar supermultiplets, or equivalently the vertices of a Scalar
 Adinkra, are valued in ${\cal V}_R$ and ${\cal V}_L$.
 In fact, the garden algebras represent special cases of real Clifford
 algebras, such that component fields, or equivalently collections of
 Adinkra vertices, that are valued in ${\cal V}_L$ or ${\cal V}_R$
 transform as some spinor representation(s) of an associated $\SO(N)$ symmetry.
 Similarly, fields or vertices valued in $\wedge {\cal EGR}({\rm d},N)$ transform
 as some $p$-form representation(s) of an $\SO(N)$ symmetry.   Thus,
 the vertices of Adinkras span representations
 whose generators act on the vertices by multiplication with
 appropriate matrices.  We refer to the
 distinct $\SO(N)$ representations carried by vertices, as their
``$R$-charge''.

 \subsection{The Clifford Algebra Superfield}
 \label{CAS}
A fundamental representation of the $N$-extended superalgebra is given by the Clifford
algebra superfield.  This involves $2^{N-1}$ bosons $\Phi_i\,^j\in {\cal U}_L^{(n)}$ and
$2^{N-1}$ fermions $\Psi_{\hat{\imath}}\,^j\in {\cal M}_R^{(n)}$, subject to the following
transformation rules:
 \brr \d\,\Phi_i\,^j &\equals&
      -i\,\e^I\,(\,L_I\,)_i\,^{\hat{k}}\,\Psi_{\hat{k}}\,^j
      \nonumber\\[.1in]
      \d\,\Psi_{\hat{\imath}}\,^j &\equals&
      \e^I\,(\,R_I\,)_{\hat{\imath}}\,^k\,\der_\tau{\Phi}_k\,^j \,.
 \label{box}\err
If we make a particular choice for the garden matrices $(\,L_I\,)_i\,^{\hat{\jmath}}$ and
$(\,R_I\,)_{\hat{\imath}}\,^j$, then it is straightforward to translate the transformation rules~(\ref{box}) into the equivalent Adinkra, using the techniques developed
in\cite{DFGHIL01} and\cite{FG1}. The resulting Adinkra is shown in~(\ref{BN}).  We can expand the
component fields in~(\ref{box}) using the bases~(\ref{fdef}) as follows:
 \brr \Phi_i\,^j &\equals&
      \sum_{p=0}^{\lfloor N/2 \rfloor }
      (\,f^{I_1\cdots I_{2\,p}}\,)_i\,^j\,
      \phi_{I_1\cdots I_{2\,p}}\,
      \nonumber\\[.1in]
      \Psi_{\hat{\imath}}\,^j &\equals&
      \sum_{p=1}^{ \lfloor N/2 \rfloor}
      (\,\tilde{f}^{I_1\cdots I_{2\,p-1}}\,)_{\hat{\imath}}\,^j\,
      \psi_{I_1\cdots I_{2\,p-1}}\, \,,
 \label{cliffsf}\err
 where $\lfloor\,\cdot\,\rfloor$ selects the integer part of its argument.
 In this way we can replace the matrix fields $\Phi_i\,^j$ and
 $\Psi_{\hat{\imath}}\,^j$ with $p$-forms on $\SO(N)$.
 The transformation rules
~(\ref{box}) imply the following corresponding rules for
 the $p$-form fields:
 \brr \d\,\phi^{[p_{\rm even}]} &\equals&
      -i\,\e^{[I_1}\,\psi^{I_2\cdots I_p]}
      +(p+1)\,i\,\e_J\,\psi^{I_1\cdots I_p\,J}
      \nonumber\\[.1in]
      \d\,\psi^{[p_{\rm odd}]} &\equals&
      -\e^{[I_1}\,\dot{\phi}^{I_2\cdots I_p]}
      +(p+1)\,\e_J\,\dot{\phi}^{I_1\cdots I_p\,J} \,.
 \label{generic}\err
Thus, the Clifford algebra superfield involves $2^{N-1}$ bosons which assemble as
even-forms on $\SO(N)$ and $2^{N-1}$ fermions which assemble as odd-forms on
$\SO(N)$.  Equation~(\ref{generic}) is equivalent to equation~(\ref{box}). The equivalence
may be proved using algebraic identities satisfied by the garden matrices, which follow
as corollaries of the garden algebra.  The Adinkra counterpart to~(\ref{generic}) is~(\ref{BN}).

Notice that $\Phi_i\,^j\in{\cal U}_L^{(n)}$, which is spanned by even-forms of the sort
defined in~(\ref{fdef}), and $\Psi_{\hat{\imath}}\,^j\in{\cal M}_R$, which is spanned by
odd-forms.  It is for this reason that the Clifford algebra superfield corresponds to the
Base Adinkra, shown in Figure \ref{chart}, rather than to the Dual Base Adinkra.  The
latter construction involves even-form fermions rather than even-form bosons.  The
algebraic counterpart to the Dual Base Adinkra is described in the following subsection.

 \subsection{The Dual Clifford Superfield}
 \label{coucou}
 An alternative way to formulate a Clifford algebra superfield
 involves $2^{N-1}$ bosons $\Phi_{\hat{\imath}}\,^j\in {\cal M}_R^{(n)}$ and $2^{N-1}$
 fermions
 $\Psi_i\,^j\in {\cal U}_L^{(n)}$, subject to the following transformation
 rules:
 \brr \d\,\Phi_{\hat{\imath}}\,^j &\equals&
      -i\,\e^I\,(\,R_I\,)_{\hat{\imath}}\,^k\,\Psi_k\,^j
      \nonumber\\[.1in]
      \d\,\Psi_i\,^j &\equals&
      \e^I\,(\,L_I\,)_i\,^{\hat{k}}\,\der_\tau{\Phi}_{\hat{k}}\,^j
      \,.
 \label{boxC}\err
The difference between~(\ref{boxC}) and~(\ref{box}) lies in the way the component degrees
of freedom are embedded in $\wedge{\cal EGR}({\rm d},N)$, as reflected by the placement of
the hatted and unhatted indices on the component fields themselves and on whether it is
the matrix $(\,L_I\,)_i\,^{\hat{\jmath}}$ or $(\,R_I\,)_{\hat{\imath}}\,^j$ which appears in the
transformation rule for the bosons or fermions.

We can expand the component fields in~(\ref{boxC}) using the bases~(\ref{fdef}) as follows:
 \brr \Phi_{\hat{\imath}}\,^j &\equals&
      \sum_{p=1}^{\lfloor N/2\rfloor}
      (\,\tilde{f}^{I_1\cdots I_{2\,p-1}}\,)_{\hat{\imath}}\,^j\,
      \phi_{I_1\cdots I_{2\,p-1}}\,
      \nonumber\\[.1in]
      \Psi_i\,^j &\equals&
      \sum_{p=0}^{\lfloor N/2\rfloor}
      (\,f^{I_1\cdots I_{2\,p}}\,)_i\,^j\,
      \psi_{I_1\cdots I_{2\,p}}\, \,,
 \label{cliffsfC}\err
where $\lfloor\,\cdot\,\rfloor$ selects the integer part of its argument.  In this way we can replace
the matrix fields $\Phi_i\,^j$ and $\Psi_{\hat{\imath}}\,^j$ with $p$-forms on $\SO(N)$.  The
transformation rules~(\ref{boxC}) imply the following corresponding rules for the $p$-form
fields:
 \brr \d\,\phi^{[p_{\rm odd}]} &\equals&
      -i\,\e^{[I_1}\,\psi^{I_2\cdots I_p]}
      +(p+1)\,i\,\e_J\,\psi^{I_1\cdots I_p\,J}
      \nonumber\\[.1in]
      \d\,\psi^{[p_{\rm even}]} &\equals&
      -\e^{[I_1}\,\dot{\phi}^{I_2\cdots I_p]}
      +(p+1)\,\e_J\,\dot{\phi}^{I_1\cdots I_p\,J} \,.
 \label{genericC}\err
Thus, the Dual Clifford algebra superfield involves $2^{N-1}$ bosons which assemble as
odd-forms on $\SO(N)$ and $2^{N-1}$ fermions which assemble as even-forms on $\SO(N)$.
Equation~(\ref{genericC}) is equivalent to equation~(\ref{boxC}).  The Adinkra counterpart to~(\ref{genericC}) is shown in~(\ref{cz2}).

 \subsection{The Conjugate Clifford Superfield, and its dual}
Two additional fundamental supermultiplets are the Klein flipped
versions of the Clifford algebra superfield and its dual, described
in Subsections~\ref{CAS} and \ref{coucou}, respectively.

The Conjugate Clifford superfield involves $2^{N-1}$ fermions $\tilde{\Psi}^{\hat{\imath}}
 \,_j\in {\cal M}_R^{(n)}$ and $2^{N-1}$ bosons $\tilde{\Phi}^i\,_j\in {\cal U}_L^{(n)}$,
subject to the following transformation rules:
 \brr \d\,\Psi^{\hat{\imath}}\,_j &\equals&
      \e^I\,(\,R_I\,)^{\hat{\imath}}\,_k\,\Phi^k\,_j
      \nonumber\\[.1in]
      \d\,\Phi^i\,_j &\equals&
      -i\,\e^I\,(\,L_I\,)^i\,_{\hat{k}}\,\der_\tau\,\Psi^{\hat{k}}\,_j \,.
 \label{boxcc}\err
In this case the fermions describe the lower components and decompose as odd-forms on
$\SO(N)$, while the bosons describe the higher components and decompose as even-forms
on $\SO(N)$.  This supermultiplet is represented by the Conjugate Base Adinkra, which is shown in~(\ref{CBN}).

The Dual Conjugate Clifford superfield involves $2^{N-1}$ fermions $\tilde{\Psi}^i\,_j\in
{\cal U}_L^{(n)}$ and $2^{N-1}$ bosons $\tilde{\Phi}^{\hat{\imath}}\,_j\in {\cal M}_R^{(n)}$,
subject to the following transformation rules:
 \brr \d\,\Psi^i\,_j &\equals&
      \e^I\,(\,L_I\,)^i\,_k\,\Phi^k\,_j
      \nonumber\\[.1in]
      \d\,\Phi^{\hat{\imath}}\,_j &\equals&
      -i\,\e^I\,(\,R_I\,)^{\hat{\imath}}\,_{\hat{k}}\,\der_\tau\,\Psi^{\hat{k}}\,_j \,.
 \label{boxccd}\err
In this case the fermions describe the lower components and decompose as even-forms on
$\SO(N)$, while the bosons describe the higher components and decompose as odd-forms
on $\SO(N)$.  This supermultiplet is represented by the Dual Conjugate Base Adinkra, which is
shown in~(\ref{CBN2}).

 \section{Scalar Multiplets}
 \label{scammed}
Generalized Clifford algebra superfields, which correspond to
generalized Base Adinkras, are reducible for $N \ge 4$.  Similarly,
unconstrained Salam-Strathdee superfields, which correspond to Top
Adinkras, are also reducible for $N \ge 4$. On the other hand, the
Scalar supermultiplets are irreducible for all $N$. A Scalar
supermultiplet involves $d$ bosonic fields $\phi_i\in{\cal V}_L\cong
\R^{d}$ and $d$ fermionic fields $\psi_{\hat{\imath}}\in {\cal
V}_R\cong \R^{d}$, where $d=d_{N}$ is the minimum value for which
$d\times d$ garden matrices exist. The supersymmetry transformation
rules are given by
 \brr \d_Q(\e)\,\phi_i &\equals&
      -i\,\e^I\,(\,L_I\,)_i\,^{\hat{\jmath}}\,\psi_{\hat{\jmath}}
      \nonumber\\[.1in]
      \d_Q(\e)\,\psi_{\hat{\imath}} &\equals&
      \e^I\,(\,R_I\,)_{\hat{\imath}}\,^j\,\der_\tau\,\phi_j \,.
 \label{s1}\err
A Scalar supermultiplet is represented by an Adinkra having $d$ fermionic vertices all at height
zero and $d$ bosonic vertices all at height minus one.%
\Ft{\label{engf}By making this
choice, we are normalizing the height of the Scalar supermultiplet by choosing canonical
dimensions for its component fields: A propagating scalar field $\phi$
has a canonical kinetic action given by $S_\phi$ $=\int d\tau\,\fr12\,
\dot{\phi}^2$; since the engineering dimension of $\tau$ is minus one, it follows that $S$
is dimensionless only if $\phi$ has dimension minus one-half.  Since
the height parameter is twice the engineering dimension, it follows
that the canonical height of a one-dimensional scalar is minus one.  Similar reasoning
may be applied to the canonical fermion action $S_\psi=\int d\tau \psi
\dot{\psi}$ to conclude that canonical propagating fermions have zero height.}  For
example, in the case ${N} = 4$, we have $d_N =4$, and the Scalar Adinkra is the following complete bipartite graph with $4+4$ vertices:
\brr
\includegraphics[width=1.8in]{Linear3.eps} 
\label{s4}\err

We now wish to examine the following question: Does there exist a superspace description of a general-$N$ Scalar Adinkra for which the vertices correlate one-to-one with the components of some unconstrained {\em {prepotential}} Salam-Strathdee superfields?  (As we saw in
Subsection~\ref{irreps}, the Scalar supermultiplet can be constructed from the Clifford algebra superfield via a projection determined by the symmetries of the Base Adinkra. However, this is not immediately helpful, as we have not yet given a superspace description of the general-$N$ Clifford algebra superfield in terms of Salam-Strathdee superfields.) In addition, we ask whether we can construct a supersymmetric action functional in terms of a superspace integral built from these prepotentials for which the propagating fields correspond precisely to the Scalar supermultiplet?
It should be kept in mind that the definitions of superfields we have been using so far in this discussion are {\em{totally}} independent of the Salam-Strathdee superfield formalism. So the answer has not been presumed in our discussion to this point.

The answer to this question has been known for some years.  It was explicitly stated,
for example in a 1982 work by Gates and Siegel\cite{GS1982}: ``Conversely, the
highest-dimension component field appearing in an action is the highest $\theta$-component
of a superfield appearing in this action.''  As it turns out, a given Adinkra can be described
using one prepotential superfield for {\em {each}} of its hooks, {\it i.e.}, its sink vertices.  The
statistics and the ${\cal GR}({\rm d},N)$ or $\SO(N)$ structure of these prepotentials are dictated
by statistics and the ${\cal GR}({\rm d},N)$ or $\SO(N)$ structures of the sink vertices on the target
Adinkra, {\it i.e.}, the Adinkra we wish to describe using the prepotentials.   Since the
prepotentials are unconstrained, these correspond to Top Adinkras, each of which has
exactly one hook.  The statistics for each prepotential are chosen such that the hook of the corresponding Top Adinkra correlates with one hook of the target Adinkra.  A fermionic
hook therefore corresponds to a bosonic prepotential in cases where $N$ is odd and to a
fermionic prepotential in cases where $N$ is even.  Similarly, a bosonic hook corresponds
to a fermionic prepotential in cases where $N$ is odd and to a bosonic prepotential in cases
where $N$ is even.  These conclusions follow because the statistics of a Salam-Strathdee
superfield coincide with the statistics of the source vertex (the lowest vertex) on the associated
Top Adinkra and because a Top Adinkra spans $N+1$ different height assignments.
Therefore the statistics of the Top Adinkra source vertex coincides with the statistics of its
hook in cases where $N$ is even and differs from the statistics of its hook in cases where
$N$ is odd.

According to this claim, the ${N} = 4$ Scalar supermultiplet requires four real fermionic superfields,
${\cal F}_{\hat{\imath}}$, as prepotentials, where the index ${\hat{\imath}}$ spans ${\cal V}_R
\cong \R^4$.  This is determined by~(\ref{s4}), where we see that the ${N} = 4$ Scalar Adinkra
has four real fermionic scalar hooks; these span ${\cal V}_R$ since the four fermion fields
corresponding to these vertices are $\psi_{\hat{\imath}}\in {\cal V}_R$.  For the case of general-$N$
Scalar supermultiplets, similar reasoning implies that $d$ real superfields ${\cal S}_{\hat{\imath}}$ should suffice.
We prove below, in Section~\ref{prepsca}, that such a prepotential construction does properly describe any Scalar
supermultiplet, and we also show how these unconstrained superfields can be used
to build supersymmetric action functionals.  The reader might wonder because $d$ unconstrained
real fermionic scalar superfields involve a total of $d\cdot 2^{N-1}+d\cdot 2^{N-1}$ component
fields whereas a Scalar supermultiplet has only $2^{N-1}+2^{N-1}$ component fields, an apparent
mismatch.  The resolution is that the excess correspond to gauge degrees of freedom; there is a particular linear
combination of the building blocks $D_{[I_1}\cdots D_{I_p]}\,\der_\tau^q\,{\cal
S}_{\hat{\imath}}$, where ${\cal S}_{\hat{\imath}}$ are the prepotentials, which describes precisely
the degrees of freedom in the Scalar supermultiplet.

The fact that this works in the general case might be surprising to some readers.  We think it is
helpful, therefore, to describe in detail the simplest case---the case $N=2$---in order to illustrate
clearly how the gauge structure appears in the prepotentials.  (In the case $N=1$ the unconstrained superfield and
the Scalar supermultiplet are identical.) Accordingly, the following section
focuses on the case $N=2$.  The general case is described in Section~\ref{prepsca}.

Our basic strategy, which ultimately gives rise to the solution described above, is predicated
on the following thoughts:  Scalar supermultiplets span two height assignments, whereas Top
Adinkras span $N+1$ height assignments.  Thus, if we wish to describe a Scalar supermultiplet
using a prepotential or a set of prepotentials, we will have to operate on these.  We explained
at length above how superspace differential operators lift vertices and can reduce the span of
an Adinkra.  A complete set of differential operators on superspace are given by
 $D_{[I_1}\cdots D_{I_p]}\,\der_\tau^q$, where $p$ and $q$ are
 integers constrained by $0\le p\le N$ and $q\ge 0$.  Our strategy
 will be to consider linear combinations
 \brr \Gamma(\{a\})_{m;I_1\cdots I_r}:=\sum_{n=1}^\lambda\,\sum_{s=0}^\infty \sum_{p=0}^N\,
      (\,a^p_s\,)_{m;I_1\cdots I_r}^{n;J_1\cdots J_p}
      D_{[J_1}\cdots D_{J_p]}\,\der_\tau^s\,{\cal S}_n \,,
 \label{Gamdef}\err
where ${\cal S}_n$ are a set of $\lambda\in\mathbb{N}$ unconstrained prepotential superfields and
$(\,a^p_s\,)_{m;I_1\cdots I_r}^{n;J_1\cdots J_p}$ are complex tensor coefficients to be
determined,\Ft{Note: Two superfields may be added only if these have the same
statistics, and describe the same  $\SO(N)$ representation, and have the
same engineering dimension.  The third of these restrictions
dramatically limits the possibilities. The summands in~(\ref{Gamdef}) have engineering dimension $\eta= \fr12\,p+s$, since the $D_I$ have dimension
one-half.  This number should be the same for each summand in  order
for $\Gamma(\{a\})$ to have definite engineering dimension.  Thus, there is only one value
of $p$ for each value of $s$ for which the coefficients can be nonvanishing.}
and  where the indices $m$ and $n$ take values in a vector space ${\cal W}$, for which
${\rm dim}\,{\cal W}=\lambda$, to be determined. (For the Scalar supermultiplets, it turns out that
${\cal W}={\cal V}_R$ and $\lambda=d$.)  The statistics of the prepotentials may vary.  In
other words, some number $b$ of the ${\cal S}_n$ could be bosonic while the remaining
$\lambda-b$ prepotentials would be fermionic.  (Clearly, $0\le b\le \lambda$.)

Our tasks are: 1) to determine whether a set of prepotentials (specified by a choice of
$\lambda$ and a choice of $b$) and a set of complex coefficients $(\,a^p_s\,)_{m;I_1\cdots
I_r}^{n;J_1\cdots J_p}$ exist such that $\Gamma(\{a\})$ includes precisely the field content
of a given target supermultiplet, in this case a Scalar supermultiplet, and 2) to use the prepotential
superfields to build a supersymmetric action functional depending only on the degrees of
freedom corresponding to the target supermultiplet.

 \section{$N=2$ Prepotentials}
 \label{n2n2}
In this section we address the question posed in the previous section, regarding the existence
of a suitable prepotential for Scalar supermultiplets, in the restricted context of $N=2$ supersymmetry.
We do not presuppose the particular solution described in the previous section but instead
arrive at this solution via methodical reasoning.

We adopt a notational convention in which the name of a component field provides information regarding the statistics (boson or fermion), the engineering dimension, and the number of
derivatives appearing on that object.  In particular, we use $B_\delta^{(m)}$ to refer
to the $m$th $\tau$ derivative of a bosonic component field having engineering dimension
$\delta$ and $F_\delta^{(m)}$ to refer to the $m$th $\tau$ derivative of a fermionic component
field having engineering dimension $\delta$.  Thus, a boson field having engineering dimension
minus one-half would be named $B_{-1/2}^{(0)}$, whereas the fourth derivative of this field
would be named $B_{-1/2}^{(4)}$.  The engineering dimension of an object can be read off
of the labels, since $[\,B_\delta^{(m)}\,]=(\delta+m)$.

To begin, we consider the simplest possibility, and involve only one
real prepotential, {\it i.e.}, we make an ansatz $\lambda=1$.  We
develop the case where this superpotential is fermionic (so that
$b=0$).\Ft{As it turns out, it is not possible to build a
local superspace action having canonical kinetic terms in the case
of even-$N$ Scalar supermultiplets using bosonic
prepotentials.  The reasons for this are explained near the end of
this section.} Using the notational convention
introduced in the previous paragraph, an unconstrained fermionic
$N=2$ superfield is given by
  \begin{figure}
 \begin{center}
 \includegraphics[width=5in]{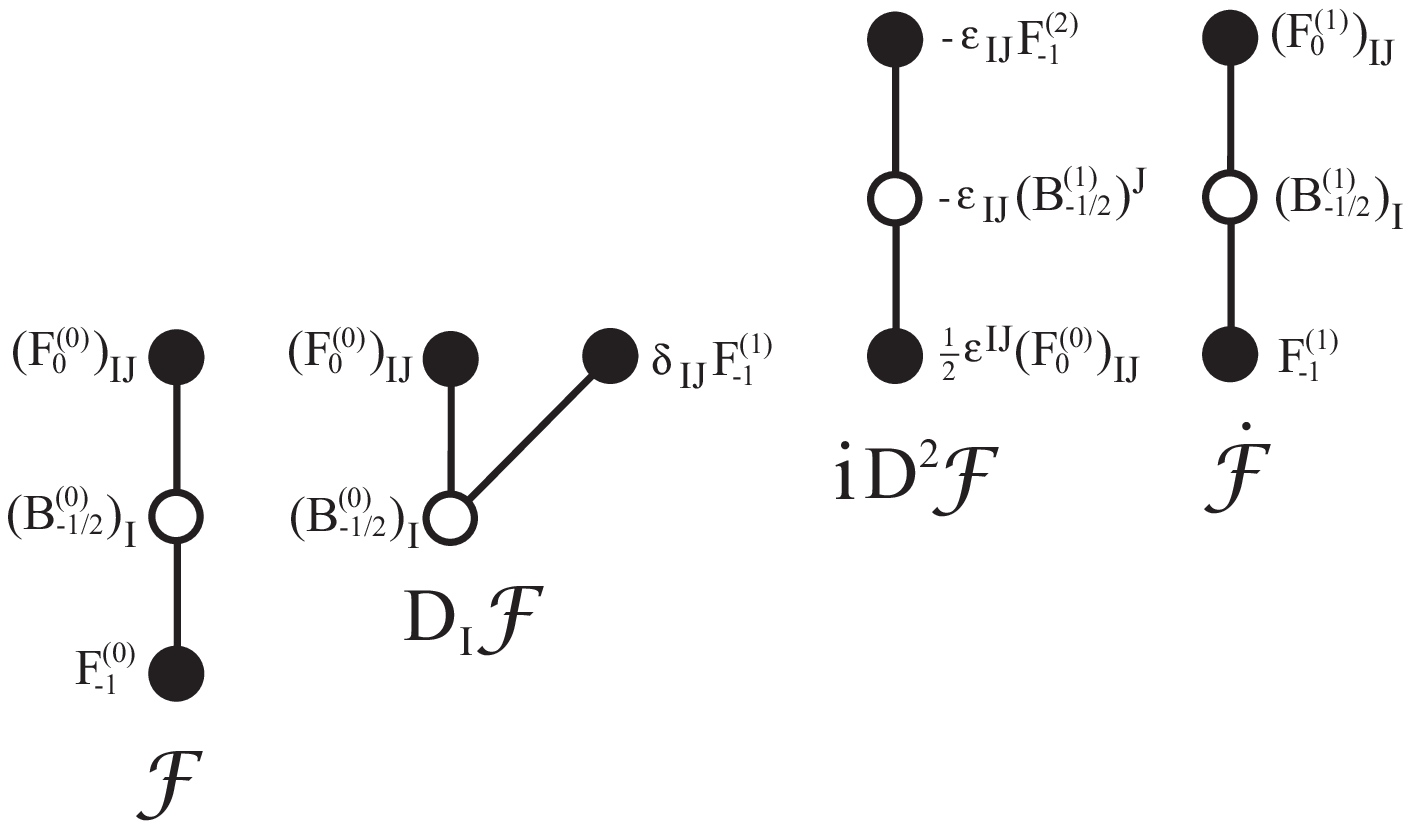}
 \caption{The Adinkras for the fermionic $N=2$ superfield
 ${\cal F}$, and derivatives thereof.  The vertical placement of the vertices in this graphic
 correlates faithfully with the vertex height assignments. This illustrates again how the
 application of a superderivative lifts vertices, how application of a top derivative, $D^2$
 in this case, swivels the Top Adinkra around its hook by $180^\circ$, and how the operator
 $\der_\tau$ lifts the ${\cal F}$ Adinkra up without swiveling.  (Thus, $\tau$ derivatives lift
 entire Adinkras while superderivatives lift vertices.)}
 \label{Ads2}
 \end{center}
 \end{figure}
 \brr {\cal F} &\equals&
      F_{-1}^{(0)}
      +\theta^I\,(\,B_{-1/2}^{(0)}\,)_I
      +\fr{1}{2!}\,i\,\theta^I\,\theta^J\,(\,F_0^{(0)}\,)_{IJ} \,.
 \label{F2}\err
We can enumerate the possible terms in a ``Gamma expansion'', defined in~(\ref{Gamdef}),
by computing various derivatives of this superfield.  Since $N=2$, there are only three possible
terms for which $q=0$.  The first involves~(\ref{F2}) itself.  The second involves the first
superderivative,
 \brr D_I\,{\cal F} &\equals&
      (\,B_{-1/2}^{(0)}\,)_I
      +i\,\theta^J\,\bpl\,(\,F_0^{(0)}\,)_{IJ}
      +\delta_{IJ}\,F_{-1}^{(1)}\,\bpr
      +\fr{1}{2!}\,i\,\ve_{JK}\,\theta^J\,\theta^K\,\bpl\,
      \ve_{IL}\,(\,B_{-1/2}^{(1)}\,)^L\,\bpr \,,
 \err
and the third involves the second superderivative,
 \brr i\,D_{[I}\,D_{J]}\,{\cal F} &\equals&
      (\,F_0^{(0)}\,)_{IJ}
      +\theta^K\,\bpl\,-2\,\delta_{K\,[I}\,(\,B_{-1/2}^{(1)}\,)_{J]}\,\bpr
      +\fr{1}{2!}\,i\,\ve_{KL}\,\theta^K\,\theta^L\,\bpl\,-\ve_{IJ}\,F_{-1}^{(2)}\,\bpr \,.
 \label{der3}\err
We have included a conventional factor of $i$ in~(\ref{der3})
because the operator $i\,D_{[I} D_{J]}$ preserves the phase of
${\cal F}$.  For convenience, we will abbreviate $\fr12\,\ve^{IJ}
\,D_ID_J$ by writing $D^2$.  The Adinkras corresponding to the
superfields ${\cal F}$, $D_I \,{\cal F}$, and $i\,D^2\,{\cal F}$ are
shown in Figure \ref{Ads2}, where the precise correspondence between
the Adinkra vertices and the superfield components are indicated.
Every other possible term in the Gamma expansion corresponds to a
$\tau$ derivative of one of these three terms, {\it i.e.}, ${\cal
F}^{(q)}$, $D_I\,{\cal F}^{(q)}$, or $i\,D^2\,{\cal F}^{(q)}$, where
$X^{(q)}:=\der_\tau^q\,X$.  Figure~\ref{Ads2} also shows the Adinkra
corresponding to ${\cal F}^{(1)}=\dot{\cal F}$; note that it has the
same graphical form as the Adinkra for $i\,D^{2}\,{\cal F}$, but
with different labels.

There are very few possibilities for forming linear combinations of the superfields described
so far.  The only way to obtain an $\SO(2)$ singlet superfield as a sum is by adding
 \brr \der_\tau\,{\cal F} &\equals&
      F_{-1}^{(1)}+\theta^I\,(\,B_{-1/2}^{(1)}\,)_I
      +\fr{1}{2!}\,i\,\theta^I\,\theta^J\,(\,F_0^{(1)}\,)_{IJ}
 \label{yop1}\err
 to some multiple of
 \brr i\,D^2\,{\cal F} &\equals&
      \fr12\,\ve^{IJ}\,(\,F_0^{(0)}\,)_{IJ}
      +\theta^I\,\bpl\,-\ve_{IJ}\,(\,B_{-1/2}^{(1)}\,)^J\,\bpr
      +\fr{1}{2!}\,i\,\theta^I\,\theta^J\,\bpl\,-\ve_{IJ}\,F_{-1}^{(2)}\,\bpr
      \,,
 \label{yop2}\err
or by adding together total derivatives $\der_\tau^q$ of both of these.  Consider first the sum of~(\ref{yop1}) and~(\ref{yop2}), using a relative coefficient of unity.  This yields
 \brr (\,i\,D^2+\der_\tau\,)\,{\cal F} &\equals&
      F_{-1}^{(1)}+\fr12\,\ve^{IJ}\,(\,F_0^{(0)}\,)_{IJ}
      \nonumber\\[.1in]
      & & +\theta^I\,\bpl\,(\,B_{-1/2}^{(1)}\,)_I
      -\ve_{IJ}\,(\,B_{-1/2}^{(1)}\,)^J\,\bpr
      \nonumber\\[.1in]
      & & -\fr{1}{2!}\,i\,\ve_{IJ}\,\theta^I\,\theta^J\,\bpl\,
      F_{-1}^{(2)}-\fr12\,\ve^{KL}\,(\,F_0^{(1)}\,)_{KL}\,\bpr \,.
 \label{twoterm}\err
This operation preserves the overall phase of ${\cal F}$.  For instance, if ${\cal F}$ is a real
superfield, satisfying ${\cal F}={\cal F}^\dagger$,  then $(\,i\,D^2+\der_\tau\,)\,{\cal F}$ is a
new real superfield built by rearranging the components of ${\cal F}$.  So the Adinkra
corresponding to~(\ref{twoterm}) is a Top Adinkra, not a Scalar Adinkra.  A similar conclusion
follows if we consider any combination $(\,a\,i\,D^2+\der_\tau\,)\,{\cal F}$, where $a$ is any
real number. Thus, this does not provide us with what we are looking for, {\it i.e.}, a superfield
corresponding to a two-height Adinkra. It might seem odd to expect that by adding together
three-height Adinkras we could obtain a two-height Adinkra.  But this is possible if the addition
serves to project out some the vertices, as we show presently.

We have determined that a linear combination of $\der_\tau\,{\cal F}$ and $i\,D^2\,{\cal F}$
might correspond to a two-height Adinkra only if the relative coefficient is not real.  It follows
that we must allow ${\cal F}$ to be a complex superfield.%
\Ft{Equivalently, we could add
another real prepotential so that we have two of these, say ${\cal F}_1$ and ${\cal F}_2$.
But these could be complexified by writing ${\cal F}={\cal F}_1+i\,{\cal F
}_2$, which ultimately amounts to the same thing.}  Thus, we have revised our original ansatz
and are now considering the case $\lambda=2$.  Now, we can form a new linear combination
of $\der_{\tau}\,{\cal F}$ and $i\,D^2{\cal F}$, this time including a relative factor of $i$ between
the two terms,
 \brr \Psi &\!\!\!:=\!\!\!&
      (\,-D^2+\der_\tau\,)\,{\cal F}.
 \label{lamdef}\err
 The component expansion for this superfield is
 \brr \Psi &\equals&
      F_{-1}^{(1)}+\fr12\,i\,\ve^{IJ}\,(\,F_0^{(0)}\,)_{IJ}
      \nonumber\\[.1in]
      & & +\theta^I\,\bpl\,(\,B_{-1/2}^{(1)}\,)_I
      -i\,\ve_{IJ}\,(\,B_{-1/2}^{(1)}\,)^J\,\bpr
      \nonumber\\[.1in]
      & & +\fr{1}{2!}\,\ve_{IJ}\,\theta^I\,\theta^J\,\bpl\,
      F_{-1}^{(2)}+\fr12\,i\,\ve^{KL}\,(\,F_0^{(1)}\,)_{KL}\,\bpr
      \,.
 \label{phi2}\err
This combination has a remarkable feature: the highest component of $\Psi$ is the $\tau$
derivative of its lowest component.  Thus, the highest component is completely determined by
data included at a lower level in the superfield.  In fact, the combinations which appear at
the lowest two levels, namely
 \brr (\,\widehat{B}_{-1/2}^{(0)}\,)_I &\!\!\!:=\!\!\!&
      (\,\delta_{IJ}-i\,\ve_{IJ}\,)\,(\,B_{-1/2}^{(0)}\,)^J
      \nonumber\\[.1in]
      \widehat{F}_0^{(0)} &\!\!\!:=\!\!\!&
      F_{-1}^{(1)}+\fr12\,i\,\ve^{IJ}\,(\,F_0^{(0)}\,)_{IJ} \,,
 \label{bf2}\err
describe a new supermultiplet spanning two engineering dimensions $d=-1/2$ and $d=0$.
Using the definitions~(\ref{bf2}) we can rewrite~(\ref{phi2}) as
 \brr \Psi &\equals&
      \widehat{F}_0^{(0)}+\theta^I\,(\,\widehat{B}_{-1/2}^{(1)}\,)_I
      +\fr12\,i\,\ve_{IJ}\,\theta^I\,\theta^J\,\widehat{F}_0^{(1)} \,.
 \label{lll}\err
Here $\widehat{F}_0^{(0)}$ is a complex scalar, and thus has two real degrees of freedom,
and $(\,\widehat{B}_{-1/2}^{(0)}\,)_I$, though it appears at first to have four real
degrees of freedom, really has two, for the following reason.

If we define $(\,P_\pm\,)_{IJ}:=\delta_{IJ}\pm i\epsilon_{IJ}$, then $(\,P_\pm\,)_I\,^J$ are
a pair of complementary projection operators, meaning $P_+{}^2=P_+$, $P_-{}^2=P_-$, $P_+P_-
=P_-P_+ = 0$, and $P_++P_-=1$.  The operator $P_+$ projects to the set of self-dual
one-forms and $P_-$ to the anti-self-dual one-forms.  Every one-form $\omega$ can be written as
$P_+\omega + P_-\omega$, where the first term is self-dual and the second term is
anti-self-dual.  The expression in~(\ref{bf2}) explicitly shows that
$\widehat{B}_{-1/2}^{(0)}$ is anti-self-dual, so the range of possibilities here is halved.
In other words, $\widehat{B}_{-1/2}^{(0)}$ satisfies the constraint
$P_+\widehat{B}_{-1/2}^{(0)}=0$.
Thus the components of~(\ref{bf2}) involve two bosonic
degrees of freedom and two fermionic degrees of freedom, or exactly half of those in
${\cal F}$.

\begin{figure}
 \begin{center}
 \includegraphics[width=4.5in]{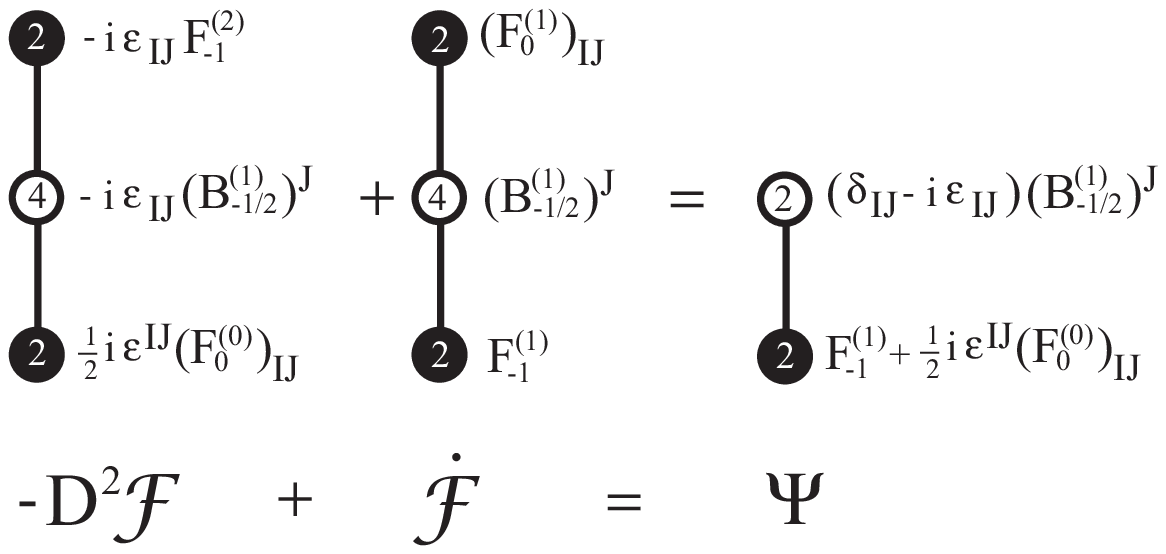}
 \caption{When adding together the Adinkras corresponding to
 the superfields $-D^2\,{\cal F}$ and $\der_\tau\,{\cal F}$,
 we take an appropriate linear combination of these superfields
 which induces a projection onto a sub-Adinkra corresponding to the
 superfield $\Psi$.  The fermionic degrees of freedom $F_{-1}^{(1)}
 -\fr12\,\ve^{IJ}\,(\,F_0^{(0)}\,)_{IJ}$ are removed
 by this process, as are the bosonic degrees of freedom
 $(\,\delta_{IJ}+i\,\ve_{IJ}\,)\,(\,B_{-1/2}^{(0)}\,)^J$,
 owing to the presence of the gauge symmetry shown in
~(\ref{gauge2}).}
 \label{lamadd}
 \end{center}
 \end{figure}

Another way to think of this situation is in terms of gauge equivalences.  It is
straightforward to see that $\widehat{F}_0^{(0)}$ and $(\,\widehat{B
}_{-1/2}^{(1)}\,)_I$, and therefore the entire superfield $\Psi$, are invariant under the
following gauge transformation:
 \brr \d\,(\,F_0^{(0)}\,)_{IJ} &\equals& \dot{\beta}_{IJ}
      \nonumber\\[.1in]
      \d\,(B_{-1/2}^{(0)}\,)_I &\equals&
      (\,\delta_{IJ}+i\,\ve_{IJ}\,)\,a^J
      \nonumber\\[.1in]
      \d\,F_{-1}^{(0)} &\equals&
      -\fr12\,i\,\ve^{IJ}\,\beta_{IJ} \,,
 \label{gauge2}\err
where $\beta_{IJ}$ is a complex two-form, describing two fermionic
degrees of freedom, and $a^J$ is a self-dual complex one-form
describing two bosonic degrees of freedom. We can use this freedom
to make a gauge choice
 \brr F_{-1}^{(0)} &\equals& 0
      \nonumber\\[.1in]
      (\,\delta_{IJ}+i\,\ve_{IJ}\,)\,(\,B_{-1/2}^{(0)}\,)^J &\equals& 0 \,.
 \err
But, the degrees of freedom removed from ${\cal F}$ by making this choice are also removed
by the operation described by~(\ref{lamdef}).  An Adinkrammatic depiction of~(\ref{lamdef}) is given
in Figure
 \ref{lamadd}.

\begin{figure}
 \begin{center}
 \includegraphics[width=3.3in]{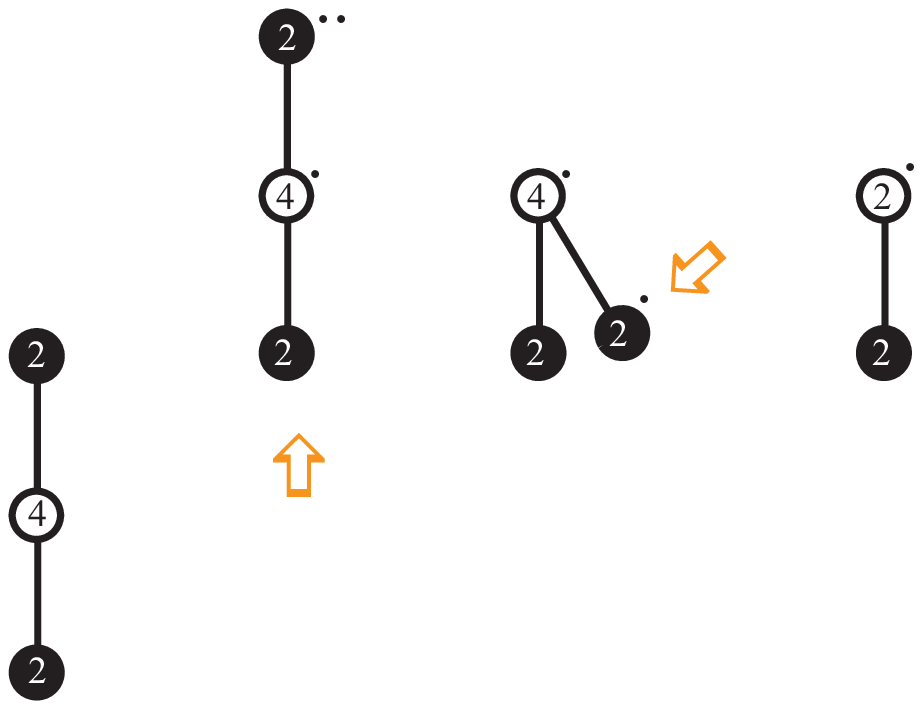}
 \caption{Schematically, the $N=2$ operator $(\,-D^2+\der_\tau\,)$ implements the
indicated operations on the Top Adinkra. First, it raises the Top Adinkra two height
units by a combination of swiveling and lifting. Then it lowers the topmost vertex, by
folding the raised Adinkra in half, removing half of the degrees of freedom in the
process.  What results is the gauge-invariant
 $\Psi$ Adinkra.}
 \label{ok2}
 \vspace{1.5in}
 \includegraphics[width=2.6in]{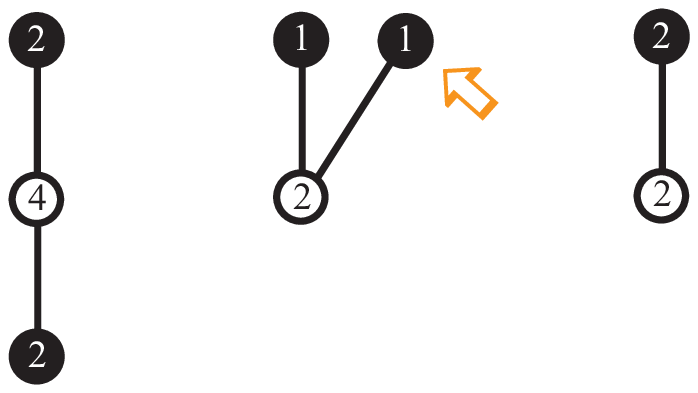}
 \caption{The $N=2$ operator $\fr12\,(\,\delta_{IJ}-i\,\ve_{IJ}\,)\,D^J$ implements the
indicated operations on the Top Adinkra. If lifts the lowermost vertex, and removes
half the degrees of freedom.   What results is the gauge-invariant $\Phi$ Adinkra.
The $\Psi$ Adinkra, shown in Figure \ref{ok2} is obtained by raising the lower vertex
in the $\Phi$ Adinkra, which is equivalent to swiveling $\Phi$ Adinkra $180^\circ$
about its hook.}
\label{op2}
\end{center}
\end{figure}

The superfield $\Psi$ describes the general solution to the
following constraint:
 \brr (\,\delta_{IJ}+i\,\ve_{IJ}\,)\,D^J\,\Psi=0 \,.
 \label{conlam}\err
Equivalently, $\Psi$ describes the projection~(\ref{lamdef}) of an unconstrained superfield
${\cal F}$.  Figure \ref{ok2} describes the Adinkrammatics associated with the projection~(\ref{lamdef}).
But the most important way to regard this supermultiplet, for the
purposes of our discussion, is via its relationship to the $N=2$
Scalar supermultiplet. The complex, anti-self-dual boson $(\,\widehat{
B}_{-1/2}^{(0)}\,)_I$ and the complex fermion $\widehat{F}_0^{(0)}$
describe the $2+2$ components corresponding to $\phi_i$ and
$\psi_{\hat{\imath}}$, as defined in~(\ref{s1}), respectively. The
precise correspondence is described immediately below in subsection
\ref{gr22}. The superfield $\Psi$, and its Adinkra, shown in Figure
\ref{op2}, correspond to a vertex-raised version of this Scalar
supermultiplet/Adinkra, since it is the derivative $(\,
\widehat{B}_{-1/2}^{(1)}\,)_I$ which contributes to these, rather
than the elemental field $(\,\widehat{B}_{-1/2}^{(0)}\,)_I$. A more
fundamental superfield which contains precisely
$(\,\widehat{B}_{-1/2}^{(0)}\,)_I$ and $\widehat{F}_0^{(0)}$ is
called $\Phi_I$, and is described in subsection \ref{n2ac}.  The
fact that this describes the $N=2$ Scalar supermultiplet is easy to
prove, since there is only one way to draw a two-height $N=2$
Adinkra where the lower components are bosons.

Note that equation~(\ref{conlam}) may be construed as a chirality
constraint, since we could regard the operator $(\,
\delta_{IJ}+i\,\ve_{IJ}\,)\,D^J$ as a complex superspace derivative.
Thus, the superfield $\Psi$ is an example of a ``chiral" superfield.
Note that in the context of $D=4$ $N=1$ supersymmetry (which is
equivalent to $D=1$ $N=4$ supersymmetry) similar superfields, along
with a differential condition analogous to~(\ref{conlam}) also
exist.

The $SO(2)$-invariant Levi-Civita tensor $\ve_{IJ}$ defines a
complex structure on the ``target space".  However, additional
algebraic structures suggested by the vector spaces defined in the
PpG diagram~(Figure \ref{ppgfig}) prove instrumental to the ability
to generalize the developments of this section to the context of
higher-$N$ supersymmetry.  In the following subsection we describe
some of this additional structure in the case $N=2$.

 \subsection{${\cal GR}(2,2)$ Structure}
 \label{gr22}
 The component transformation rules can be determined from
~(\ref{F2}) using
 $\d_Q(\e)=-i\,\e^I\,Q_I$,
 where $\e^I$ is an $SO(2)$ doublet of real supersymmetry parameters
 and $Q_I=i\,\der_I+\theta_I\,\der_\tau$ is the local superspace
 supersymmetry generator.  Accordingly, the superfield ${\cal F}$
 transforms as
 \brr \d_Q(\e)\,{\cal F}=\e^I\,(\,\der_I-i\,\theta_I\,\der_\tau\,)\,{\cal
      F} \,.
 \err
 Via explicit computation using~(\ref{F2}), this tells us
 \brr \d_Q\,F_{-1}^{(0)} &=&
      \e^I\,(\,B_{-1/2}^{(0)}\,)_I
      \nonumber\\[.1in]
      \d_Q\,(\,B_{-1/2}^{(0)}\,)_I &=&
      i\,\e^J\,(\,F_0^{(0)}\,)_{IJ}
      +i\,\e_I\,F_{-1}^{(1)}
      \nonumber\\[.1in]
      \d_Q\,(\,F_0^{(0)}\,)_{IJ} &=&
      -2\,\e_{[I}\,(\,B_{-1/2}^{(1)}\,)_{J]} \,.
 \label{sft}\err
 We can use~(\ref{sft}) to determine the
 corresponding transformations of the ``gauge-invariant" degrees of
 freedom defined in~(\ref{bf2}), with the result given by
 \brr \d_Q\,(\,\widehat{B}_{-1/2}^{(0)}\,)_I &=& i\,(\,\e_I-i\,\ve_{IJ}\,\e^J\,)\,
      \widehat{F}_0^{(0)}
      \nonumber\\[.1in]
      \d_Q\,\widehat{F}_0^{(0)} &=&
      (\,\e_I+i\,\ve_{IJ}\,\e^J\,)\,\der_\tau\,(\,\widehat{B}_{-1/2}^{(0)}\,)^I
      \,.
 \label{t6}\err
 This illustrates explicitly that the gauge-invariant
 fields do properly comprise a supersymmetry representation in and
 of themselves.
 Now consider the following independent real combinations, which
 suggestively package the $2+2$ invariant degrees of freedom,
 \brr \phi_1 &:=& {\rm Re}\,(\,\widehat{B}_{-1/2}^{(0)}\,)_2
      -{\rm Im}\,(\,\widehat{B}_{-1/2}^{(0)}\,)_1
      \nonumber\\[.1in]
      \phi_2 &:=&
      {\rm Im}\,(\,\widehat{B}_{-1/2}^{(0)}\,)_2
      +{\rm Re}\,(\,\widehat{B}_{-1/2}^{(0)}\,)_1
      \nonumber\\[.1in]
      \psi_{\hat{1}} &:=& {\rm Re}\,\widehat{F}_0^{(0)}
      \nonumber\\[.1in]
      \psi_{\hat{2}} &:=& {\rm Im}\,\widehat{F}_0^{(0)} \,.
 \err
 In terms of these, the transformation rules~(\ref{t6}) become
 \brr \d_Q\,\phi_1 &=&
      -i\,\e^1\,\psi_{\hat{2}}
      +i\,\e^2\,\psi_{\hat{1}}
      \nonumber\\[.1in]
      \d_Q\,\phi_2 &=&
      i\,\e^1\,\psi_{\hat{1}}
      +i\,\e^2\,\psi_{\hat{2}}
      \nonumber\\[.1in]
      \d_Q\,\psi_{\hat{1}} &=&
      \e^1\,\dot{\phi}_2
      +\e^2\,\dot{\phi}_1
      \nonumber\\[.1in]
      \d_Q\,\psi_{\hat{2}} &=&
      -\e^1\,\dot{\phi}_1
      +\e^2\,\dot{\phi}_2 \,.
 \label{pp2}\err
 Using the index conventions described in section \ref{Clifford},
 the ``physical" degrees of freedom, $\phi_i$ and
 $\psi_{\hat{\imath}}$, may be assigned values
 in ${\cal V}_L$ and ${\cal V}_R$, respectively, thereby
 exposing a natural ${\cal GR}(2,2)$ structure associated with this
 supermultiplet.  This is made all the
 more explicit if we make the following particular basis choice for
 the $N=2$ garden matrices,
 \brr L_1 &=& \ba{cc}0&1\\-1&0\ea
      \hspace{.4in}
      R_1 =\ba{cc}0&1\\-1&0\ea
      \nonumber\\[.1in]
      L_2 &=& \ba{cc}-1&0\\0&-1\ea
      \hspace{.4in}
      R_2 = \ba{cc}1&0\\0&1\ea \,.
 \err
Using these, the transformation rules~(\ref{pp2}) can be written
more concisely as
  \brr \d_Q(\e)\,\phi_i &=&
      -i\,\e^I\,(\,L_I\,)_i\,^{\hat{\jmath}}\,\psi_{\hat{\jmath}}
      \nonumber\\[.1in]
      \d_Q(\e)\,\psi_{\hat{\imath}} &=&
      \e^I\,(\,R_I\,)_{\hat{\imath}}\,^j\,\der_\tau\,\phi_j \,,
 \label{s2n2}\err
 which, as we recognize from~(\ref{s1}), precisely describes a Scalar
 supermultiplet.

 \subsection{An $N=2$ Invariant Action}
 \label{n2ac}
 A manifestly supersymmetric action can be written as follows:
 \brr S=\int d\tau \, d^2\theta \bpl\,
      \fr12\,i\,{\cal F}^\dagger\,\dot{\Psi}+{\rm h.c.}\,\bpr \,.
 \label{acok2}\err
Using the component expansion for ${\cal F}$, given in~(\ref{F2}) and the component
expansion for $\Psi$, given in~(\ref{phi2}), then performing the theta integration, we
determine the action in terms of the component fields,
 \brr S &\equals& \int d\tau \bpl\,
      -\fr12\,(\,\widehat{B}_{-1/2}^{(1)}\,)_I^\dagger
      (\,\widehat{B}_{-1/2}^{(1)}\,)^I
      -i \, \fr12\,(\,\widehat{F}_0^{(0)\,\dagger}\,\widehat{F}_0^{(1)}
      -\widehat{F}_0^{(1)\,\dagger}\widehat{F}_0^{(0)}\,)\,\bpr
      \,,
 \label{action2}\err
where $(\,\widehat{B}_{-1/2}\,)_I$ and $\widehat{F}_0$ are the gauge-invariant
combinations given in~(\ref{bf2}).  This demonstrates that~(\ref{acok2}) provides for a
canonical kinetic action for the component fields and also that this result is invariant
under the gauge transformation~(\ref{gauge2}).  Thus,~(\ref{acok2}) depends only on
the 2+2 gauge-invariant degrees of freedom defined in~(\ref{bf2}) rather than on the
full $4+4$ degrees of freedom described by ${\cal F}$.  By way of contrast, the
prepotential superfield ${\cal F}$ includes the physical as well as the spurious, gauge
degrees of freedom.

We could repeat the above analysis using bosonic prepotentials rather than fermionic
prepotentials.  However, it proves impossible in that case to use gauge invariant superfield
combinations to build a bosonic action analogous to~(\ref{action2}) which involves canonical
kinetic terms for the component fields.  The diligent reader can verify that the
same arguments and computations would then apply, but the bosonic and fermionic components
would be switched.  The resulting action~(\ref{action2}) would then be trivial.  Altering~(\ref{acok2}) turns out not to help, and this fact can be seen from dimensional arguments.

We have seen that the lowest component of $\Psi$ corresponds to the gauge-invariant
physical fermions.  We notice that the gauge-invariant physical bosons are the lowest
components of the anti self-dual part of $D_I\,{\cal F}$.  Accordingly, we define
 \brr \Phi_I &\!\!\!:=\!\!\!& \fr12\,(\,\delta_{IJ}-i\,\ve_{IJ}\,)\,D^J\,{\cal F} \,.
 \label{phitwo}\err
We then notice that
 \brr \Psi &\equals& -\ve^{IJ}\,D_I\,\Phi_J \,.
 \label{psitwo}\err
as is readily verified using~(\ref{lamdef}) along with the identity $D_ID_J=D_{[I}D_{J]}+
i\,\delta_{IJ}\,\der_\tau$.  Equations~(\ref{phitwo}) and~(\ref{psitwo}) describe the
gauge-invariant superfield analogs of the physical component fields described by
the prepotential ${\cal F}$.  The map from ${\cal F}$ to $\Phi_I$ is described Adinkrammatically
in Figure \ref{op2}.  The $\Phi_I$ Adinkra contains precisely the physical degrees of
freedom in this supermultiplet.  The $\Psi$ Adinkra, by way of contrast, is missing the zero
mode of the physical bosons.

 \section{Scalar Prepotentials for General $N$}
 \label{prepsca}
The strategy employed in the previous section ought to generalize to cases where  $
N>2$.  To determine how, it is helpful to reflect on how we managed to succeed in that
case.  This is made transparent by looking at Figure \ref{lamadd}, which we can re-write
in more streamlined form as follows:
 \brr
 \includegraphics[width=2.5in]{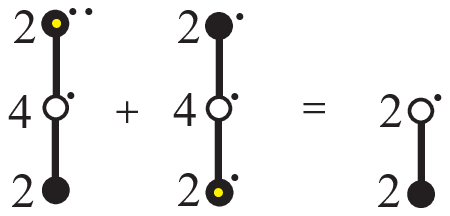} 
 \label{short2}\err
where the first term on the left-hand side is the Adinkra corresponding to $D^2\,{\cal F
}^{(0)}$, the second term is the Adinkra corresponding to ${\cal F}^{(1)}$, ignoring for
the moment numerical coefficients.  We have distinguished one fermionic vertex by adding
a yellow spot, which manifests the de facto orientation of the vertex chain.  This is helpful
for purposes of enabling inter-term vertex comparisons. (The spotted vertex is the source
of the ${\cal F}^{(0)}$ Adinkra; the fact that this vertex appears at the top of the $D^2\,{\cal
F}$ Adinkra rather than at  the bottom reflects the fact that the operator $D^2$ swivels the
 ${\cal F}$ Adinkra $180^\circ$ about its hook.)  Now we see that the vertex-wise addition
 of the Adinkras on the left-hand side of~(\ref{short2}) exhibits a promising feature: the
same two terms appear in the vertex sum at the lowest level as appear in the vertex
sum at the highest level, albeit with one extra dot at the higher level.  This suggests that
one might be able to choose the relative numerical coefficient between the two superfields corresponding
to these Adinkras in just such a way that their sum includes precisely the
same information at two different levels of the superfield.  This would then render the
highest vertex of the Adinkra sum superfluous.  This would also necessarily implement
a gauge projection, since the sum of two fields containing two degrees of freedom can
carry only two degrees of freedom itself.  Furthermore, since supersymmetry automatically
ensures a balance between fermionic and bosonic degrees of freedom, the gauge
projection on the fermionic vertices would necessarily be accompanied by a similar gauge
projection on the bosonic vertices.  What we showed above, via explicit computation, was
that all of this is, in fact, tractable.

Let's examine how this reasoning might generalize to higher $N$.  In general, a Top
Adinkra, which corresponds to a real prepotential superfield, say ${\cal F}^{(0)}$, spans
$N+1$ distinct height assignments.  We can attempt to reduce the span of this Adinkra
by adding ${\cal F}^{(N/2)}=\der_\tau^{N/2}{\cal F}^{(0)}$ to a swiveled version of ${
\cal F}^{(0)}$, corresponding to $D^N{\cal F}^{(0)}$, where $D^N=\fr{1}{N!}\,\ve^{I_1
\cdots I_N}D_{I_1}\cdots D_{I_N}$.  The purpose of the $N/2$ derivatives on ${\cal
F}^{(N/2)}$ is to ensure that the engineering dimension of this term coincides properly
with that of $D^{N}{\cal F}^{(0)}$.  Based on our previous analysis, we suspect that
this could succeed only if we used instead {\it at least} a pair of real superfields
${\cal F}_1$ and ${\cal F}_2$, or, to be more general, some set of superfields ${\cal
F}_n$ where $n=1,...,\lambda$, rather than a single real superfield  ${\cal F}$.  But the
best we could realistically hope for in this case would be to find a suitable coefficient
between each ${\cal F}_n^{(N/2)}$ term and the corresponding $D^N{\cal F}_n^{(0)}$
term so as to reduce the span from $N+1$ distinct height assignments to $N$, whereas
if we want to describe a Scalar supermultiplet, we have to reduce the span all the way to
two.

The problem then is how to include more terms in our $\Gamma$ expansion~(\ref{Gamdef}) in such a way
that not only do the engineering dimensions of each term
match up properly, but so too do the $\SO(N)$ tensor structures.
For example, both ${\cal F}^{(N/2)}$ and $D^N{\cal F}^{(0)}$ are $\SO(N)$ invariant,
but none of the other
possible terms $D_{[I_1}\cdots D_{I_p]}\,{\cal F}^{(N-p)/2}$, where $0<p<N$, are
$\SO(N)$ invariant, nor are contractions of these with the Levi-Civita symbol $\ve^{I_1
\cdots I_N}$.  It is here that the garden algebra saves the day.  Objects
precisely of the sort defined in~(\ref{fdef}) are indicated for this purpose.  These are
$p$-forms on $\SO(N)$ that are also operators on $\wedge {\cal GR}({\rm d},N)$.  We can
use these objects to add tensorial balance to more possible terms.  But if we do this,
it becomes necessary that the prepotentials take values in ${\cal V}_R$ or ${\cal V}_L$,
since elements of $\wedge {\cal GR}({\rm d},N)$ act on elements of ${\cal V}_L\oplus {\cal
V}_R$.  This, of course plays right into our hand, since, ultimately, the component fields
of the Scalar supermultiplets should take their values in precisely these vector spaces.  In
this way, we could, for instance, form an $\SO(N)$ singlet superfield by adding up terms
of the following sort:
 \brr \ve^{I_1\cdots I_N}\,D_{I_N}\cdots D_{I_{2\,s+2}}\,
      (\,f_{I_{2\,s+1}\cdots
      I_1}\,)_i\,^{\hat{\imath}}\,\der_\tau^s{\cal S}_{\hat{\imath}} \,.
 \label{pter}\err
 to define a superfield
 $\Phi_i:=(\,{\cal O}_\phi\,)_i\,^{\hat{\imath}}{\cal S}_{\hat{\imath}}$
 which would include only the fields of the Scalar supermultiplet.
 Alternatively, we could add up terms of the following sort:
 \brr \ve^{I_1\cdots I_N}\,D_{I_N}\cdots D_{I_{2\,s+1}}\,
      (\,\tilde{f}_{I_{2\,s}\cdots
      I_1}\,)_{\hat{\imath}}\,^{\hat{\jmath}}\,\der_\tau^s{\cal S}_{\hat{\jmath}} \,.
 \label{fter}\err
to define another superfield $\Psi_{\hat{\imath}}:=(\,{\cal O}_\psi\,)_{\hat{\imath}}\,^{
\hat{\jmath}}{\cal S}_{\hat{\jmath}}$ which would also include only the fields of the
Scalar supermultiplet albeit organized differently than in the field $\Phi_i$. It is natural
(and ultimately successful) to assume that there exist superfields $\Phi_i$ and
$\Psi_{\hat{\imath}}$, defined precisely in this way, with the properties that the lowest
components coincide precisely with the component fields $\phi_i$ and $\psi_{\hat{
\imath}}$ defined in~(\ref{s1}).

Based on the above discussion, we presuppose that a Scalar supermultiplet is described
by a set of $d$ unconstrained prepotential superfields ${\cal S}_{\hat{\imath}}$.  For
reasons similar to those described in Section~\ref{n2n2}, it turns out that when $N$ is
an even integer, the prepotential will be fermionic  while in cases where $N$ is an odd
integer, the prepotential will be bosonic.  Accordingly, when $N$ is even we will write
${\cal S}_{\hat{\imath}}:={\cal F}_{\hat{\imath}}$ and when $N$ is odd we will write
${\cal S}_{\hat{\imath}}:={\cal B}_{\hat{\imath}}$.  In terms of the prepotentials, we define
related superfields by
 \brr \Phi_i &\equals& (\,{\cal O}_\phi\,)_i\,^{\hat{\imath}}\,{\cal S}_{\hat{\imath}}
      \nonumber\\[.1in]
     \Psi_{\hat{\imath}} &\equals& (\,{\cal O}_\psi\,)_{\hat{\imath}}\,^{\hat{\jmath}}\,
     {\cal S}_{\hat{\jmath}}
      \,.
 \label{p1}\err
where $(\,{\cal O}_\phi\,)_i\,^{\hat{\imath}}$ and $(\,{\cal O}_\psi\,)_{\hat{\imath}}\,^{\hat{
\jmath}}$ are operators determined such that $\Phi_i\,|=\phi_i$ and $\Psi_{\hat{\imath
}}\,|=\psi_{\hat{\imath}}$, where $\phi_i$ and $\psi_{\hat{\imath}}$ are the particular
component fields appearing in~(\ref{s1}).  By making this definition, we imply that the
supersymmetry transformation rules induced on $\phi_i$ and $\psi_{\hat{\imath}}$ by
virtue of the fact that these are the lowest components of the superfields defined in~(\ref{p1}), via
the realization of a supersymmetry transformation on superspace,
correspond precisely with the component transformation rules given in~(\ref{s1}).
Among other things, this also implies
 \brr D_I\,\Phi_i &\equals&
      -i\,(\,L_I\,)_i\,^{\hat{\imath}}\,\Psi_{\hat{\imath}}
      \nonumber\\[.1in]
      D_I\,\Psi_{\hat{\imath}} &\equals&
      (\,R_I\,)_{\hat{\imath}}\,^i\,\der_\tau\,\Phi_i \,.
 \label{sst}\err
 Using~(\ref{sst}), it is straightforward to determine
 \brr D_{[I_1}\cdots D_{I_p]}\,\Psi_{\hat{\imath}}=
      \left\{\begin{array}{r}
      (-i)^{p/2}\,(\,\tilde{f}_{I_p\cdots I_1}\,)_{\hat{\imath}}\,^{\hat{\jmath}}\,
      \der_\tau^{p/2}\,\Psi_{\hat{\jmath}}
      \hspace{.2in};\,\,p\,\,{\rm even}\,,\\[.15in]
      (-i)^{(p-1)/2}\,(\,\tilde{f}_{I_p\cdots I_1}\,)_{\hat{\imath}}\,^j\,
      \der_\tau^{(p+1)/2}\,\Phi_j
      \hspace{.2in};\,\,p\,\,{\rm odd}\,. \end{array}\right.
 \label{usez}\err
 By substituting the definitions~(\ref{p1}) into~(\ref{sst}), and
 using the fact that the prepotentials ${\cal S}_{\hat{\imath}}$ are
 unconstrained, we obtain the following operator equations
 as corollaries of~(\ref{sst}),
 \brr -i\,(\,L_I\,)_i\,^{\hat{k}}\,(\,{\cal
      O}_\psi\,)_{\hat{k}}\,^{\hat{\jmath}}
      &\equals& D_I\,(\,{\cal O}_\phi\,)_i\,^{\hat{\jmath}}
      \nonumber\\[.1in]
      (\,R_I\,)_{\hat{\imath}}\,^k\,\der_\tau\,(\,{\cal
      O}_\phi\,)_k\,^{\hat{\jmath}}
      &\equals& D_I\,(\,{\cal O}_\psi\,)_{\hat{\imath}}\,^{\hat{\jmath}}
      \,.
 \err
Now, if we contract the first of these equations from the left with $(\,R_J\,)_{\hat{
l}}\,^i$ and then symmetrize on the indices $I$ and $J$, and use the garden
algebra~(\ref{garden}), we determine
 \brr (\,{\cal O}_\psi\,)_{\hat{\imath}}\,^{\hat{\jmath}} &\equals&
      -i\,\fr{1}{N}\,(\,R^I\,)_{\hat{\imath}}\,^k\,D_I\,(\,{\cal
      O}_\phi\,)_k\,^{\hat{\jmath}} \,.
 \label{consist}\err
Operating with~(\ref{consist}) on the prepotential ${\cal S}_{\hat{\imath}}$ allows
us to re-write this consistency condition as
 \brr \Psi_{\hat{\imath}} &\equals&
       -i\,\fr{1}{N}\,(\,R^I\,D_I\,\Phi\,)_{\hat{\imath}} \,.
 \label{conpsi}\err
Thus, the superfields $\Psi_{\hat{\imath}}$ can be expressed in a simple way, in
terms of the superfields $\Phi_i$.

How, then, can we determine the superfield $\Phi_i$, or, equivalently, the operator
$(\,{\cal O}_\phi\,)_i\,^{\hat{\imath}}$?  As it turns out, this problem is part and parcel
of the problem of finding an invariant action.  That problem, in turn, can be solved by
straightforward computation, by postulating that the logical choice for a manifestly
supersymmetric action, expressed as an integration over superspace of a particular
locally-defined superfield expression built using the ${\cal S}_{\hat{\imath}}$, is
equivalent to a demonstrably supersymmetric component action built using the
scalar component fields $\phi_i$ and $\psi_{\hat{\imath}}$.  Details are given
presently.

 \subsection{Invariant Actions}
A manifestly supersymmetric action can be written as
 \brr S=\int d\tau \, d^N\theta\,{\cal L}
 \err
where ${\cal L}$ is a locally-defined superfield ${\cal
L}$agrangian, built using the available superfields ${\cal
S}_{\hat{\imath}}$, $\Phi_i$ and/or $\Psi_{\hat{\imath}}$. We seek a
free action, which implies that ${\cal L}$ is bilinear in these
fields. Furthermore, our action should also be invariant under
$\wedge {\cal GR}({\rm d},N)$, so that indices $i$ and
$\hat{\imath}$ must be contracted.   There are exactly three
possible terms in this regard which also provide for a dimensionless
action.  The first is proportional to $\Phi^i\der_\tau^{\fr{4-N}{2}}
\Phi_i$, the second is proportional to ${\cal
S}^{\hat{\imath}}\,\der_\tau^{\fr{2+N}{2}}{\cal S}_{\hat{\imath}}$,
and the third is proportional to ${\cal
S}^{\hat{\imath}}\,\der_\tau\,\Psi_{\hat{\imath}}$.  In each case
the appropriate power of $\der_\tau$ is determined by dimensional
analysis.\Ft{ Since $[\,d\tau\,]=-1$ and $[\,d^N\theta\,]=\fr12\,N$,
it follows that $[\,S\,]=0$ only if $[\,{\cal L}\,]=1-\fr12\,N$. The
engineering dimensions of $\Phi_i$ and $\Psi_{\hat{\imath}}$ follow
from the requirements $\Phi_i\,|=\phi_i$ and
$\Psi_{\hat{\imath}}\,|=\psi_{\hat{\imath}}$, coupled with the fact
that the engineering dimension of a propagating boson is
$[\,\phi_i\,]=-\fr12$ and that of a propagating fermion is
$[\,\psi_{\hat{\imath}}\,]=0$, as explained in footnote \ref{engf}.
Since $(\,{\cal O}_\phi\,)_i\,^{\hat{\imath}}$ and $(\,{\cal
O}_\psi\,)_{ \hat{\imath}}\,^{\hat{\jmath}}$ are built using terms
of the sort~(\ref{pter}) and~(\ref{fter}), respectively, it follows
that $[\,{\cal S}_{\hat{\imath}}\,]=-\fr12\,N$. These facts suffice
for determining the appropriate power of $\der_\tau$ appearing in
the superfield products involving ${\cal S}_{\hat{\imath}}$,
$\Psi_{\hat{\imath}}$, and $\Phi_i$.}  The first  two possibilities
will, in general involve too many $\tau$ derivatives to provide for
a canonical kinetic component action. Accordingly, as a
well-motivated ansatz, we write the following,
 \brr S &\equals& i^{1-\alpha}\cdot i^{\bigl\lfloor\fr{N}{2}\bigr\rfloor}\,\int d\tau\,d^N\theta\,\bpl\,
        \fr12\,{\cal S}^{\hat{\imath}}\,
        \der_\tau\,\Psi_{\hat{\imath}}\,\bpr \,.
  \label{coml}\err
where $\alpha=0$ if $N$ is even and $\alpha=1$ if $N$ is odd, and where $\Psi_{\hat{
\imath}}=(\,{\cal O}_\psi\,{\cal S}\,)_i$, as described above.  The purpose of the
$N$-dependent phase in~(\ref{coml}) is to ensure that the action is real.  Now impose
that~(\ref{coml})  is equivalent to
  \brr S=\int d\tau\,\bpl\,\fr12\,\dot{\phi}^i\,\dot{\phi}_i
       -\fr12\,i\,\psi^{\hat{\imath}}\,\dot{\psi}_{\hat{\imath}}\,\bpr \,,
       \label{ScaLACT}
  \err
which is demonstrably invariant under~(\ref{s1}), owing to the relationship
$L_I=-R_I^T$.\Ft{The requirement that~(\ref{coml}) be invariant under~(\ref{s1}) is the underlying
motivation for the criterion $L_I=-R_I^T$.}  It is straightforward, using standard superspace
techniques, and a little bit of algebra, to re-write~(\ref{coml}) as
  \brr S
      &\equals&  \int d\tau\,\left.\bpl\,
      -\fr12\,(\,{\cal O}_\phi\,{\cal S}\,)^i\,\ddot{\Phi}_i
      -\fr12\,i\,(\,{\cal O}_\psi\,{\cal
      S}\,)^{\hat{\imath}}\,\dot{\Psi}_{\hat{\imath}}\,\bpr\,\right| \,.
 \err
where $(\,{\cal O}_\phi\,)_i\,^{\hat{\jmath}}$ and $(\,{\cal O}_\psi\,)_{\hat{\imath}}\,^{\hat{
\jmath}}$ are determined as
  \brr (\,{\cal O}_\phi\,)_i\,^{\hat{\imath}} &\equals&
       i^{1-\alpha}\cdot i^{\bigl\lfloor\fr{N}{2}\bigr\rfloor}\,\frac{1}{N!}\,\ve^{I_1\cdots
       I_N}\,\sum_{s=0}^{\bigl\lfloor\fr{N-1}{2}\bigr\rfloor}\,
       {N\choose 2\,s+1}\,
      D_{I_N}\cdots D_{I_{2\,s+2}}\,
      (\,f_{I_{2\,s+1}\cdots
      I_1}\,)_i\,^{\hat{\imath}}\,(\,i\,\der_\tau)^s
      \nonumber\\[.1in]
      (\,{\cal O}\,_\psi\,)_{\hat{\imath}}\,^{\hat{\jmath}} &\equals&
      i^{2-\alpha}\cdot i^{\bigl\lfloor\fr{N}{2}\bigr\rfloor}\,\frac{1}{N!}\,\ve^{I_1\cdots I_N}\,
      \sum_{s=0}^{\bigl\lfloor\fr{N}{2}\bigr\rfloor}\,
      {N\choose 2\,s}\,
      D_{I_N}\cdots D_{I_{2\,s+1}}\,
      (\,\tilde{f}_{I_{2\,s}\cdots
      I_1}\,)_{\hat{\imath}}\,^{\hat{\jmath}}\,(\,i\,\der_\tau\,)^s \,.
 \label{projectors}\err
An explicit derivation of this is shown explicitly in Appendix \ref{derx}.  It is gratifying
to check that the operators in~(\ref{projectors}) properly satisfy the consistency condition~(\ref{consist}).

\begin{figure}
 \begin{center}
 \includegraphics[width=7in]{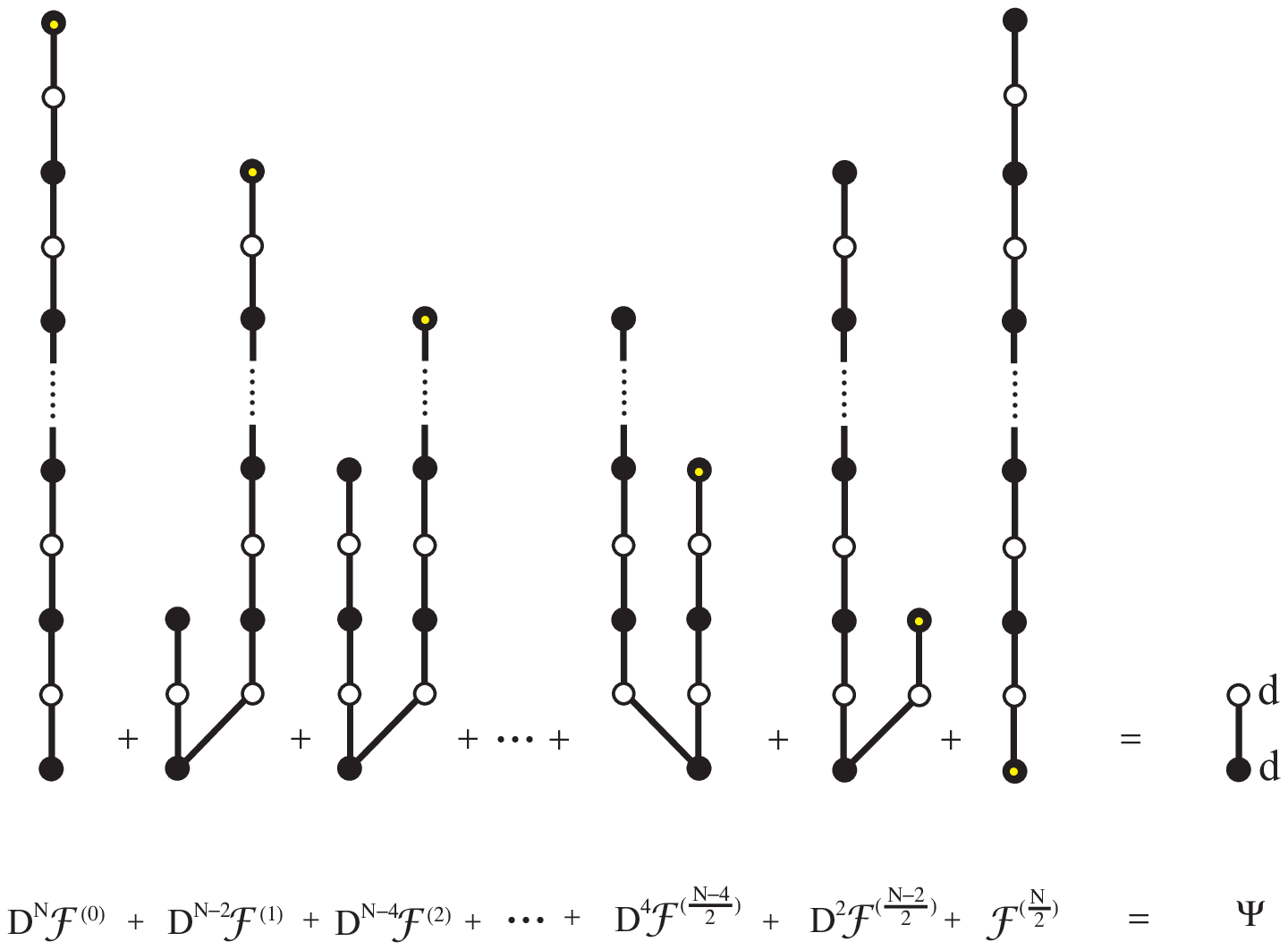}
 \caption{An Adinkrammatic picture of the $\Psi$ projection
in the case where $N$ is even.  (The odd-$N$ diagram is similar.) This is a generalization
of the $N=2$ version appearing in Figure \ref{lamadd}.  A yellow dot has been placed in a
vertex to distinguish one end of the chain, for purposes of vertex comparison.  Most of the vertices
in this Figure should have dots, illustrating that they have some number of $\tau$ derivatives.
But these have been suppressed in this rendering so as to minimize clutter. Similarly, all of the
$\wedge{\cal GR}({\rm d},N)$ indices and vertex multiplicities have been suppressed. (In addition to the binomial coefficient vertex multiplicities corresponding to the groupings into $p$-forms,
each of the Adinkras on the left hand side appears with an overall multiplicity of $d$, corresponding to the $d$ dimensions of ${\cal V}_{R}$.) The purpose
of this diagram is to give a course-grained picture of how vertices line up, level-by-level, when
the various terms in~(\ref{projectors}) are added together, taking an appropriate linear combination of the corresponding fields.}
 \label{psiproj}
 \end{center}
 \end{figure}

\begin{figure}
 \begin{center}
 \includegraphics[width=5in]{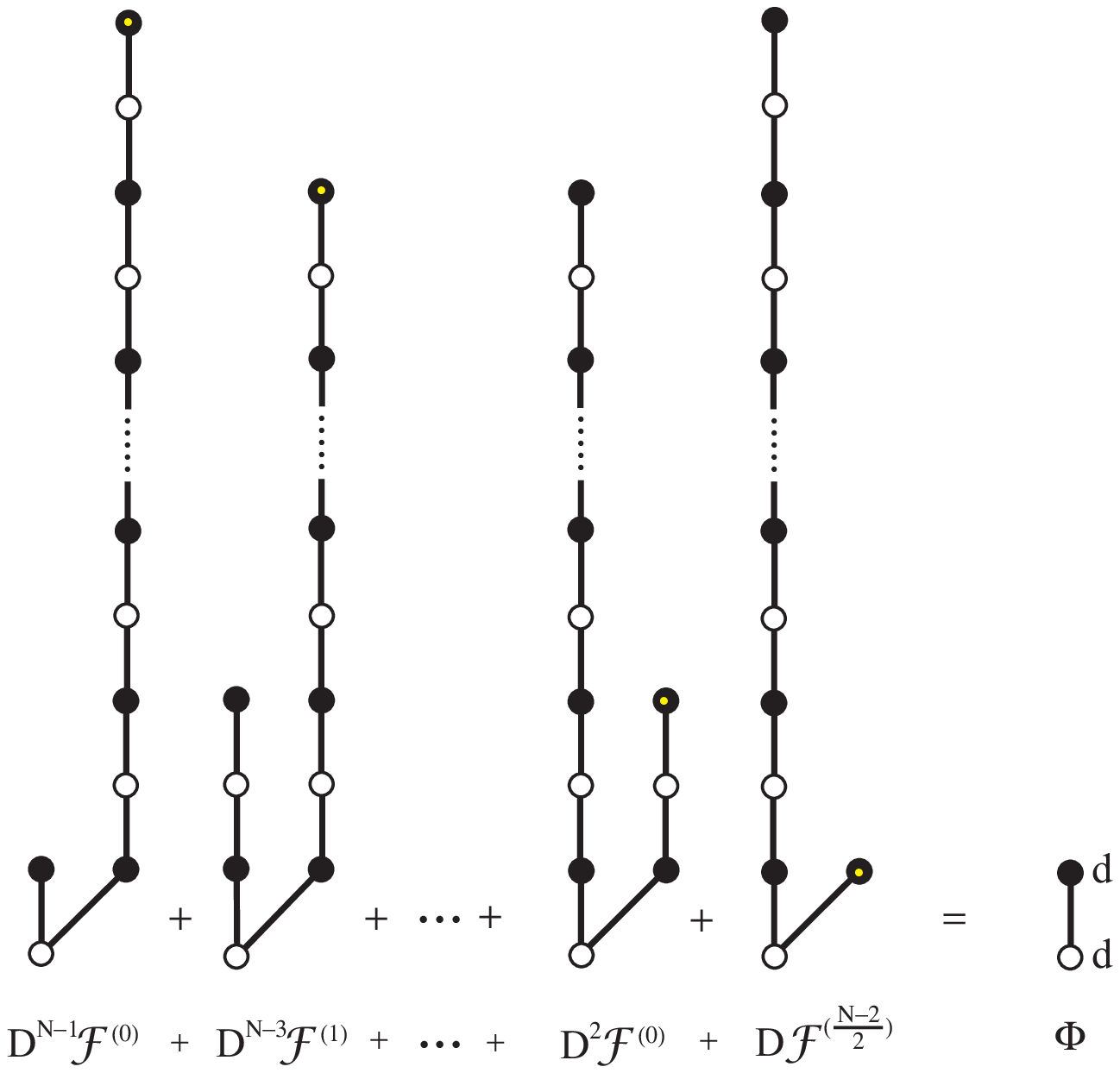}
 \caption{An Adinkrammatic picture of the $\Phi$ projection in
the case where $N$ is even. (The odd-$N$ diagram is similar.) A yellow dot has been placed
in a vertex to distinguish one end of the chain, for purposes of vertex comparison.  Most of the
vertices in this Figure should have dots, illustrating that they have some number of $\tau$
derivatives.  But these have been suppressed in this rendering so as to minimize clutter.
Similarly, all of the $\wedge{\cal GR}({\rm d},N)$ indices and vertex multiplicities have been
suppressed. (In addition to the binomial coefficient vertex multiplicities corresponding to the groupings into $p$-forms,
each of the Adinkras on the left hand side appears with an overall multiplicity of $d$, corresponding to the $d$ dimensions of ${\cal V}_{R}$.) The purpose of this diagram is to give a course-grained picture of how vertices
line up, level-by-level, when the various terms in~(\ref{projectors}) are added together,
taking an appropriate linear combination of the corresponding fields.}
 \label{phiproj}
 \end{center}
 \end{figure}

A graphical depiction which illustrates how it is that the projectors $(\,{\cal O}_\psi\,)_{
\hat{\imath}}\,^{\hat{\jmath}}$ and $(\,{\cal O}_\phi\,)_i\,^{\hat{\jmath}}$ can work their magic
is given in Figures \ref{psiproj} and \ref{phiproj}, in the case where $N$ is even and the
prepotentials ${\cal S}_{\hat{\imath}}:={\cal F}_{\hat{\imath}}$ are fermionic.  To (superficially)
see how this works, consider Figure~\ref{psiproj}.  The very first and the very last terms in
on the left-hand side of this figure correspond (schematically) to $D^N{\cal F}^{(0)}$ and
${\cal F}^{(N/2)}$.  We explained above how the relative coefficient between these terms
can be tuned so that the degrees of freedom corresponding to the sum of the uppermost
vertices is merely a differentiated version of the same sum appearing in the sum of the
lowermost vertices.  Therefore, the uppermost vertices are projected out upon summation.
As an extra bonus, bosons at the second to the top level are also projected away automatically
as a consequence of supersymmetry. The puzzle facing us previously was how could we
continue to diminish the span of the summed Adinkra.  This is resolved by looking at the
second term and the second to last term on the left-hand side of the figure.  These correspond
to $D^{N-2}\, {\cal F}^{(1)}$ and $D^2{\cal F}^{(\fr{N-2}{2})}$, respectively.  These Adinkra
terms already have their spans reduced by two as compared to the  outside terms
considered previously.  As a consequence, when these new inside terms are included in
the sum they cannot undo any of the projections already accomplished at the highest Adinkra
levels.  But the second Adinkra and the second to last Adinkra can now have their relative
coefficient tuned to as to implement a projection on fermionic and bosonic vertices at lower heights.  Continuing this process inward, the coefficients can be tuned so as to implement a zippering
action, removing vertices all the way down to the lowest two levels, precisely what is needed to
describe the Scalar Adinkra.

  \begin{table}
 \begin{center}
 \begin{tabular}{|c||c|c|c|c|c|c|c|c|}
 \hline
 &&&&&&&&\\[-.2in]
 $N$ & 1&2&3&4&5&6&7&8\\
 \hline
 \hline
 &&&&&&&&\\[-.2in]
 $d_N$ & 1&2&4&4&8&8&8&8\\
 \hline
 &&&&&&&&\\[-.2in]
 $g_N$ & 0&4&24&56&240&496&1008&2032\\
 \hline
 \end{tabular}
 \caption{The number of gauge degrees of freedom $g_N$ included in the
 Scalar supermultiplet superfields grows large as $N$ increases.}
 \label{gauge}
 \end{center}
 \end{table}

  \subsection{Gauge Transformations}
 The supersymmetric action~(\ref{coml}) can be re-written as
 \brr S=\int d\tau\,d^N\theta\,\fr12\,{\cal S}^{\hat{\imath}}\,{\Tilde {\cal K}}_{
      \hat{\imath}}\,^{\hat{\jmath}}\,{\cal S}_{\hat{\jmath}}
      \,,
  \label{Act}
 \err
 where ${\Tilde {\cal K}} = {\cal O}_\psi\,\der_\tau$ or, more
 specifically,
 \brr {\Tilde {\cal K}}_{\hat{\imath}}\,^{\hat{\jmath}} &\equals&
      i^{1-\alpha}\cdot i^{\bigl\lfloor\fr{N}{2}\bigr\rfloor}\,\frac{1}{N!}\,\ve^{I_1\cdots I_N}\,
      \sum_{s=0}^{\bigl\lfloor\fr{N}{2}\bigr\rfloor}\,
      {N\choose 2\,s}\,
      D_{I_N}\cdots D_{I_{2\,s+1}}\,
      (\,\tilde{f}_{I_{2\,s}\cdots
      I_1}\,)_{\hat{\imath}}\,^{\hat{\jmath}}\,(\,i\,\der_\tau\,)^{s+1}
       \label{K-R}
 \err
is the superspace kinetic operator acting on Salam-Strathdee superfields which are
elements of ${\cal V}_R$.

A given Scalar supermultiplet includes $d+d$ degrees of freedom.
These are explicitly seen in equation~(\ref{ScaLACT}).  It is also
seen from this expression that there are {\em {no gauge}} symmetries
associated with this component action.  Nevertheless, as we shall
now show, the action in~(\ref{coml}) describes a gauge theory.   We
have shown how the component fields can be packaged using a set of
$d$ unconstrained prepotential superfields ${\cal
S}_{\hat{\imath}}$. However, each prepotential includes
$2^{N-1}+2^{N-1}$ degrees of freedom.  Thus, each Scalar
supermultiplet prepotential construction includes
 \brr g_N=(\,2^N-2\,)\,d
 \err
gauge degrees of freedom which do not appear in the action.  The corresponding gauge
structure can be described by the transformation ${\cal S}_{\hat{\imath}}\to {\cal S}_{
\hat{\imath}} +\delta\,{\cal S}_{\hat{\imath}}$, where
 \brr
 {\Tilde {\cal K}}_{\hat{\imath}}\,^{\hat{\jmath}}\,\delta\,{\cal
 S}_{\hat{\jmath}} &\equals& 0 \,.
  \label{gsymm}
 \err
Thus, there is a portion of ${\cal S}_{\hat{\imath}}$ which is annihilated by the kinetic operators
${\Tilde {\cal K}}_{\hat{\imath}}\,^{\hat{\jmath}}$.  It is possible to use the gauge freedom parametrized
by $\delta\,{\cal S}_{\hat{\imath}}$ to choose a gauge in which the {\em {only}} component
fields that occur in ${\cal S}_{\hat{\imath}}$ are the $d$ bosons and $d$ fermions in~(\ref{ScaLACT}).  This defines the so-called ``Wess-Zumino'' gauge for the prepotentials.
With the realization that  ${\cal S}_{\hat{\imath}}$ is a gauge field, it follows that the superfields
$\Phi_i$ and $\Psi_{\hat{\imath}}$ defined in~(\ref{p1}) are field strength superfields ({\it i.e.}
invariants under the transformation ${\cal S}_{\hat{\imath}}\to {\cal S}_{\hat{\imath}} +\delta\,{\cal
S}_{\hat{\imath}}$).

These kinetic energy operators also lead to the construction of projection operators (see
appendix B for a simple and well-known example of this process).  The construction of these
projection operators begins by noting that
 \brr
   {\Tilde {\cal K}}_{\hat{i}}\,^{\hat{j}}  \,  {\Tilde {\cal K}}_{\hat{j}}\,^{\hat{k}}
 &\equals& [\partial_{\tau}]^{{\fr 12 } N + 1} \,
  {\Tilde {\cal K}}_{\hat{i}}\,^{\hat{k}}
 \label{squareK-R}
 \err
 (and the similarity between~(\ref{MaxWell2}) and~(\ref{squareK-R}) is obvious).  Thus,
 it is natural to define a projection operator via
   \brr
 [ \,  {\Tilde {\cal P}}{}^{(SM)} {}]_{\hat{i}}\,^{\hat{k}}
 &\equals& [\partial_{\tau}]^{-{\fr 12 } N - 1} \,
  {\Tilde {\cal K}}_{\hat{i}}\,^{\hat{k}}        {~~~~~~~~~~}
  \nonumber\\[.1in]
  &\equals&
   i^{1-\alpha}\cdot i^{\bigl\lfloor\fr{N}{2}\bigr\rfloor}\,\frac{1}{N!}\,\ve^{I_1\cdots I_N}\,
   \sum_{s=0}^{\bigl\lfloor\fr{N}{2}\bigr\rfloor}\,
   {N\choose 2\,s}\,
   D_{I_N}\cdots D_{I_{2\,s+1}}\,
   (\,\tilde{f}_{I_{2\,s}\cdots
   I_1}\,)_{\hat{\imath}}\,^{\hat{\jmath}}\,(\,i\,\der_\tau\,)^{s -{\fr 12 } N }\, \, \, .  {~~~~~~}
 \label{acc10}\err
This projection operator permits another characterization of the gauge variation
of the prepotential.  The gauge variation can be written in the form
\brr
\delta\,{\cal S}_{\hat{\jmath}} &\equals&  {\{} \, \, \d{}_{\hat{j}}\,^{\hat{k}} ~-~ [ {\Tilde {\cal
P}}{}^{(SM)} {}]_{\hat{j}}\,^{\hat{k}} \, {\}} \, \Lambda_{\hat{k}} ~\equiv~   {\{} \, \,
[ {\Tilde {\cal P}}{}^{(nSM)} {}]_{\hat{j}}\,^{\hat{k}}  \, {\}} \, \Lambda_{\hat{k}}
~~.
 \label{gaugevar}
 \err
where $\Lambda_{\hat{k}} $ is a superfield not subject to any restrictions.  The
condition in~(\ref{gsymm}) is satisfied due to the equations in~(\ref{squareK-R})
and~(\ref{acc10}).

We thus reach the conclusion that the projection operator defined by~(\ref{K-R}) and~(\ref{acc10}) is associated with the general $N$ version of the Adinkra that appears in~(\ref{s4}).  The general version of this Adinkra has a number of $N$ distinct colored links
connecting $d_N$ bosonic vertices and $d_N$ fermionic vertices.  All the bosonic vertices have a
common height and all the fermionic vertices have a common height.

The operator~(\ref{K-R}) is clearly defined to act upon any Salam-Strathdee superfield
element of ${\cal V}_R$.  On the other hand, a similar kinetic energy operator ${{\cal
K}}_{{\imath}} \,^{{\jmath}}$ and projection operator $[{ {\cal P}}{}^{(SM)} {}]_{\hat{i}
}\,^{\hat{k}}$ defined by
  \brr {{\cal K}}_{{\imath}}\,^{{\jmath}} &\equals&
      i^{1-\alpha}\cdot i^{\bigl\lfloor\fr{N}{2}\bigr\rfloor}\,\frac{1}{N!}\,\ve^{I_1\cdots I_N}\,
      \sum_{s=0}^{\bigl\lfloor\fr{N}{2}\bigr\rfloor}\,
      {N\choose 2\,s}\,
      D_{I_N}\cdots D_{I_{2\,s+1}}\,
      (\,{f}_{I_{2\,s}\cdots
      I_1}\,)_{{\imath}}\,^{{\jmath}}\,(\,i\,\der_\tau\,)^{s+1}
       \label{K-L}
 \err
    \brr
 [ \,  { {\cal P}}{}^{(SM)} {}]_{{i}}\,^{{k}}
 &\equals&
   i^{1-\alpha}\cdot i^{\bigl\lfloor\fr{N}{2}\bigr\rfloor}\,\frac{1}{N!}\,\ve^{I_1\cdots I_N}\,
   \sum_{s=0}^{\bigl\lfloor\fr{N}{2}\bigr\rfloor}\,
   {N\choose 2\,s}\,
   D_{I_N}\cdots D_{I_{2\,s+1}}\,
   (\,{f}_{I_{2\,s}\cdots
   I_1}\,)_{{\imath}}\,^{{\jmath}}\,(\,i\,\der_\tau\,)^{s -{\fr 12 } N }\, \, \, .  {~~~~~~}
 \label{acc5}\err
can act upon any Salam-Strathdee superfield element of ${\cal V}_L$.

 \section{Conclusions}
We have illustrated quite explicitly how the dynamics associated with arbitrary-$N$
Scalar supermultiplets can be described using superspace actions involving unconstrained
prepotential superfields.  The prepotentials typically include many spurious degrees
of freedom upon which their action~(\ref{Act}) does not depend,
and which describe an inherent gauge structure.
We have also described a methodology, based on these developments,
which should, in principle, enable one to construct prepotential descriptions of any
one-dimensional supermultiplet.  This methodology associates one prepotential
superfield to each hook, {\it i.e.}, sink, on the Adinkra corresponding to the
supermultiplet, where the ${\cal GR}({\rm d},N)$ or $\SO(N)$ assignments of these
supermultiplets correlate with the corresponding assignments of the Adinkra hooks.

The discovery of the explicit forms of the operators $[ {\Tilde
{\cal P}}{}^{(SM)} {}]_{\hat{j}}\,^{ \hat{k}} $ and $[ {\Tilde {\cal
P}}{}^{(nSM)} {}]_{\hat{j}}\,^{\hat{k}} $ are new results and among
the most important in our longer program of using the mathematical
structure of garden algebras and Adinkras to penetrate the still
unknown complete structure of irreducible representations of
Salam-Strathdee superspace.  The former of these operators yields an
irreducible representation while the latter does not.  It remains a
major task to understand completely the structure of the
representations which remain after projection with $[ {\Tilde {\cal
P}}{}^{(nSM)} {}]_{\hat{j}}\,^{\hat{k}} $.  These operators are
examples of 1D superprojectors similar to those introduced
in\cite{GS1982}.   In fact, we may specialize to the case of 1D, $N$
= 4, the superprojector $[ {\Tilde {\cal P}}{}^{(SM)}
{}]_{\hat{j}}\,^{\hat{k}} $ above.  This result may then be compared
with the dimensional reduction on a 0-brane of the 4D, $N$ = 1
superprojectors given in equation (3.11.18) of\cite{ggrs}.  Such an
investigation will be undertaken at some future date in order, at
least in this special case, to unravel the representations contained
in $[ {\Tilde {\cal P}}{}^{(nSM)} {}]_{\hat{j}}\,^{\hat{k}} $.
Summarizing this aspect of the present work, we may say that we have
presented the first existence proof for extending the concept of
superprojectors to 1D arbitrary $N$-extended Salam-Strathdee
superfields.  Further exploration of this topic is of vital
importance to our future studies.

Our work can also be used to highlight another issue for future study.  We have shown
by starting from an action of the form of~(\ref{coml}), it is possible to reach one of the form
of~(\ref{Act}).  In the first of these actions, $S^{\hat i}$ represents a Top Adinkra while
$\Psi_{\hat i}$ represents a Base Adinkra.  This two objects have other names in the
conventional discussion of superfield theories.  The former are known as ``unconstrained
prepotentials'' while the latter are called ``superfield field strengths''.   It is a fact, that in
every successful quantization of a supersymmetrical theory {\em {in which supersymmetry
is manifest in all steps}}, there always occur Top Adinkras that allow the passage from the
analogs of~(\ref{coml}) to~(\ref{Act}).   Thus, Top Adinkras are vital in all known manifestly supersymmetrical
quantization procedures.  This naturally raises a question, ``If the Base
Adinkra is replaced by some other Adinkra, is it always possible to begin with the
analog~(\ref{coml}) and arrive at the analog of~(\ref{Act})?''  If the answer is ``No'', then
such a theory cannot be quantized in a manner that keeps supersymmetry manifest
by any known method.  We believe that such a question is very relevant to the issue
of off-shell central charges, a very old topic in the supersymmetry literature.

\begin{flushright}
 \parbox{4in}{\leavevmode\llap{``}%
  \sl The heart's a startin',\\[-1.5mm]
      And this crown comment, the action so meant\\[-1.5mm]
      To be used but could not, now spry to foment!''\\[-1mm]
       --- Shawn Benedict Jade}
\end{flushright}

\paragraph{\bf Acknowledgments:}
 The research of S.J.G.\ is supported in part by the National Science
Foundation Grant PHY-0354401.
 T.H.\ is indebted to the generous support of the Department of
Energy through the grant DE-FG02-94ER-40854.

 \appendix

 \section{Explicit Computation}
 \label{derx}
 This Appendix includes an explicit derivation of the projection
 operators appearing in~(\ref{projectors}), in the specific case where $N$ is even, so
 that the prepotentials are fermionic.  Thus ${\cal
 S}_{\hat{\imath}}:={\cal F}_{\hat{\imath}}$.  The calculation for
 odd $N$ is similar.

 It is well known that a superspace integration $\int d^N\theta$
 is equivalent to taking the $\theta^I\to 0$ limit of the $N$th superspace
 derivative, {\it i.e.}, $\int d^N\theta {\cal L}=
 (-1)^{\lfloor N/2\rfloor}\,D^N\,{\cal
 L}\,|$, where $D^N:=\fr{1}{N!}\,\ve^{I_1\cdots I_N}\,D_{I_1}\cdots
 D_{I_N}$.
  Accordingly, we can re-write~(\ref{coml}) as $S=\int
 d\tau\,L$, where the component Lagrangian is given by
 \brr L &\equals& (-1)^{\lfloor N/2\rfloor}\,i^{1-\alpha}\,i^{\bigl\lfloor\fr{N}{2}\bigr\rfloor}\,\fr12\,D^N\,(\,{\cal
      S}^{\hat{\imath}}\,\dot{\Psi}_{\hat{\imath}}\,)\,|
 \err
 Now, if we distribute the $N$ derivatives in $D^N$ by operating to
 the right, we obtain
 \brr L &\equals& i^{1-\alpha}\,i^{\bigl\lfloor\fr{N}{2}\bigr\rfloor}\,\fr12\,\fr{1}{N!}\,\ve^{I_1\cdots
       I_N}\,\sum_{p=0}^N
      (-1)^{p\,(1-\alpha)}\,{N\choose p}\,
      (\,D_{I_N}\cdots D_{I_{p+1}}\,{\cal S}^{\hat{\imath}}\,)\,
      (\,D_{I_p}\cdots D_{I_1}\,\dot{\Psi}_{\hat{\imath}}\,)\,|
 \err
 If we then use the results~(\ref{usez}), we can re-write this as
 \brr L &\equals&
      i^{1-\alpha}\,i^{\bigl\lfloor\fr{N}{2}\bigr\rfloor}\,\fr12\,\fr{1}{N!}\,\ve^{I_1\cdots I_N}\,\bpl\,\sum_{p\,\,{\rm odd}}
      (-1)^{1-\alpha}\,{N\choose p}\,(\,D_{I_N}\cdots D_{I_{p+1}}\,{\cal
      S}^{\hat{\imath}}\,)\,(-i)\,^{(p-1)/2}\,(\,\tilde{f}_{I_1\cdots I_p}\,)_{\hat{\imath}}\,^i
      \der_\tau^{(p+1)/2}\,\dot{\Phi}_i
      \nonumber\\[.1in]
      & & \hspace{1.6in}
      +\sum_{p\,\,{\rm even}}
      {N\choose p}\,(\,D_{I_N}\cdots D_{I_{p+1}}\,{\cal
      S}^{\hat{\imath}}\,)\,(-i)^{p/2}\,(\,\tilde{f}_{I_1\cdots
      I_p}\,)_{\hat{\imath}}\,^{\hat{\jmath}}\,
      \der_\tau^{p/2}\,\dot{\Psi}_{\hat{\jmath}}\,\bpr \,|
 \err
 Integrating by parts, this becomes
 \brr L &\cong& i^{1-\alpha}\,i^{\bigl\lfloor\fr{N}{2}\bigr\rfloor}\,\fr12\,\fr{1}{N!}\,\ve^{I_1\cdots I_N}\,\bpl\,
      (-1)^{1-\alpha}\,\sum_{p\,\,{\rm odd}}
      i^{(p-1)/2}\,{N\choose p}\,(\,\der_\tau^{(p-1)/2}\,D_{I_N}\cdots D_{I_{p+1}}\,{\cal
      S}^{\hat{\imath}}\,)\,(\,\tilde{f}_{I_1\cdots I_p}\,)_{\hat{\imath}}\,^i
      \,\ddot{\Phi}_i
      \nonumber\\[.1in]
      & & \hspace{1.4in}
      +\sum_{p\,\,{\rm even}}\,
      i^{p/2}\,
      {N\choose p}\,(\,\der_\tau^{p/2}\,D_{I_N}\cdots D_{I_{p+1}}\,{\cal
      S}^{\hat{\imath}}\,)\,(\,\tilde{f}_{I_1\cdots
      I_p}\,)_{\hat{\imath}}\,^{\hat{\jmath}}\,
      \dot{\Psi}_{\hat{\jmath}}\,\bpr \,|
 \err
 Using a symmetry property, $(\,\tilde{f}_{I_1\cdots
 I_p}\,)_{\hat{\imath}}\,^j=-(\,f_{I_p\cdots I_1}\,)^j\,_{\hat{\imath}}$, which follows
 from the definitions~(\ref{fdef}) and the property $L_I=-R_I^T$, this becomes
 \brr L
      &\equals&  i^{1-\alpha}\,i^{\bigl\lfloor\fr{N}{2}\bigr\rfloor}\,\fr12\,\fr{1}{N!}\,\ve^{I_1\cdots I_N}\,\bpl\,
      -(-1)^{1-\alpha}\,\sum_{p\,\,{\rm odd}}
      i^{(p-1)/2}\,{N\choose p}\,(\,\der_\tau^{(p-1)/2}\,D_{I_N}\cdots D_{I_{p+1}}\,{\cal
      S}^{\hat{\imath}}\,)\,(\,f_{I_p\cdots I_1}\,)^i\,_{\hat{\imath}}
      \,\ddot{\Phi}_i
      \nonumber\\[.1in]
      & & \hspace{1.5in}
      +\sum_{p\,\,{\rm even}}\,
      i^{p/2}\,
      {N\choose p}\,(\,\der_\tau^{p/2}\,D_{I_N}\cdots D_{I_{p+1}}\,{\cal
      S}^{\hat{\imath}}\,)\,(\,\tilde{f}_{I_p\cdots
      I_1}\,)^{\hat{\jmath}}\,_{\hat{\imath}}\,
      \dot{\Psi}_{\hat{\jmath}}\,\bpr \,|
 \err
 Redefining the dummy indices which are summed over, we can re-write
 this as
 \brr L
      &\equals& i^{1-\alpha}\,i^{\bigl\lfloor\fr{N}{2}\bigr\rfloor}\,\fr12\,\fr{1}{N!}\,\ve^{I_1\cdots
      I_N}\,\bpl\,-(-1)^{1-\alpha}\,\sum_{s=0}^{\bigl\lfloor\fr{N-1}{2}\bigr\rfloor}
      i^{s}\,{N\choose 2\,s+1}\,(\,\der_\tau^{s}\,D_{I_N}\cdots D_{I_{2\,s+2}}\,{\cal
      S}^{\hat{\imath}}\,)\,(\,f_{I_{2\,s+1}\cdots I_1}\,)^i\,_{\hat{\imath}}
      \,\ddot{\Phi}_i
      \nonumber\\[.1in]
      & & \hspace{1.5in}
      -i\,\sum_{s=0}^{\bigl\lfloor\fr{N}{2}\bigr\rfloor}\,
      i^{s+1}\,
      {N\choose 2\,s}\,(\,\der_\tau^{s}\,D_{I_N}\cdots D_{I_{2\,s+1}}\,{\cal
      S}^{\hat{\imath}}\,)\,(\,\tilde{f}_{I_{2\,s}\cdots
      I_1}\,)^{\hat{\jmath}}\,_{\hat{\imath}}\,
      \dot{\Psi}_{\hat{\jmath}}\,\bpr \,|
 \label{penult}\err
 This form is organized as
 \brr L
      &\equals&  \bpl\,
      -\fr12\,(\,{\cal O}_\phi\,{\cal S}\,)^i\,\ddot{\Phi}_i
      -\fr12\,i\,(\,{\cal O}_\psi\,{\cal
      S}\,)^{\hat{\imath}}\,\dot{\Psi}_{\hat{\imath}}\,\bpr\,| \,.
 \label{ult}\err
 By comparing~(\ref{penult}) to~(\ref{ult}), we can read off the
 definitions of $(\,{\cal O}_\phi\,)_i\,^{\hat{\jmath}}$ and
 $(\,{\cal O}_\psi\,)_{\hat{\imath}}\,^{\hat{\jmath}}$, with the
 result given in~(\ref{projectors}).

 \section{From Maxwell Theory to Projectors}
 \label{projec}
The purpose of this appendix is to demonstrate within the simplest known gauge
theory--Maxwell theory--that the presence of the kinetic energy term in the action of
necessity leads to the existence of projection operators.  In order to illustrate this
property, it may be useful to review this process in the more familiar arena of 4D
non-supersymmetric Maxwell theory.  The usual action can be written in the form
\brr  {\cal S}_{Maxwell} &\equals& -\, \fr 14 \, \int d^4 x ~
F^{{\2a} \, {\2b} } \, F_{{\2a} \, {\2b} }
~=~  \fr 12 \, \int d^4 x ~ A^{\2a}  [ \,  \delta{}_{\2a}
{}^{\2b} \, \partial^{\2c} \, \partial_{\2c}
~-~  \partial_{\2a}  \, \partial^{\2b}
\, ] A_{\2b}  \nonumber\\[.1in]
& \!\!\!\equiv\!\!\!&
 \fr 12 \, \int d^4 x ~ A^{\2a}  \, {\cal K} _{\2a} {}^{\2b} \,
 A_{\2b}
\label{MaxWell1}
 \err
 and a simple calculation reveals
 \brr
  {\cal K} _{\2a} {}^{\2b} \,  {\cal K} _{\2b} {}^{\2c}
  ~=~  \partial^{\2d} \, \partial_{\2d} \, {\cal K} _{\2a}
  {}^{\2c}
 ~=~ {\bo} \, {\cal K} _{\2a}
  {}^{\2c}   ~~~.
 \label{MaxWell2}
 \err
 This implies that a new operator may be defined
 \brr
  {{\cal P}^{(T)}} _{\2a} {}^{\2b} ~=~ \frac{1}{\, \bo \, } \,
   \, {\cal K} _{\2a}
  {}^{\2b}   ~~~\to~~
 {{\cal P}^{(T)}} _{\2a} {}^{\2b} \,  {{\cal P}^{(T)}} _{\2b}
 {}^{\2c}  ~=~  {{\cal P}^{(T)}} _{\2a}
  {}^{\2c}   ~~~,
 \label{MaxWell3}
 \err
and it is also well known that there exist solutions $\d A_{\2a} $ which satisfy
$ {\cal K} _{\2a} {}^{\2b} \, \d A_{\2b} $ $=$ $0$.  These
solutions can be written as
\brr
\d A_{\2b} ~=~  {\big [} \,{\d} _{\2b} {}^{\2c} ~-~
 {{\cal P}^{(T)}} _{\2b} {}^{\2c} \, {\big]} \, \Lambda{}_{\2c}
 ~\equiv~
 {{\cal P}^{(L)}} _{\2b} {}^{\2c} \,  \Lambda{}_{\2c}
  ~~~,
 \label{MaxWell4}
 \err
which can be seen to be solution upon using the second result in~(\ref{MaxWell3}).  Upon
use of the definition of $ {{\cal P}^{(T)}}$, this takes the form
\brr
\d A_{\2b} ~=~  {\big [} \,{\d} _{\2b} {}^{\2c} ~-~ [ \,  \delta{}_{\underline
b} {}^{\2c} ~-~    \frac{1}{\, \bo \, } \,  \, \partial_{\2b}  \, \partial^{\2c}
\, ]  \, {\big]} \, \Lambda{}_{\2c} ~=~ \partial_{\2b} {\big \{ } \,
 \frac{1}{\, \bo \, }  \, \, \partial^{\2c}
\,  \, \Lambda{}_{\2c}   \, {\big\} }
     ~~~,
 \label{MaxWell5}
 \err
After making a field definition $\Lambda_{\2c}  \,=\,   \partial_{\2c} \, \Lambda$, this
takes the familiar form of a Maxwell gauge transformation $\d A_{\2b} =
\partial_{\2b} \Lambda$.  The operators $ {{\cal P}^{(T)}}$ and $ {{\cal P}^{(L)}}$
are projection operators since
\brr
&~& {{\cal P}^{(L)}} _{\2a} {}^{\2b} \,  {{\cal P}^{(L)}} _{\2b} {}^{\2c}
  ~=~  {{\cal P}^{(L)}} _{\2a}
  {}^{\2c}   ~~~,~~~
 {{\cal P}^{(L)}} _{\2a} {}^{\2b} \,  {{\cal P}^{(T)}} _{\2b} {}^{\2c}
  ~=~ 0  ~=~  {{\cal P}^{(T)}} _{\2a} {}^{\2b} \,  {{\cal P}^{(L)}} _{\2b} {}^{\2c}   ~~~,  \nonumber\\[.1in]
& ~&
 {{\cal P}^{(L)}} _{\2a} {}^{\2b} \, +\, {{\cal P}^{(T)}} _{
\2a} {}^{\2b}  ~=~  \d _{\2a} {}^{\2b}  ~~~.
 \label{PROJs}
 \err
The first projector $ {{\cal P}^{(T)}}$ is known as the ``transverse'' projector and the second
$ {{\cal P}^{(L)}}$ is known as the ``longitudinal'' projector.  The usual Maxwell action can
thus be written as
\brr  {\cal S}_{Maxwell} &\equals&  \fr 12 \, \int d^4 x ~ A^{\2a}  \,
 {\bo} \, \, {{\cal P}^{(T)}} _{\2a} {}^{\2b} \,
 A_{\2b}
\label{MaxWell6}
 \err

Although we picked Maxwell theory to show the relation between the kinetic energy operator
in a gauge theory and the presence of projection operators, any theory defined over an ordinary
manifold may be chosen as the starting point of a similar discussion.  Projection operators
are a ubiquitous feature of gauge theory.  The only feature that may surprise the reader
is that in the context of a supersymmetrical theory, even if none of the component fields in
the theory (as those in the Scalar supermultiplet) are gauge fields, none the less, these fields
are the components of a {\em {gauge superfield}}.

\newpage
 \Refs{References}{[00]}

 \Bib{DFGHIL01} C.~Doran, M.~Faux, S.~J.~Gates, Jr., T.~H{\"u}bsch, K.~Iga, G.~Landweber:
 {\em On Graph Theoretic Identifications of
 Adinkras, Supersymmetry Representations and Superfields},
  math-ph/0512016.
 \Bib{DFGHIL00} C.~Doran, M.~Faux, S.~J.~Gates, Jr., T.~H{\"u}bsch, K.~Iga, G.~Landweber:
 {\em  Off-Shell Supersymmetry and Filtered Clifford Supermodules},
  math-ph/0603012.
 \Bib{FG1} M.~Faux and S.~J.~Gates, Jr.:
 {\em Adinkras: A Graphical Technology for Supersymmetric Representation Theory},
  Phys.~Rev.~{\bf D71}~(2005),~065002, hep-th/0408004.
 \Bib{GK} S.~J.~Gates, Jr. and S.~V.~Ketov: {\em $2D$ $(4,4)$ Hypermultiplets (II): Field Theory Origins of Duality}, Phys.~Lett.~{\bf B418}~(1998),~no.~1--2,~119--124, hep-th/9504077.
 \Bib{GR1} S.~J.~Gates, Jr. and L.~Rana:
 {\em A Theory of Spinning Particles for Large $N$-Extended Supersymmetry},
  Phys.~Lett.~{\bf B352}~(1995),~no.~1--2,~50--58, hep-th/9504025.
  \Bib{GR2} S.~J.~Gates, Jr. and L.~Rana:
 {\em A Theory of Spinning Particles for Large $N$-Extended Supersymmetry (II)},
  Phys.~Lett.~{\bf B369}~(1996),~no.~3--4,~262--268, hep-th/9510051.
 \bibitem{enuf}
 S.~J.~Gates, Jr., W.~D.~Linch, III and J.~Phillips:
 {\em When Superspace Is Not Enough},
 hep-th/0211034.
 \Bib{ggrs} S.~J.~Gates, Jr., M.~T.~Grisaru, M.~Ro${\check{\rm c}}$ek, and
 W.~Siegel:
 {\em Superspace or One Thousand and One Lessons in Supersymmetry},
 The Banjamin/Cummings Pub.~Co.~Inc., Reading, MA, 1983.
 \Bib{KRT}  Z.~Kuznetsova, M.~Rojas, and F.~Toppan:
 {\em Classification of irreps and invariants of the $N$-extended Supersymmetric
 Quantum Mechanics},
  JHEP~{\bf0603}~(2006),~098, hep-th/0511274.
 \Bib{DF} D.S.~Freed: {\em Five lectures on supersymmetry}, American Mathematical Society, Providence, RI, 1999.
\Bib{PD} P.~Deligne, P.~Etingof, D.S.~Freed, L.C.~Jeffrey, D.K.~Kazhdan, J.W.~Morgan, D.R.~Morrison, and E.~Witten, Eds.: {\em Quantum fields and strings: a course for mathematicians. Vol. 1, 2}, American Mathematical Society, Providence, RI, 1999.
 \Bib{GS1982} S.~J.~Gates, Jr. and W.~Siegel:
 {\em Linearized $N=2$ Superfield Supergravity},
 Nucl.~Phys.~{\bf B189}~(1982)~39.

 \endRefs

 \end{document}